\documentclass[%
 aip,
 amsmath,amssymb,
 reprint,%
]{revtex4-1}

\usepackage[english]{babel}

\usepackage{graphicx}
\usepackage{dcolumn}
\usepackage{bm}

\usepackage[utf8]{inputenc}
\usepackage[T1]{fontenc}
\usepackage{mathptmx}
\usepackage{etoolbox}
\usepackage{gnuplottex}
\usepackage{comment}
\usepackage{wrapfig}
\usepackage{fixltx2e}
\usepackage{hyperref}
\usepackage{ stmaryrd }
\usepackage{soul}
\makeatletter
\def\convertto#1#2{\strip@pt\dimexpr #2*65536/\number\dimexpr 1#1}
\makeatother

\usepackage{tikz}
\usetikzlibrary{external}
\usetikzlibrary{arrows,decorations.pathmorphing}
\usepackage{graphics}
\usepackage{mwe,tikz}
\usepackage[percent]{overpic}

\usepackage{cancel} 
\usepackage{pgfplots}
\pgfplotsset{compat=1.18}
\usepackage[font=small,labelfont=bf,justification=justified, format=plain]{caption}
\usepackage{caption}
\usepackage{subcaption}
\usepackage{sansmath} 

\usepackage{pict2e} 

\usepackage{paralist}       
\usepackage{xcolor}         
\usepackage{amssymb }
\usepackage{latexsym}                  
\usepackage{keyval}                    
\usepackage{moreverb}                  
\usepackage{gnuplottex}
\usepackage{amsmath}
\usepackage{physics}
\usepackage{chemfig}
\definecolor{myorange}{RGB}{3,145,155}
\definecolor{myred}{HTML}{db5856}
\definecolor{myblue}{HTML}{8093F1}
\definecolor{1blue1}{RGB}{1,4,93}
\definecolor{2blue2}{RGB}{50,150,150}

\usepackage{amsfonts}

\usepackage{caption}

\usepackage{ragged2e}

\makeatletter
\def\@email#1#2{%
 \endgroup
 \patchcmd{\titleblock@produce}
  {\frontmatter@RRAPformat}
  {\frontmatter@RRAPformat{\produce@RRAP{*#1\href{mailto:#2}{#2}}}\frontmatter@RRAPformat}
  {}{}
}%
\makeatother
\begin{document}

\preprint{AIP/123-QED}

\title{Simulating spectrally resolved exciton--exciton-interaction pump--probe spectroscopy}
\author{Kate\v{r}ina Charv\'{a}tov\'{a}}%
\affiliation{Faculty of Mathematics and Physics, Charles University, Prague, Czech Republic}

\author{Matteo Bruschi}%
\affiliation{Faculty of Mathematics and Physics, Charles University, Prague, Czech Republic}

\author{Pavel Mal\'{y}*}%
\email{pavel.maly@matfyz.cuni.cz}%
\affiliation{Faculty of Mathematics and Physics, Charles University, Prague, Czech Republic}


\date{\today}

\begin{abstract}
Exciton--exciton-interaction pump–probe spectroscopy (EEI–PP) captures the fifth-order signal of light-matter interaction, directly probing exciton–exciton annihilation. This allows studying excitation energy transport even in extended, nearly isoenergetic landscapes. Although spectrally resolved fifth-order signals have already been explored in simplified cases, in larger systems only spectrally integrated signals have been analyzed. 
We present a microscopic theoretical description for spectrally resolved EEI–PP response with realistic lineshapes, revealing how single- and multi-exciton state structure shape the spectral features and dynamics. The description is applicable to large molecular systems such as squaraine polymers, facilitating a direct comparison with experimental spectra. By a decomposition of the total signal into individual excitation pathways, we demonstrate that the spectral resolution is needed to distinguish otherwise intertwined single- and multi-exciton dynamics, such as single-exciton relaxation and exciton--exciton annihilation.

\end{abstract}

\maketitle

\section{\label{sec:introduction}Introduction}
In many molecular aggregates, such as squaraine polymers,\cite{lambert2015energy}  molecules are strongly coupled. Upon absorption, the system transitions into an electronic excited state, forming an exciton.\cite{valkunas_photosynthetic_2000, May2011, Valkunas2013} Due to the excitonic coupling, the exciton is delocalized over several monomer units of the polymer. This enables efficient energy transfer through the system, which is an essential aspect of materials for organic solar cells.\cite{lambert2015energy, giannini2022exciton, carrillo2025progress}  Such systems can be described in two representations. In the diabatic site basis, excitations are localized on individual molecules, corresponding to states where one molecule is excited, while all others remain in their ground state. In contrast, the adiabatic exciton basis comprises eigenstates of a system taking into account the exciton delocalization between the individual molecules. These two bases coincide in the limit of negligible coupling. See Fig. \ref{fig:site_exciton_basis} for a detailed description of the used notation.

The probability of generating multiple excitons in the same molecular aggregate increases with excitation intensity, enabling the study of interactions between them.\cite{maly2023separating, bruschi_multi-exciton_2026} Excitons transfer through the aggregate and, upon encounter, exciton--exciton annihilation (EEA) can occur, leading to a loss of one of the excitons.\cite{valkunas1995nonlinear} This process can be illustrated in the site basis (see Fig. \ref{fig:EEA_scheme}a). Initially, two molecules in the system simultaneously occupy their first excited electronic state. During EEA, energy is transferred from a donor molecule, leaving it in the ground state, to an acceptor molecule, exciting it to a higher excited electronic state.\cite{valkunas1995nonlinear, Malyshev1999, Ryzhov2001} The other molecule quickly transitions back to the first excited state by internal conversion (see Fig. \ref{fig:EEA_scheme}b). Therefore, this process leads to a loss of excitation, possibly described in the multiexcitonic basis as a transfer from a double-excited state to a single-excited state (see Fig. \ref{fig:EEA_scheme}c).\cite{renger_multiple_1997, renger_theory_1997, bruggemann_microscopic_2001, bruggemann_exciton_2003} The effective rate of annihilation depends on several factors, such as the structure of the molecular complex and the energy of the higher excited states.\cite{Tempelaar2017}

\begin{figure}[]
{\sffamily
\sansmath
\begin{tikzpicture}[scale=0.9]
\draw[gray] (-2,3) -- (-2,-7.5);
\draw[gray] (-0.7,2.5) -- (-0.7,-7.5);
\draw[gray] (7.55,3) -- (7.55,-7.5);
\draw[gray] (5.2,3) -- (5.2,-7.5);

\draw[gray] (-2,3) -- (7.55,3);
\node [] at (2,2.75) {site basis};
\node [] at (6.4,2.75) {exciton basis};
\draw[gray] (-2,2.5) -- (7.55,2.5);
\draw[] (0,1) -- (1.,1);
\draw[] (0,1.5) -- (1.,1.5);
\draw[] (0,2) -- (1.,2);

\node [] at (-0.3,1.) {$|g_i\rangle$};
\node [] at (-0.3,1.5) {$|e_i\rangle$};
\node [] at (-0.3,2.) {$|f_i\rangle$};

\draw[fill] (0.5,1.5) circle (3pt);

\draw[] (2,1) -- (3.,1);
\draw[] (2,1.5) -- (3,1.5);
\draw[] (2,2) -- (3,2);

\node [] at (1.7,1.) {$|g_j\rangle$};
\node [] at (1.7,1.5) {$|e_j\rangle$};
\node [] at (1.7,2.) {$|f_j\rangle$};

\draw[fill] (2.5,1) circle (3pt);

\draw[] (4,1) -- (5.,1);
\draw[] (4,1.5) -- (5.,1.5);
\draw[] (4,2) -- (5.,2);

\node [] at (3.7,1.) {$|g_k\rangle$};
\node [] at (3.7,1.5) {$|e_k\rangle$};
\node [] at (3.7,2.) {$|f_k\rangle$};

\draw[fill] (4.5,1) circle (3pt);

\node [] at (-1.35,1.5) {$|i\rangle$};

\draw[gray] (-2,0.5) -- (7.55,0.5);
\draw[] (0,-1) -- (1.,-1);
\draw[] (0,-0.5) -- (1.,-0.5);
\draw[] (0,0) -- (1.,0);

\node [] at (-0.3,-1.) {$|g_i\rangle$};
\node [] at (-0.3,-0.5) {$|e_i\rangle$};
\node [] at (-0.3,0.) {$|f_i\rangle$};

\draw[fill] (0.5,-0.5) circle (3pt);

\draw[] (2,-1) -- (3.,-1);
\draw[] (2,-0.5) -- (3,-0.5);
\draw[] (2,0) -- (3,0);

\node [] at (1.7,-1.) {$|g_j\rangle$};
\node [] at (1.7,-0.5) {$|e_j\rangle$};
\node [] at (1.7,0) {$|f_j\rangle$};

\draw[fill] (2.5,-0.5) circle (3pt);

\draw[] (4,-1) -- (5.,-1);
\draw[] (4,-0.5) -- (5.,-0.5);
\draw[] (4,0) -- (5.,0);

\node [] at (3.7,-1.) {$|g_k\rangle$};
\node [] at (3.7,-0.5) {$|e_k\rangle$};
\node [] at (3.7,0.) {$|f_k\rangle$};

\draw[fill] (4.5,-1) circle (3pt);

\node [] at (-1.35,-0.5) {$|ij\rangle$};

\draw[gray] (-2,-1.5) -- (5.2,-1.5);
\draw[] (0,-3) -- (1.,-3);
\draw[] (0,-2.5) -- (1.,-2.5);
\draw[] (0,-2.0) -- (1.,-2);

\node [] at (-0.3,-3.) {$|g_i\rangle$};
\node [] at (-0.3,-2.5) {$|e_i\rangle$};
\node [] at (-0.3,-2) {$|f_i\rangle$};

\draw[fill] (0.5,-2.0) circle (3pt);

\draw[] (2,-3) -- (3.,-3);
\draw[] (2,-2.5) -- (3,-2.5);
\draw[] (2,-2) -- (3,-2);

\node [] at (1.7,-3.) {$|g_j\rangle$};
\node [] at (1.7,-2.5) {$|e_j\rangle$};
\node [] at (1.7,-2) {$|f_j\rangle$};

\draw[fill] (2.5,-3) circle (3pt);

\draw[] (4,-3) -- (5.,-3);
\draw[] (4,-2.5) -- (5.,-2.5);
\draw[] (4,-2) -- (5.,-2);

\node [] at (3.7,-3.) {$|g_k\rangle$};
\node [] at (3.7,-2.5) {$|e_k\rangle$};
\node [] at (3.7,-2) {$|f_k\rangle$};

\draw[fill] (4.5,-3) circle (3pt);

\node [] at (-1.35,-2.5) {$|f_i\rangle$};

\draw[gray] (-2,-3.5) -- (7.55,-3.5);
\draw[] (0,-5) -- (1.,-5);
\draw[] (0,-4.5) -- (1.,-4.5);
\draw[] (0,-4.0) -- (1.,-4);

\node [] at (-0.3,-5.) {$|g_i\rangle$};
\node [] at (-0.3,-4.5) {$|e_i\rangle$};
\node [] at (-0.3,-4) {$|f_i\rangle$};

\draw[fill] (0.5,-4.5) circle (3pt);

\draw[] (2,-5) -- (3.,-5);
\draw[] (2,-4.5) -- (3,-4.5);
\draw[] (2,-4) -- (3,-4);

\node [] at (1.7,-5.) {$|g_j\rangle$};
\node [] at (1.7,-4.5) {$|e_j\rangle$};
\node [] at (1.7,-4) {$|f_j\rangle$};

\draw[fill] (2.5,-4.5) circle (3pt);

\draw[] (4,-5) -- (5.,-5);
\draw[] (4,-4.5) -- (5.,-4.5);
\draw[] (4,-4) -- (5.,-4);

\node [] at (3.7,-5.) {$|g_k\rangle$};
\node [] at (3.7,-4.5) {$|e_k\rangle$};
\node [] at (3.7,-4) {$|f_k\rangle$};

\draw[fill] (4.5,-4.5) circle (3pt);

\node [] at (-1.35,-4.5) {$|ijk\rangle$};

\draw[gray] (-2,-5.5) -- (5.2,-5.5);
\draw[] (0,-7) -- (1.,-7);
\draw[] (0,-6.5) -- (1.,-6.5);
\draw[] (0,-6.0) -- (1.,-6);

\node [] at (-0.3,-7.) {$|g_i\rangle$};
\node [] at (-0.3,-6.5) {$|e_i\rangle$};
\node [] at (-0.3,-6) {$|f_i\rangle$};

\draw[fill] (0.5,-6.0) circle (3pt);

\draw[] (2,-7) -- (3.,-7);
\draw[] (2,-6.5) -- (3,-6.5);
\draw[] (2,-6) -- (3,-6);

\node [] at (1.7,-7.) {$|g_j\rangle$};
\node [] at (1.7,-6.5) {$|e_j\rangle$};
\node [] at (1.7,-6) {$|f_j\rangle$};

\draw[fill] (2.5,-6.5) circle (3pt);

\draw[] (4,-7) -- (5.,-7);
\draw[] (4,-6.5) -- (5.,-6.5);
\draw[] (4,-6) -- (5.,-6);

\node [] at (3.7,-7.) {$|g_k\rangle$};
\node [] at (3.7,-6.5) {$|e_k\rangle$};
\node [] at (3.7,-6) {$|f_k\rangle$};

\draw[fill] (4.5,-7) circle (3pt);

\node [] at (-1.35,-6.5) {$|f_ij\rangle$};

\draw[gray] (-2,-7.5) -- (7.55,-7.5);


\draw[thick,domain=5.4:7.4,samples=200,smooth]
    plot (\x,{1. + 0.1*sin(720*(\x-5.7)/1.1)});

\fill[red,opacity=0.3] (6.4,1.) circle (0.30);

\node [] at (6.4,2.) {$|\alpha\rangle$};

\node [] at (6.4,-0.5) {$|\beta\rangle$};

\draw[thick,domain=5.4:7.4,samples=200,smooth]
    plot (\x,{-1.5 + 0.1*sin(720*(\x-5.7)/1.1)});

\fill[red,opacity=0.3] (5.9,-1.5) circle (0.3);
\fill[red,opacity=0.3] (6.9,-1.5) circle (0.3);

\draw[thick,domain=5.4:7.4,samples=200,smooth]
    plot (\x,{-2.5 + 0.1*sin(720*(\x-5.7)/1.1)});

\fill[orange,opacity=0.35] (6.4,-2.5) circle (0.30);

\node [] at (6.4,-4.2) {$|\xi\rangle$};

\draw[thick,domain=5.4:7.4,samples=200,smooth]
    plot (\x,{-5.2 + 0.1*sin(720*(\x-5.7)/1.1)});

\fill[red,opacity=0.3] (5.7,-5.2) circle (0.28);
\fill[red,opacity=0.3] (6.4,-5.2) circle (0.28);
\fill[red,opacity=0.3] (7.1,-5.2) circle (0.28);

\draw[thick,domain=5.4:7.4,samples=200,smooth]
    plot (\x,{-6.0 + 0.1*sin(720*(\x-5.7)/1.1)});

\fill[red,opacity=0.3] (5.9,-6.0) circle (0.3);
\fill[orange,opacity=0.35] (6.9,-6.0) circle (0.30);

\draw[thick,domain=5.4:7.4,samples=200,smooth]
    plot (\x,{-6.8 + 0.1*sin(720*(\x-5.7)/1.1)});

\fill[blue,opacity=0.35] (6.4,-6.8) circle (0.3);

\end{tikzpicture}
}
\caption{\justifying Notation for single-, double- and triple-excited states in site (left column) and exciton basis (right column). In the site basis, states are denoted by Latin letters $i$,$j$,$k$. Single-excited states are denoted as $| i \rangle $, where the $i$-th molecule is in its first excited state and the other molecules are in the ground state. Double-excited states in the site basis correspond to either states $|ij\rangle$, where two molecules $i$ and $j$ are simultaneously in their first excited state, or $|f_i\rangle$, where one molecule is in its higher-excited state and the others are in their ground state. Triple-excited states in site basis correspond to either state $|ijk\rangle$, where three molecules are excited, or to state $|f_ij\rangle$ with the $i$-th molecule in the higher-excited state and the $j$-th molecule in first-excited state. Instead, states in the exciton basis are indicated using Greek letters $\alpha$ for single-excited state, $\beta$ for double-excited state and $\xi$ for triple-excited state.}
\label{fig:site_exciton_basis}
\end{figure}
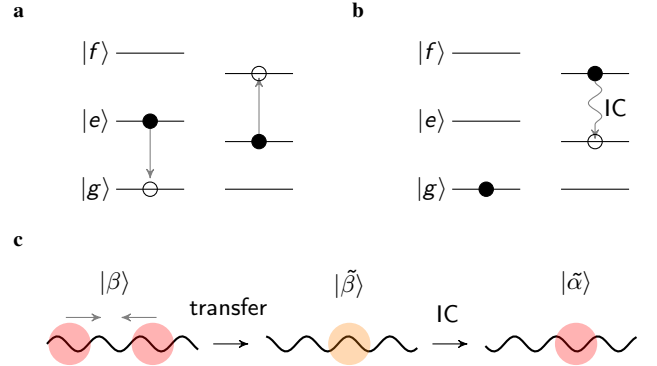
\begin{figure}[]
{\sffamily
\sansmath
\begin{subfigure}[t]{0.25\textwidth}
\caption{}
\vspace{0pt} 
\begin{tikzpicture}[scale=0.9]
\draw[] (0,1) -- (1.,1);
\draw[] (0,2) -- (1.,2);
\draw[] (0,3) -- (1.,3);

\draw[] (1.6,1) -- (2.6,1);
\draw[] (1.6,1.7) -- (2.6,1.7);
\draw[] (1.6,2.7) -- (2.6,2.7);

\draw[->,> = stealth', shorten > = 1pt,gray] (0.5,2) -- (0.5,1.12);
\draw[->,> = stealth', shorten > = 1pt,gray] (2.1,1.7) -- (2.1,2.68);

\node [] at (-0.3,1.) {$|g\rangle$};
\node [] at (-0.3,2) {$|e\rangle$};
\node [] at (-0.3,3.) {$|f\rangle$};

\draw[fill] (0.5,2) circle (3pt);
\draw[fill] (2.1,1.7) circle (3pt);
\draw[] (0.5,1) circle (3pt);
\draw[] (2.1,2.7) circle (3pt);
\end{tikzpicture}
\end{subfigure}%
\begin{subfigure}[t]{0.25\textwidth}
\caption{}
\begin{tikzpicture}[scale=0.9]
\draw[] (0,1) -- (1.,1);
\draw[] (0,2) -- (1.,2);
\draw[] (0,3) -- (1.,3);

\draw[] (1.6,1) -- (2.6,1);
\draw[] (1.6,1.7) -- (2.6,1.7);
\draw[] (1.6,2.7) -- (2.6,2.7);

\draw[->,> = stealth', shorten > = 1pt,auto,decorate, decoration={snake, pre length=3pt,post length=3pt},gray] (2.1,2.7) -- (2.1,1.7);

\node [] at (-0.3,1.) {$|g\rangle$};
\node [] at (-0.3,2) {$|e\rangle$};
\node [] at (-0.3,3.) {$|f\rangle$};
\node [] at (2.4,2.2) {IC};

\draw[] (2.1,1.7) circle (3pt);
\draw[fill] (0.5,1) circle (3pt);
\draw[fill] (2.1,2.7) circle (3pt);
\end{tikzpicture}
\end{subfigure}\\ 
\noindent
\begin{subfigure}[t]{0.90\textwidth}
\caption{}
\centering
\begin{tikzpicture}[scale=1]

\draw[thick,domain=0:2,samples=50]
    plot (\x,{0.0 + 0.1*sin(720*(\x-0.)/1.1)});

\draw[->,> = stealth', shorten > = 1pt,black] (2.2,0.0) -- (2.7,0.0);

\draw[thick,domain=2.9:4.9,samples=50]
    plot (\x,{0.0 + 0.1*sin(720*(\x-0)/1.1)});

\draw[->,> = stealth', shorten > = 1pt,black] (5.1,0.0) -- (5.6,0.0);

\draw[thick,domain=5.8:7.8,samples=50]
    plot (\x,{0.0 + 0.1*sin(720*(\x-0)/1.1)});

\fill[red,opacity=0.3] (0.3,0.0) circle (0.28);
\fill[red,opacity=0.3] (1.4,0.0) circle (0.28);
\node [] at (0.9,0.8) {$|\beta\rangle$};
\draw[->,> = stealth', shorten > = 1pt,gray] (0.25,0.35) -- (0.75,0.35);
\draw[->,> = stealth', shorten > = 1pt,gray] (1.45,0.35) -- (0.95,0.35);

\fill[orange,opacity=0.3] (4,0.0) circle (0.28);
\node [] at (4,0.8) {$|\tilde{\beta}\rangle$};

\fill[red,opacity=0.3] (7,0.0) circle (0.28);
\node [] at (7,0.8) {$|\tilde{\alpha}\rangle$};

\node [white] at (15,0) {IC};
\node [] at (2.4,0.5) {transfer};
\node [] at (5.3,0.4) {IC};
\end{tikzpicture}
\end{subfigure} 
}
\caption{\justifying Exciton-exciton annihilation (EEA) process can be interpreted either in the site basis (top row) or in the multiexciton basis (bottom row). In the top row, we report the EEA process for two three-level molecules: (a) two molecules in the first-excited state $\ket{e}$ can interact such that one transitions to the ground state $\ket{g}$ while the other to a higher-excited state $\ket{f}$. (b) The molecule in the $\ket{f}$ state undergoes rapid internal conversion (IC), resulting in the overall loss of one exciton. On the bottom row, (c) we report the scheme for EEA in the exciton basis, where the annihilation is given by the transition between double- to single-excited states.}
\label{fig:EEA_scheme}
\end{figure}

Because annihilation decreases the number of excitons in the sample, it plays the role of an additional decay channel. Exciton--exciton annihilation was therefore first studied by measuring the dependence of the fluorescence yield on excitation pulse intensity.\cite{sundstrom1988annihilation, stiel1988j} The same process is responsible for losses in organic solar cells\cite{tzabari2013exciton} and also mixes otherwise independent signals in action-detected spectroscopy techniques, leading to an additional ground-state bleach signal.\cite{maly_signatures_2018, bruschi2023unifying, bolzonello2023nonlinear, javed2024photosynthetic} 

Because the EEA occurs upon interaction of two excitons, it can function as a sensitive local probe of excitation dynamics. By measuring a nonlinear spectroscopic signal arising from the EEA, we can observe exciton transfer and determine its diffusion coefficient.\cite{dostal2018direct, maly2023separating,valkunas1995nonlinear, luttig2023high} This phenomenon is the basis of new techniques, exciton--exciton-interaction pump--probe spectroscopy (EEI--PP) and two-dimensional electronic spectroscopy (EEI--2DES).\cite{dostal2018direct, maly2023separating} EEI--PP and EEI--2DES signals are based on fifth-order nonlinear polarization and are sensitive to exciton-pair interactions. For purpose of this text, we will denote them $\text{PP}^{(5)}$ and $\text{2DES}^{(5)}$ respectively. These techniques have been used very recently to probe excitation transfer in molecular aggregates.\cite{dostal2018direct,maly2020wavelike,kriete2019interplay, maly2023separating, luttig2026simple, shi2025annihilation,zhang_probing_2025,binzer2026shot-to-shot} Beyond the excitation diffusion, the fifth-order signal was theoretically shown to be sensitive to delocalization of the excitation among several molecules, as well as to the interplay of structural and energetic disorder.\cite{luttig2021anisotropy} 

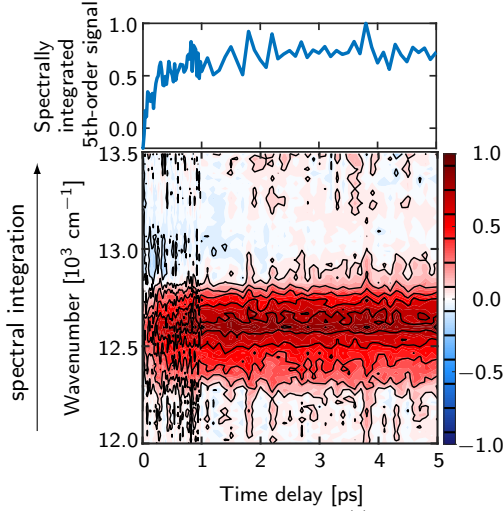
\begin{figure}[t!]
{\sffamily
\sansmath
\vspace{2.5cm}
\hspace{1cm}
  \begin{overpic}[width=0.22\textwidth]{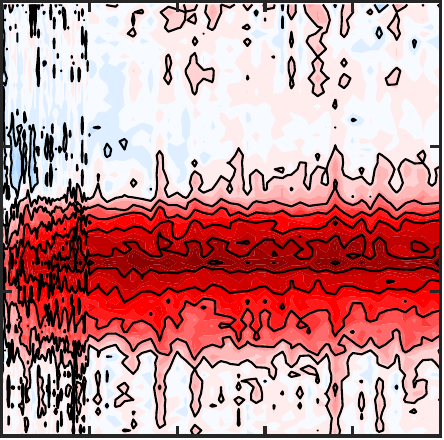}
     \put(-0.1,99){\includegraphics[width=0.22\textwidth]{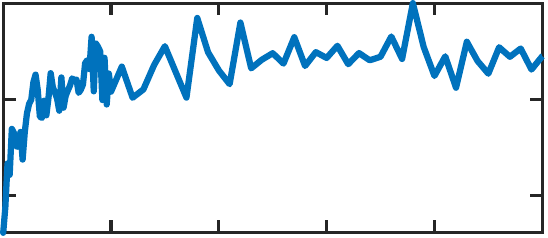}}
    \put(-11,104){\footnotesize  0.0}
    \put(-11,121){\footnotesize  0.5}
    \put(-11,139){\footnotesize  1.0}
    \put(-36,105){\footnotesize  \rotatebox{90}{Spectrally}}
    \put(-28,105){\footnotesize  \rotatebox{90}{integrated}}
    \put(-20,100){\footnotesize  \rotatebox{90}{5th-order signal}}
      \put(-27,10){\footnotesize  \rotatebox{90}{Wavenumber [$10^{3}$ cm$^{-1}$]}}
    \put(-14.5,95.75){\footnotesize  \rotatebox{0}{$13.5$}}
    \put(-14.5,63){\footnotesize  \rotatebox{0}{$13.0$}}
    \put(-14.5,30.5){\footnotesize  \rotatebox{0}{$12.5$}}
    \put(-14.5,-0.5){\footnotesize  \rotatebox{0}{$12.0$}}
  \put(-1,-7){\footnotesize  0}
  \put(18.5,-7){\footnotesize  1}
  \put(38,-7){\footnotesize  2}
  \put(58,-7){\footnotesize 3}
  \put(78,-7){\footnotesize  4}
  \put(98,-7){\footnotesize  5}
  \put(26,-18.5){\footnotesize  Time delay [ps]}
  \put(101.,0){\includegraphics[width=0.0105\textwidth]{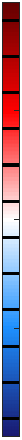}}
  \put(111,96){\footnotesize   1.0}
  \put(111,71.5){\footnotesize   $0.5$}
  \put(111,47){\footnotesize  $0.0$}
  \put(106,22.5){\footnotesize  $-0.5$}
  \put(106,-2){\footnotesize  $-1.0$} 
  \put(-35,5){\color{black}\vector(0,1){90}}
    \put(-48,15){ \rotatebox{90}{spectral integration}} 
     \end{overpic}
     \vspace{0.5cm}
     }
\caption{\justifying Experimental example of PP$^{(5)}$ spectra of polymer $[\text{SQA}-\text{SQB}]_{19}$. Spectral resolution provides additional information about the composition of the nonlinear signal. The positive peak near 12500\;cm$^{-1}$ grows with waiting time, whereas the negative peak near 13000\;cm$^{-1}$ decays. Explaining the origin of these two features is one of the main goals of the present work.}
\label{Fig:5PP_spectral_resolution}
\end{figure}

$\text{2DES}^{(5)}$ has been simulated for small systems such as squaraine dimers and trimers,\cite{Heshmatpour2020,suss2020wave,suss2020exciton,bubilaitis2024signatures,mayershofer2026nonlinear} where excitations interact immediately without transport. For larger molecular systems, EEA can be diffusion-limited; namely, excitation transport must occur for EEA to take place.\cite{maly2020wavelike} In such cases, only spectrally integrated fifth-order signals have been simulated to date.\cite{maly2020wavelike,luttig2021anisotropy,maly2023separating} From these, one can determine the diffusion coefficient under certain assumptions, such as equal oscillator strengths, not generally valid in excitonic systems,\cite{maly2020wavelike} but all the spectral information about lineshapes is lost. This is a severe limitation in utilizing the full potential of $\text{PP}^{(5)}$ and $\text{2DES}^{(5)}$ spectroscopies and their application to complex systems, such as photosynthetic light-harvesting complexes. Exemplary data of fifth-order transient spectrum, $\text{PP}^{(5)}(\omega,t)$, of a squaraine co-polymer is shown in Fig. \ref{Fig:5PP_spectral_resolution}.\cite{maly2023separating} At the top, the spectrally integrated signal is reported, which has been used to infer an effective annihilation rate.\cite{maly2020wavelike} The transient spectrum, however, clearly shows much richer information, such as the opposite-sign (negative) transient peak around $13000\text{ cm}^{-1}$, with very different kinetics from the rising main peak. What is the origin of this peak, why does it have opposite sign, and could it report on two-exciton state energies and dynamics? To address these questions and utilize the full potential of fifth-order transient spectroscopy, a microscopic theory of their signals is needed.

In this article, we develop such theoretical description, presenting a model for simulating the fifth-order pump--probe spectra and discussing its properties. In contrast to the common assumption that fifth-order signal reports mainly on EEA,\cite{shi2025annihilation, zhang_probing_2025} we demonstrate that, even for the simplest case of a molecular dimer, the fifth-order signal mixes annihilation and relaxation dynamics. We verify the validity of our approach by its application to an experimental example of a squaraine co-polymer $[\text{SQA}-\text{SQB}]_{19}$ consisting of 19 dimeric units, each composed of a squaraine A (SQA) and a squaraine B (SQB) molecule.\cite{maly2023separating}
Importantly, our approach goes beyond spectrally integrated descriptions and allows us to disentangle excitonic relaxation and annihilation under general conditions at the level of excitation pathways.
The chosen microscopic model explicitly includes higher-lying excited states as a key component in the EEA process. It incorporates a realistic description of lineshapes, providing additional degrees of freedom sensitive to the structure of the molecular complexes and properties of higher excited states, especially in comparison with the spectrally integrated signal. The perturbative approach used offers the possibility of distinguishing individual pathway contributions, revealing the origin of spectral features otherwise inaccessible,\cite{guan2025Quasiclassical} such as that in Fig. \ref{Fig:5PP_spectral_resolution}.

\section{Theoretical description of spectroscopic signal and microscopic model}
\label{Sec:model} 
To introduce our model for simulating fifth-order pump--probe spectra, we outline the formalism for spectroscopic response to light-matter interaction and address some common approximations. Consequently, we define our microscopic model used for simulations, including the description of waiting-time dynamics and spectral lineshapes.

\subsection{Spectroscopic response}
\label{Sec:model_spectra}
\input{pictures/fig_Feynman}

The spectroscopic signal is proportional to the linear $\boldsymbol{P}^{\text{L}}(t)$ and nonlinear polarization $\boldsymbol{P}^{\text{NL}}(t)$ induced in the material by its interaction with light. This response of the system can be naturally divided into contributions of different orders\cite{mukamel}  
\begin{equation}
     \boldsymbol{P}_{\text{induced}}(t)=\underbrace{\boldsymbol{P}^{(1)}(t)}_{\boldsymbol{P}^{\text{L}}(t)}+\underbrace{\boldsymbol{P}^{(3)}(t)+\boldsymbol{P}^{(5)}(t)+...}_{\boldsymbol{P}^{\text{NL}}(t)}.
\end{equation}
The order separation is possible both theoretically and experimentally, using the dependence on the electric field of the excitation pulses.\cite{maly2023separating, luttig2026simple, bruschi_multi-exciton_2026}

In the language of perturbation theory, the $n$-th order signal corresponds to $n$ light–matter interactions. In our model, we used the semiclassical and dipole approximation to describe the interaction between the system and the pulses 
\begin{equation}
    \hat{H}_{\text{int}}(t)=-\hat{\bm{\mu}}\cdot\pmb{\mathcal{E}}(t).
    \label{Eq:interaction_hamiltonian_in_dipoleA}
\end{equation}
In this setting, the interaction is mediated through the transition dipole moment operator of the system $\hat{\bm{\mu}}=\sum_{\iota,\gamma}\bm{\mu}_{\iota\gamma}|\iota\rangle\langle \gamma|$, which induces transitions between states $|\iota\rangle$ and $|\gamma\rangle$ with a different number of excitations. $\iota, \gamma =g, \alpha,\beta,\xi$ are indices of ground, single-, double- and triple excited states in multiexcitonic basis, respectively. Here, $ \pmb{\mathcal{E}} (t)$ is a classical electric field given by the total electric field of all pulses in the experiment. For the case of pump--probe spectroscopy, where we have one pump and one probe pulses, the total electric field can be expressed as
\begin{equation}
    \pmb{\mathcal{E}} (t) =  \pmb{\mathcal{E}}^{\text{Pu}}(t)+\pmb{\mathcal{E}}^{\text{Pr}}(t).
\end{equation}
For sensitivity to exciton--exciton annihilation, which we want to study, we need to create at least two excitons in the sample. Therefore, the lowest-order signal resulting in a double-excited state population is the fifth-order, in which the sample interacts with the electric field five times: four times with the pump and once with the probe pulse.\cite{luttig2021anisotropy} To connect the microscopic light–matter interaction with the experimentally measured signal, we express the fifth-order polarization as a convolution of the macroscopic sample response function $\overleftrightarrow{S}(\tau_5,\tau_4,\tau_3, \tau_2, \tau_1)$ and the electric field $\pmb{\mathcal{E}}(t)$ at the time of each interaction
\begin{equation}
\begin{split}
&\boldsymbol{P}^{(5)}(t)= \int^\infty_{0}d\tau_5\cdots\int^\infty_{0}d\tau_1 \overleftrightarrow{S}^{(5)}(\tau_5,\tau_4,\tau_3, \tau_2, \tau_1) \\
&\times \pmb{\mathcal{E}}(t-\tau_5) \pmb{\mathcal{E}}(t-\tau_5-\tau_4)\pmb{\mathcal{E}}(t-\tau_5-\tau_4-\tau_3)\\
&\times \pmb{\mathcal{E}}(t-\tau_5-\tau_4-\tau_3-\tau_2) \pmb{\mathcal{E}}(t-\tau_5-\tau_4-\tau_3-\tau_2-\tau_1),
\end{split}
\label{Eq:fifth_order_polarization}
\end{equation}
where the double-sided arrow indicates a tensor, in this case of rank six.
The fifth-order signal, proportional to the induced fifth-order polarization,\cite{maly2020wavelike} $\pmb{\mathcal{E}}^{(5)}_{\text{signal}}(t) \propto i \bm{P}^{(5)}(t)$, is detected through its interference with the probe pulse
$ I_{\text{signal}}^{(5)}(\omega) \propto 2\text{Re}\{\pmb{\mathcal{E}}^*_{\text{Pr}}(\omega)\cdot\pmb{\mathcal{E}}^{(5)}_{\text{signal}}(\omega)\}$.\cite{mukamel, luttig2021anisotropy}

To understand the features in the experimental data, we simulated third- and fifth-order signals using our previous theoretical model\cite{maly2020wavelike,luttig2021anisotropy} extended to include realistic lineshapes. To maintain reasonable computational cost, we simulated the nonlinear response using the doorway-window approach, which allows us to describe the excitation and detection processes independently.\cite{mukamel} Furthermore, we assume the impulsive limit and rotating-wave approximation.\cite{mukamel} To remove orientational anisotropy, we employed a specific experimental configuration where the probe pulse is oriented at the magic angle (54.7°) relative to the pump polarization.\cite{hamm2011concepts} For this setting, the intensity of the signal can be described in the impulsive limit as\cite{luttig2021anisotropy}
\begin{equation}
    I_{\text{signal}} \propto \sum_{a,b,c,d,e,f}  \frac{1}{45} \delta_{ab} (\delta_{cd}\delta_{ef}+\delta_{ce}\delta_{df}+\delta_{cf}\delta_{de}) s^{(5)}_{abcdef}.
\end{equation}
The microscopic response function\cite{mukamel,luttig2021anisotropy}
\begin{equation}
\begin{split}
&\overleftrightarrow{s}^{(5)}(\tau_5,\tau_4,\tau_3, \tau_2, \tau_1)=\left(\frac{i}{\hbar}\right)^5 \Tr\{\hat{\boldsymbol{\mu}}\mathcal{U}_0(\tau_5)[\hat{\boldsymbol{\mu}},\mathcal{U}_0(\tau_4)\times \\
&\times[\hat{\boldsymbol{\mu}},\mathcal{U}_0(\tau_3)[\hat{\boldsymbol{\mu}},\mathcal{U}_0(\tau_2)[\hat{\boldsymbol{\mu}},\mathcal{U}_0(\tau_1)[\hat{\boldsymbol{\mu}},\hat{\rho}(t_0)]]]]]\}.
\end{split}
\label{Eq:response_function}
\end{equation}
represents the dynamics of the system and the bath (such as the surrounding solvent) described by the total density matrix $\hat{\rho}$ upon light-matter interaction. At time $t_0$, the system is in the ground-state equilibrium as described by the total density matrix $\hat{\rho}(t_{0})$. Each light-matter interaction is represented by a transition dipole moment operator $\hat{\bm{\mu}}$ inside a commutator. After each interaction, the total state evolves according to a time-evolution superoperator $\mathcal{U}_0$ in the Liouville space of the molecular system and its bath. Finally, the last application of the transition dipole moment operator, followed by the total trace, yields the signal emission.

Contributions to the response function can be graphically depicted using the so-called (double-sided) Feynman diagrams.\cite{mukamel} In these diagrams, each light-matter interaction is represented by an arrow. Due to the structure of the commutator, the interactions from the right carry a negative sign. Therefore, the overall sign for the diagram is given by $(-1)^\ell$, where $\ell$ is the number of interactions on the right side. In Fig. \ref{Fig:FD_EEI2DES}, we report Feynman diagrams that contribute to the rephasing fifth-order response. In 2DES, there are two pump pulses delayed by time $T_{\text{Pu}}$ and a probe pulse which follows after an additional delay time $T$. At fifth-order, the first pump pulse interacts twice, resulting in a double-quantum coherence $| g \rangle \langle \beta |$ during the delay time $T_{\text{Pu}}$. Then, the second pump also interacts twice, resulting in a population in the ground, single-, or double-excited state manifold. 
Therefore, provided it is the highest non-vanishing nonlinear signal, the fifth-order signal can be selected based on the excitation frequency, considering the spectra positioned at $ \sim 2 \omega_{0}$, in contrast to the third-order signal which appears at $ \sim \omega_{0}$ frequency.\cite{dostal2018direct} Nevertheless, we are mainly interested in the pump--probe setting, in which pump pulses are overlapped and arrive at the same time on the sample ($T_{\text{Pu}} = 0$). Therefore, we cannot resolve the signal along the excitation frequency, which is effectively equivalent to integrate over it, according to the projection slice theorem\cite{hamm2011concepts}. Even though now different orders of the signal overlap, they can be separated by intensity cycling.\cite{maly2023separating, krich_separating_2025} Moreover, additional excitation pathways contribute to $\text{PP}^{(5)}$ due to time-ordering, compared to the $\text{2DES}^{(5)}$ signal. This forms the bottom (excitation) part of the diagram.
The upper (detection) parts of the diagrams are of the same type as in PP, depicting absorption or stimulated emission caused by a single interaction with the probe pulse. As a result, a coherence is generated between states in adjacent excitonic manifolds, which determines the detection frequency at $\sim \omega_{0}$.
The bottom and upper parts, connected with evolution of the population over time $T$, give us seven diagrams that form $\text{PP}^{(5)}$ and $\text{2DES}^{(5)}$, and three additional diagrams (marked by subscript I in Fig. \ref{Fig:FD_EEI2DES}), which significantly contribute only to the $\text{PP}^{(5)}$ signal. For each of these ten unique diagrams, there are similar diagrams that give an equivalent result (reported in Fig. S1-5 in the SI). These can be effectively accounted for by including an integer pre-factor (orange numbers in Eq. \ref{EEI2D_pathways}).\cite{maly2020wavelike} The coefficients differ between $\text{2DES}^{(5)}$ and $\text{PP}^{(5)}$ because the overlap of the pump and probe pulses not only introduces three additional pathways, but also changes the amplitudes of the existing pathways. The microscopic response functions associated with each class of pathways are expressed in the exciton basis as

\begin{widetext}
\begin{equation}
\begin{aligned}
{\text{NGSB}^{(5)}}_{abcdef}&=\sum_{\alpha,\dot{\alpha},\tilde{\alpha},\beta}\mu_{\tilde{\alpha}g}^a\mu_{\tilde{\alpha}g}^b\mu_{\dot{\alpha}g}^c(\textcolor{orange}{8}\mu_{\alpha g}^d\mu_{\dot{\alpha}g}^e+\textcolor{orange}{2}\mu_{\beta \dot{\alpha}}^d\mu_{\beta\alpha}^e)\mu_{\alpha g}^f\mathcal{U}_{g,g}(T)\mathcal{L}^{\text{abs}}_{\tilde{\alpha}g}(\omega)\;\;\;\;\;\;\;\;\;\;\;\;\;\;\;\;\;\;\;(\mathcal{O}(N^3))\\
{\text{NSE}^{(5)}}_{abcdef}&=\textcolor{orange}{8}\sum_{\alpha,\dot{\alpha},\tilde{\alpha},\beta}\mu_{\tilde{\alpha}g}^a\mu_{\tilde{\alpha}g}^b\mu_{\dot{\alpha}g}^c(\mu_{\alpha g}^d\mu_{\dot{\alpha}g}^e+\mu_{\beta \dot{\alpha}}^d\mu_{\beta \alpha}^e)\mu_{\alpha g}^f\mathcal{U}_{\tilde{\alpha},\dot{\alpha}}(T)\mathcal{L}^{\text{em}}_{\tilde{\alpha}g}(\omega)\;\;\;\;\;\;\;\;\;\;\;\;\;\;\;\;\;\;\;\;(\mathcal{O}(N^2))\\
{\text{SE}_2^{(5)}\text{(2E)}}_{abcdef}&=\textcolor{orange}{-6}\sum_{\alpha,\dot{\alpha},\tilde{\alpha},\beta,\dot{\beta}}\mu_{\dot{\beta} \tilde{\alpha}}^a\mu_{\dot{\beta} \tilde{\alpha}}^b\mu_{\beta \dot{\alpha}}^c\mu_{\dot{\alpha}g}^d\mu_{\beta \alpha}^e\mu_{\alpha g}^f\mathcal{U}_{\dot{\beta},\beta}(T)\mathcal{L}^{\text{em}}_{\dot{\beta} \tilde{\alpha}}(\omega)\;\;\;\;\;\;\;\;\;\;\;\;\;\;\;\;\;\;\;\;\;\;\;\;\;\;\;\;\;\;\;\;\;\;\;(\mathcal{O}(N^2))\\
{\text{SE}_2^{(5)}\text{(EEA)}}_{abcdef}&=\textcolor{orange}{-6}\sum_{\alpha,\dot{\alpha},\tilde{\alpha},\beta}\mu_{\tilde{\alpha}g}^a\mu_{\tilde{\alpha}g}^b\mu_{\beta \dot{\alpha}}^c\mu_{\dot{\alpha}g}^d\mu_{\beta \alpha}^e\mu_{\alpha g}^f\mathcal{U}_{\tilde{\alpha},\beta}(T)\mathcal{L}^{\text{em}}_{\tilde{\alpha}g}(\omega)\;\;\;\;\;\;\;\;\;\;\;\;\;\;\;\;\;\;\;\;\;\;\;\;\;\;\;\;\;\;\;\;\;\;\;\;\;\;\;(\mathcal{O}(N^2))\\
{\text{NESA}^{(5)}}_{abcdef}&=\textcolor{orange}{-8}\sum_{\alpha,\dot{\alpha},\tilde{\alpha},\beta,\dot{\beta}}\mu_{\dot{\beta} \tilde{\alpha}}^a\mu_{\dot{\beta} \tilde{\alpha}}^b\mu_{\dot{\alpha}g}^c(\mu_{\alpha g}^d\mu_{\dot{\alpha}g}^e+\mu_{\beta \dot{\alpha}}^d\mu_{\beta \alpha}^e)\mu_{\alpha g}^f\mathcal{U}_{\tilde{\alpha},\dot{\alpha}}(T)\mathcal{L}^{\text{abs}}_{\dot{\beta} \tilde{\alpha}}(\omega)\;\;\;\;\;\;\;\;\;\;\;\;(\mathcal{O}(N^3))\\
{\text{ESA}_2^{(5)}\text{(2E)}}_{abcdef}&=\textcolor{orange}{6}\sum_{\alpha,\dot{\alpha},\tilde{\alpha},\beta,\dot{\beta},\xi}\mu_{\xi \dot{\beta}}^a\mu_{\xi \dot{\beta}}^b\mu_{\beta \dot{\alpha}}^c\mu_{\dot{\alpha}g}^d\mu_{\beta \alpha}^e\mu_{\alpha g}^f\mathcal{U}_{\dot{\beta},\beta}(T)\mathcal{L}^{\text{abs}}_{\xi \dot{\beta}}(\omega)\;\;\;\;\;\;\;\;\;\;\;\;\;\;\;\;\;\;\;\;\;\;\;\;\;\;\;\;\;\;\;\;\;\;\;\;(\mathcal{O}(N^3))\\
{\text{ESA}_2^{(5)}\text{(EEA)}}_{abcdef}&=\textcolor{orange}{6}\sum_{\alpha,\dot{\alpha},\tilde{\alpha},\beta,\dot{\beta}}\mu_{\dot{\beta} \tilde{\alpha}}^a\mu_{\dot{\beta} \tilde{\alpha}}^b\mu_{\beta \dot{\alpha}}^c\mu_{\dot{\alpha}g}^d\mu_{\beta \alpha}^e\mu_{\alpha g}^f\mathcal{U}_{\tilde{\alpha},\beta}(T)\mathcal{L}^{\text{abs}}_{\dot{\beta} \tilde{\alpha}}(\omega)\;\;\;\;\;\;\;\;\;\;\;\;\;\;\;\;\;\;\;\;\;\;\;\;\;\;\;\;\;\;\;\;\;\;\;\;\;\;(\mathcal{O}(N^3)).\\
\label{EEI2D_pathways}
\end{aligned}
\end{equation}
\end{widetext}

We can break the expression for each pathway contribution into three products: transition dipole moments $\mu_{\iota\gamma}^x$ projected along the pulse polarization direction $x$, the time-evolution superoperator $\mathcal{U}_{\iota,\gamma}(T)$, and the (normalized) lineshape $\mathcal{L}_{\iota\gamma}(\omega)$, where $\iota$ and $\gamma$ stand for ground, single-, double- or triple-excited state in multiexciton basis. The transition dipole moments determine the optical selection rules and the polarization dependence. The time-evolution superoperator $\mathcal{U}(T)$ describes population dynamics within the single- and double-excited states manifolds, including transfer, annihilation, and relaxation, during the waiting time $T$ (see Sec. \ref{Sec:model_dynamics}). The absorption/emission lineshape $\mathcal{L}^{\text{abs/em}}(\omega)$ encodes coherence dephasing and spectral broadening due to the interaction with the environment (see Sec. \ref{Sec:model_lineshapes}).

Fifth-order pump--probe signal contains standard third-order-like contributions, however, here with opposite sign (highlighted with the word negated) -- negated ground state bleach (NGSB$^{(5)}$), negated stimulated emission (NSE$^{(5)}$) and negated excited-state absorption (NESA$^{(5)}$), each consisting of two types, I or II.
In addition to that, pathways involving double-excited state populations contribute as well, which provide direct access to exciton–exciton interactions during the delay time $T$. In particular, EEA-related pathways (SE$_2^{(5)}$(EEA) and ESA$_2^{(5)}$(EEA)) are sensitive to annihilation, while two-exciton pathways (SE$_2^{(5)}$(2E) and ESA$_2^{(5)}$(2E)) reflect transport within the double-excited manifold. This separation of contributions will be useful later for interpreting spectrally resolved fifth-order pump--probe signals. Notice that the nomenclature used for the pathways differs from our previous works\cite{maly2020wavelike,luttig2021anisotropy} and reflects efforts to unify it with that used in other works.\cite{rose2023interpretations, krich_separating_2025, luttig2026simple}

The total contribution of the excitation pathways depends on the number of molecules $N$, as shown in brackets at the end of each expression in Eq. \ref{EEI2D_pathways}. The scaling dependences are obtained based on considerations in the site basis. \cite{maly2023separating} Notice how each kind of pathway exhibits a different scaling with the system size. For example, the number of NGSB$^{(5)}$ and NESA$^{(5)}$ pathways scales as $\propto N^3$, compared to the number of NSE$^{(5)}$ contributions which scales as $\propto N^2$. However, due to their sign, some contributions can partially cancel each other, as for NGSB$^{(5)}$ and NESA$^{(5)}$. Furthermore, the presence of excitonic delocalization results in the redistribution of the transition dipole moments. This effectively changes the weight of certain pathways, potentially enhancing/diminishing their contribution compared to others. As we will see, spectral features emerge from a delicate balance between the different kinds of pathways. Moreover, the dynamics is also non-trivially influenced by the system size. Consequently, the spectrum of a larger system cannot, in general, be reconstructed by simply rescaling the individual excitation-pathway contributions obtained for a smaller system according to their dependence on $N$.

\subsection{System Hamiltonian}
\label{Sec:model_hamiltonian}
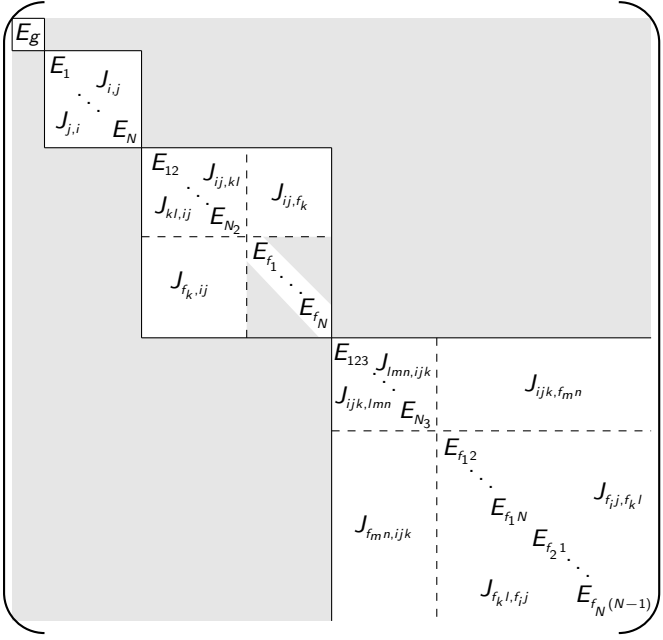
\begin{figure}
{\sffamily
\sansmath
\[
\begin{tikzpicture}[ xscale=1.07,
  yscale=1.07,
  every node/.style={minimum size=8mm, anchor=center} every node/.style={minimum size=8mm, anchor=center}, 
  roundmatrix/.style={draw, thick, rounded corners=3pt}]

  \fill[gray!20] (-0.5,3.9) rectangle (7.0,4.3);
  \fill[gray!20] (0.7,2.7) rectangle (7.0,3.9);
  \fill[gray!20] (3.05,0.35) rectangle (7.0,3.9);
  \fill[gray!20] (-0.9,-3.15) rectangle (-0.5,3.9);
  \fill[gray!20] (-0.5,-3.15) rectangle (0.7,2.7);
  \fill[gray!20] (-0.5,-3.15) rectangle (3.05,0.35);
  \fill[fill=gray!20] (2.,0.35)--(2.,1.3)--(2.9,0.35);
  \fill[fill=gray!20] (3.05,1.6)--(2.2,1.6)--(3.05,0.75);
  
  \draw[thick] (-0.5,4.5) to[out=180,in=90] (-1,4) -- (-1,-2.8) to[out=270,in=180] (-0.5,-3.3);
  \draw[thick] (6.6,4.5) to[out=0,in=90] (7.1,4) -- (7.1,-2.8) to[out=270,in=0] (6.6,-3.3);

  \draw[] (-0.9,3.9) -- (-0.9,4.3);
  \draw[] (-0.5,3.9) -- (-0.5,4.3);
  \draw[] (-0.9,3.9) -- (-0.5,3.9);
  \draw[] (-0.9,4.3) -- (-0.5,4.3);
  \node at (-0.7,4.1) {$E_g$};

  \draw[] (-0.5,3.9) -- (-0.5,2.7);
  \draw[] (0.7,3.9) -- (0.7,2.7);
  \draw[] (-0.5,3.9) -- (0.7,3.9);
  \draw[] (-0.5,2.7) -- (0.7,2.7);
  \node at (-0.3,3.7) {$E_{_1}$};
  \node at (0.3,3.5) {$J_{_{i,j}}$};
  \node at (0.05,3.35) {$\ddots$};
  \node at (0.5,2.9) {$E_{_N}$};
  \node at (-0.2,3.0) {$J_{_{j,i}}$};

  \draw[] (0.7,2.7) -- (3.05,2.7);
  \draw[] (0.7,2.7) -- (0.7,0.35);
  \draw[dashed] (0.7,1.6) -- (3.05,1.6);
  \draw[dashed] (2.,0.35) -- (2.,2.7);

  \draw[] (3.05,0.35) -- (3.05,2.7);
  \draw[] (0.7,0.35) -- (3.05,0.35);
  \node at (1.7,2.35) {$J_{_{ij,kl}}$};
  \node at (1.1,1.95) {$J_{_{kl,ij}}$};
  \node at (2.55,2.1) {$J_{_{ij, f_k}}$};
  \node at (1.3,1.) {$J_{_{f_k, ij}}$};

  \node at (1.0,2.5) {$E_{_{12}}$};
  \node at (1.4,2.2) {$\ddots$};
  \node at (1.75,1.8) {$E_{_{N_2}}$};
  \node at (2.25,1.35) {$E_{_{f_{_{1}}}}$};
  \node at (2.55,1.1) {$\ddots$};
  \node at (2.85,0.65) {$E_{_{f_{_{N}}}}$};

  \draw[] (3.05,0.35) -- (7.0,0.35);
  \draw[dashed] (3.05,-0.8) -- (7.0,-0.8);
  \draw[] (3.05,0.35) -- (3.05,-3.15);
  \draw[dashed] (4.35,0.35) -- (4.35,-3.15);
  \node at (3.3,0.15) {$E_{_{123}}$};
  \node at (3.45,-0.4) {$J_{_{ijk,lmn}}$};
  \node at (3.95,-0.) {$J_{_{lmn,ijk}}$};
  \node at (3.7,-0.1) {$\ddots$};
  \node at (4.1,-0.6) {$E_{_{N_3}}$};
  \node at (4.65,-1.05) {$E_{_{f_12}}$};
  \node at (4.9,-1.3) {$\ddots$};
  \node at (5.25,-1.8) {$E_{_{f_{1}N}}$};
  \node at (6.6,-1.6) {$J_{_{f_ij,f_kl}}$};
  \node at (5.2,-2.8) {$J_{_{f_kl,f_ij}}$};
  \node at (5.75,-2.2) {$E_{_{f_{_{2}}1}}$};
  \node at (6.1,-2.4) {$\ddots$};
  \node at (6.55,-2.9) {$E_{_{f_{_{N}}(N-1)}}$};
  \node at (3.7,-2.) {$J_{_{f_mn,ijk}}$};
  \node at (5.8,-0.25) {$J_{_{ijk,f_mn}}$};

\end{tikzpicture}
\]
 }
    \caption{\justifying The Hamiltonian for the electronic states of the molecular system has a block-diagonal structure. There are four blocks: a single-element ground-state block, followed by blocks of single-, double-, and triple-excited states. Double excited state block is further divided into block of $N_2=\frac{(N-1)N}{2}$ two-exciton states and a block of $N_f=N$ higher-excited states. In the triple-excited state block there are $N_3=\frac{(N-2)(N-1)N}{6}$ three-exciton states and $N_f(N-1)=N(N-1)$ $fe$ states. The zero elements are indicated by the gray regions, reflecting the absence of couplings between different excitation manifolds.\cite{seibt2016optical, mikalvciute2025complete} Additionally, we do not assume any coupling between different $f$ states.}
    \label{fig:hamiltonian}
\end{figure}

The molecular complex is modeled using the Frenkel-exciton model.\cite{May2011, Valkunas2013} In this case, the $i$-th molecule is represented as a three-level system, composed of a ground $\ket{g}$, a first-excited $\ket{e_{i}}$, and a higher-excited $\ket{f_{i}}$ state (see Fig.\ref{fig:site_exciton_basis}). The inclusion of the higher-excited state is necessary to obtain a microscopic model of exciton--exciton annihilation. Therefore, the Hamiltonian of the $i$-th chromophore is given by
\begin{equation}
\begin{split}
     \hat{H}_i = &|g_i\rangle E_g \langle g_i| + |e_i \rangle (E_i  + \Delta \hat{V}_i) \langle e_i | + \\
     &| f_i \rangle ( E_{f_i} + \Delta \hat{V}_{f_i}) \langle f_i |
\end{split}
\end{equation}
where $E_{g}$, $E_{i}$, and $E_{f_{i}}$ are the energies of the ground, single-, and higher-excited states, respectively.
For simplicity, we take the ground state as a reference by setting $E_{g} = 0$. We assume that the energy of the first-excited state of the $i$-th molecule is $E_i=\bar{E}_i + \delta E_i$, where $\bar{E}_i$ is the average site energy and $\delta E_i$ represents static disorder, which is sampled from a normal distribution with zero mean and finite standard deviation. Instead, the energy of higher-excited state is defined as $E_{f_i}=E_i + \Delta E_{f_i} + \delta E_{f_i} $, given by the sum of the energy of the first-excited state, the energy gap $\Delta E_{f_i}$ and static disorder $\delta E_{f_i}$. The system-bath interaction is mediated by the energy-gap operators $\Delta \hat{V}_{i}$ and $\Delta \hat{V}_{f_i}$ for the states $|e_i \rangle$ and $|f_i\rangle$, respectively, as will be described in detail in Sec. \ref{Sec:model_bath}.   

The transition dipole moments for the electronic transition between ground and first-excited states $\bm{\mu}_{g_{i},e_{i}}$, as well as their spatial positions, are chosen to reflect the structure and other properties of the sample studied. We will further use the collective basis for the description of the transition dipole moments ${\mu}_{g,i} \equiv {\mu}_{g_{i},e_{i}}$. For simplicity, the transition dipole moments from the first to higher-excited state are assumed to be proportional to that from the ground to the first-excited state as $\boldsymbol{\mu}_{i,f_i}=F_\mu\boldsymbol{\mu}_{g,i}$, where $F_\mu$ is a scaling factor introduced to account for differences between the strength of these two transitions.\cite{maly2020wavelike,luttig2021anisotropy}

The structure of the Hamiltonian is block diagonal, as shown in Fig. \ref{fig:hamiltonian}.
The excitonic couplings between the first-excited states are calculated from their transition dipole moments and their positions using dipole-dipole interaction, forming the single-excited block of the Hamiltonian. Energies, excitonic couplings, and transition dipole moments of double- and triple-excited states in the site basis are derived from those of single-excited states. We demonstrate this on the example of two-exciton state $|ij \rangle = |i \rangle|j \rangle$ in the site basis. Its energy is assumed to be equal to the sum of that of individual excited states $E_{ij}=E_i+E_j$, while the excitonic coupling between different two-exciton states is assumed to take the form
\begin{equation}
    J_{ij,kl}=J_{i,k}\delta_{jl}+J_{i,l}\delta_{jk}+J_{j,k}\delta_{il}+J_{j,l}\delta_{ik}.
\end{equation}
where the coupling is present only between states that share one excitation on the same molecule.
The transition dipole moment between states $|k \rangle$ and $|ij \rangle$ is given by:
\begin{equation}
\boldsymbol{\mu}_{k,ij}=\delta_{jk}\boldsymbol{\mu}_{g,i}+\delta_{ik}\boldsymbol{\mu}_{g,j}.
\end{equation}
The coupling between two-exciton $|ij\rangle$ state and higher-excited $f_k$ state instead is given by:
\begin{equation}
    J_{ij,f_k}=F_J J_{i,j} (\delta_{ik}+ \delta_{jk}), 
\end{equation}
with $F_J$ as a scaling factor. Therefore, excitonic coupling between two-exciton state and higher-excited state is present only if the states share an excitation of the same molecule. Notice that we exclude the coupling between different f states ($J_{f_i,f_j}=0$).
The triple-excited block of the Hamiltonian can be constructed analogously; see Table 1 in the SI for the detailed expressions.

The Hamiltonian is diagonalized and used to calculate transition rates, rotationally averaged transition dipole moments, and lineshapes. The eigenstates obtained from the diagonalization of the Hamiltonian form the multiexciton basis and are denoted by Greek letters, as graphically depicted in Fig. \ref{fig:site_exciton_basis}. 

\subsection{System-bath interaction}
\label{Sec:model_bath}
The influence of the environment (solvent and vibrations) on the system is completely characterized by the spectral density.
The spectral function of the bath is given by the Fourier transform of the time-correlation function of the energy-gap operator $\Delta \hat{V}$\cite{mukamel} 
\begin{equation}
\begin{split}
        \tilde{C}(\omega)= \int_{-\infty}^{+\infty} dt e^{i\omega t} \langle \Delta \hat{V}(t) \Delta \hat{V}(0) \rangle.
\end{split}
\end{equation}
The spectral function can also be expressed in terms of its odd part $\tilde{C}''(\omega)$ called spectral density
\begin{equation}
\begin{split}
        \tilde{C}(\omega)=\left(1+\coth\frac{\hbar\omega}{2k_BT}\right)\tilde{C}''(\omega),
\end{split}
\end{equation}
where $T$ is the temperature of the bath and $k_B$ the Boltzmann constant. The spectral density $ \tilde{C}''(\omega)$ is usually described using the Brownian oscillator models.\cite{mukamel} In the following, we consider an overdamped mode, representing the contribution of the solvent and low-frequency phonons, and an underdamped mode, describing high-frequency intramolecular vibrations
\begin{equation}
    \tilde{C}''(\omega)=\frac{2\lambda\Lambda}{\omega^2+\Lambda^2}+\frac{2\lambda_v\Omega^2\omega\gamma}{(\Omega^2-\omega^2)^2+\omega^2\gamma^2},
\end{equation}
where $\lambda$ is the reorganization energy and $\Lambda$ is the inverse correlation time of the overdamped mode, while $\Omega$ is the frequency, $\gamma$ is the damping constant and $\lambda_v=S\Omega$ is the reorganization energy of the underdamped mode.

The system-bath interaction is assumed to induce fluctuations of the site energies. In this case, the energy-gap operator is diagonal $\Delta \hat{V}_{ij} = \Delta \hat{V}_{ii} \delta_{ij}$.
Furthermore, the bath associated with different molecules is assumed to be uncorrelated, thus
\begin{equation}
    \langle \Delta \hat{V}_{ii}(t) \Delta \hat{V}_{jj}(0) \rangle = \langle \Delta \hat{V}_{ii}(t) \Delta \hat{V}_{ii}(0) \rangle \delta_{ij}.
\end{equation}
Instead, the energy-gap operator associated with two-exciton (or three-exciton) states is assumed to be the sum of those of single-excited states
\begin{equation}
    \Delta \hat{V}_{ij,ij}= \Delta \hat{V}_{ii} + \Delta \hat{V}_{jj}.
\end{equation}
In the following, we assume that the time-correlation function of higher-excited states is proportional to that of the first-excited state
\begin{equation}
    \langle \Delta \hat{V}_{f_if_i}(t) \Delta \hat{V}_{f_jf_j}(0) \rangle = \langle \Delta \hat{V}_{ii}(t) \Delta \hat{V}_{ii}(0) \rangle \delta_{ij} \Phi^2
\end{equation}
through the scaling parameter $\Phi$. Furthermore, we assume that higher-excited states partially share the same bath as the first-excited states 
\begin{equation}
    \langle \Delta \hat{V}_{f_if_i}(t) \Delta \hat{V}_{jj}(0) \rangle = \langle \Delta \hat{V}_{ii}(t) \Delta \hat{V}_{ii}(0) \rangle \delta_{ij} \Phi \Phi_{\text{corr}}.
\end{equation}
where $\Phi_{\text{corr}}$ quantifies the degree of correlation.

\subsection{Population dynamics}
\label{Sec:model_dynamics}

Due to the excitonic coupling and system-bath interaction, populations are transferred between different states of the system.
In this case, rates related to population transfer and relaxation within each manifold are obtained using Redfield theory in the secular approximation.\cite{May2011, Valkunas2013}
The detailed expressions related to population transfer rates are reported in Sec. 1.3 of the SI.
Instead, the EEA process is composed of two steps: the first part is the transition from a two-exciton state to a higher-excited state, which was calculated by standard Redfield theory for a double-excited manifold in the exciton basis, because the higher-excited states are coupled with the two-exciton states. The second step is the internal conversion from the higher-excited to first-excited states, which is modeled using the Lindblad theory.\cite{bruggemann_exciton_2003} The associated rate between double- and single-excited states in the multiexciton basis is
\begin{equation}
    k_{\alpha\beta}=\sum_i k_{\text{IC}} |c^{\alpha}_{i}|^2|c^{\beta}_{f_i}|^2, 
\end{equation}
where $k_{\text{IC}}$ is the rate of internal conversion, while $c^{\alpha}_{i}$ and $c^{\beta}_{f_i}$ are the transformation coefficients from the site to the exciton basis. The time-evolution superoperator $\mathcal{U}(T)$ which appears in Eqs. \eqref{EEI2D_pathways} is then obtained as an exponential of the rate matrix including blocks of single- and double-excited-state dynamics. 

\subsection{Spectral lineshapes}
\label{Sec:model_lineshapes}
The peaks appearing in the spectrum are generally broadened due to the effect of the system-bath interaction, which induces coherence dephasing and population transfer in the system.
In our methodology, the lineshapes are obtained from second-order cumulant expansion.\cite{mukamel} In accordance with Refs. \citenum{renger2002relation}, the absorption lineshape for $|\iota\rangle\langle\gamma|$ coherence in multiexciton basis is given by 
\begin{equation}
\begin{split}
    \mathcal{L}^{\text{abs}}_{\iota\gamma}(\omega)=& \Re \int_{0}^{\infty} dt\ e^{i\omega t}  \\
    &\times e^{-i(\omega_{\iota\gamma}+2\lambda_{\iota\iota}-2\lambda_{\iota\gamma}) t-(g_{\iota\iota}(t)+g_{\gamma\gamma}(t)-2g_{\iota\gamma}(t))-\Gamma_{\iota\gamma} t}
\end{split}
  \end{equation}
while the emission lineshape is
\begin{equation}
\begin{split}
  \mathcal{L}^{\text{em}}_{\iota\gamma}(\omega)=&  \Re \int_{0}^{\infty} dt \ e^{i\omega t} \\
  &\times e^{-i(\omega_{\iota\gamma}-2\lambda_{\gamma\gamma}+2\lambda_{\iota\gamma}) t-(g^*_{\iota\iota}(t)+g^*_{\gamma\gamma}(t)-2g^*_{\iota\gamma}(t))-\Gamma_{\iota\gamma} t}  
\end{split}
\end{equation}
where $\omega_{\iota\gamma}$ is the transition frequency between the states.
The reorganization energy can be expressed from the spectral density\cite{mukamel}
\begin{equation}
    \lambda=\int^\infty_{0} \frac{d\omega}{\pi\omega} \tilde{C}''(\omega)
\end{equation}
which quantifies the system-bath interaction, and the lineshape function is obtained from the time-correlation function\cite{mukamel} 
\begin{equation}
\begin{split}
    g_{\iota\gamma}(t)&=\int^t_0 dt'' \int^{t''}_0 dt' \langle \Delta \hat{V}_{\iota\iota}(t')V_{\gamma\gamma}(0) \rangle.
\end{split}
\end{equation}
In contrast, $\Gamma_{\iota\gamma}$ is the lifetime dephasing due to population decay. We approximated it with $\Gamma_{\iota\gamma} \approx \frac{k_{\iota\iota}+k_{\gamma\gamma}}{2}$, which comes from a Lindblad treatment of the coherence evolution.\cite{chang_accuracy_2015}  In this case, we are neglecting coherence-transfer contributions, which can in principle extend the decoherence time beyond that determined by lifetime-induced broadening.\cite{jean_1995} Nevertheless, for the strongly coupled system considered here, we expect the resulting error to be small.

The signal for each pathway is then composed according to Eq. \ref{EEI2D_pathways} and saved as a single realization.  In case of systems with energy and structural disorder, we have to average over  this disorder. Therefore this process is repeated several hundred times, always starting with the generation of a new slightly perturbed structure and building the corresponding Hamiltonian. The final signal is then obtained by averaging over all realizations.

\section{Fifth-order signal behavior}


\subsection{Weakly coupled dimer}
\label{Sec:dimer}
To build intuition about the spectral features appearing in the fifth-order signal and to validate the methodology, we first analyze a minimal model system: a weakly coupled heterodimer with two single-excited states $|i\rangle$ and $|j\rangle$ and one two-exciton state $|ij\rangle$. Due to weak coupling, we can assume that the exciton basis coincides with the site basis (see Fig.\ref{fig:site_exciton_basis}), which we will use further on. In this model, we neglect direct contributions of higher-excited states $\ket{f_{n}}$ to the signal and use them only to enable the EEA process. In the analytical approach, we therefore include EEA explicitly with rate $k_A$. 

Here, the energy of the two-exciton state is equal to the sum of the energies of the two first-excited states $E_{ij}=E_i+E_j$. Likewise, the transition dipole moment from single excited state $|j\rangle$ to two-exciton state $|ij\rangle$ is equivalent to the transition dipole moment from ground state $|g\rangle$ to single-excited state $|i\rangle$, so that $ \boldsymbol{\mu}_{ij,j}= \boldsymbol{\mu}_{i,g}\equiv \boldsymbol{\mu}_i$. Expression for lineshape of coherence $|ij\rangle\langle j|$ can be simplified analogously to get $\mathcal{L}_{ij,j}(\omega)=\mathcal{L}_{i,g}(\omega)$.  
Meanwhile transition from single-excited state $|i\rangle$ to two-exciton state $|ij\rangle$ corresponds to transition from ground state $|g\rangle$ to single-excited state $|j\rangle$ and therefore $ \boldsymbol{\mu}_{ij,i}=\boldsymbol{\mu}_{j,g}\equiv \boldsymbol{\mu}_j$ and  $\mathcal{L}_{ij,i}(\omega)=\mathcal{L}_{j,g}(\omega)$. 

\begin{figure}[htbp]
{\sffamily
\sansmath
\centering
\begin{subfigure}[t]{0.25\textwidth}
\centering
    \begin{tikzpicture}[scale=0.9]
\draw[] (0,1) -- (2.6,1);
\draw[] (0,2.25) -- (1.,2.25);
\draw[] (1.6,1) -- (2.6,1);
\draw[] (0,2.95) -- (2.6,2.95);

\draw[] (1.6,1.7) -- (2.6,1.7);

\draw[scale =1,->, thick, color= myblue, bend left=20] (0.9,2.4) to (1.55,1.8);
\draw[->,> = stealth', thick, shorten > = 1pt, myblue] (0.2,2.95) -- (0.2,2.25);
\draw[->,> = stealth', thick, shorten > = 1pt, myblue] (2.5,2.95) -- (2.5,1.7);

\draw[<->,> = stealth',  shorten > = 1pt,black] (0.5,2.3) -- (0.5,2.95);
\draw[<->,> = stealth',  shorten > = 1pt,black] (2.3,1.75) -- (2.3,2.95);

\draw[<->,> = stealth', shorten > = 1pt,black] (2.1,1.05) -- (2.1,1.7);
\draw[<->,> = stealth', shorten > = 1pt,black] (0.5,1.05) -- (0.5,2.25);

\node [] at (-0.3,1.) {$|g\rangle$};
\node [] at (-0.3,2.25) {$|i\rangle$};
\node [] at (3,1.7) {$|j\rangle$};
\node [] at (3,2.95) {$|ij\rangle$};

\node [ myblue] at (1.15,2.0) {$k_T$};
\node [ myblue] at (-0.05,2.75) {$\frac{k_A}{2}$};
\node [ myblue] at (2.8,2.3) {$\frac{k_A}{2}$};
\node [] at (2.45,1.3) {$\mu_{j}$};
\node [] at (0.2,1.6) {$\mu_{i}$};
\node [] at (0.85,2.6) {$\mu_{j}$};
\node [] at (2,2.3) {$\mu_{i}$};

\node [white] at (2,3.5) {$\mu_{i}$};
\node [white] at (-1,2.6) {$\mu_{i}$};
\node [] at (-0.3,3.8) {a};
\node [white] at (-0.3,-0.24) {a};
\end{tikzpicture}
\label{fig:dimer_scheme}
\end{subfigure}%
\begin{subfigure}[t]{0.25\textwidth}
\begin{tikzpicture}
\node [] at (1.3,2.7) {$\mu_i>\mu_j$};
\node [] at (-0.5,2.9) {b};
\begin{axis}[
    width=0.95\textwidth,
    height=0.9\textwidth,
    xlabel={\footnotesize Time delay [ps]},
    ylabel={\footnotesize  Fifth-order signal},
    tick label style={font=\footnotesize},
    legend style={
        font=\scriptsize,
        draw=none,
        fill=none,
        at={(1.0,0.55)},
        anchor=north east,
        /tikz/every even column/.append style={column sep=2pt},
    },
    legend image post style={xscale=0.6},
    domain=0:15,
    samples=800,
    ymin=120.0,
    ymax=270,
]

\pgfmathsetmacro{\Kax}{1}
\pgfmathsetmacro{\Kay}{1}
\pgfmathsetmacro{\Kaz}{1}
\pgfmathsetmacro{\Ktx}{1.7}
\pgfmathsetmacro{\Kty}{1}
\pgfmathsetmacro{\Ktz}{0.3}
\pgfmathsetmacro{\ma}{1.5}
\pgfmathsetmacro{\mb}{1.0}

\pgfmathsetmacro{\Ka}{1/1}

\pgfmathsetmacro{\fOneMax}{
1485/4 + (135/2)*2/exp(2) - 108/exp(2)
}

\pgfmathsetmacro{\fTwoMax}{80}

\pgfmathsetmacro{\fThreeMax}{1765/4}

\addplot[
    very thick,
     myblue
]
{
+((16*\ma^2 + \mb^2*(48-(24*\Kax)/(\Kax - \Ktx)))*exp(-\Ktx*x)-\mb^2*(48-(24*\Kax)/(\Kax - \Ktx))*exp(-\Kax*x))*\ma^4-((16*\ma^4 + \ma^2*\mb^2*(48 -(24*\Kax)/(\Kax - \Ktx)))*exp(-\Ktx*x) +(\ma^2*\mb^2*24*\Kax)/(\Kax - \Ktx)*exp(-\Kax*x)- 16*(\mb^4 + 3*\ma^2*\mb^2 + \ma^4))*\mb^2
};
\addlegendentry{$k_T/k_A = 1.7$}

\addplot[
    very thick,
     myred
]
{
+16*\ma^6*exp(-\Kty*x) -24* \ma^4*\mb^2*\Kty*x*(exp(-\Kty*x))-16*\ma^4*\mb^2*exp(-\Kty*x)-48*\ma^2*\mb^4*exp(-\Kty*x) +\ma^2*\mb^4*24*\Kty*x*exp(-\Kty*x)+ 16*(\mb^6 + 3*\ma^2*\mb^4 + \ma^4*\mb^2)
};
\addlegendentry{$k_T/k_A = 1.0$}

\addplot[
    very thick,
    orange
]
{
+((16*\ma^2 + \mb^2*(48-(24*\Kaz)/(\Kaz - \Ktz)))*exp(-\Ktz*x)-\mb^2*(48-(24*\Kaz)/(\Kaz - \Ktz))*exp(-\Kaz*x))*\ma^4-((16*\ma^4 + \ma^2*\mb^2*(48 -(24*\Kaz)/(\Kaz - \Ktz)))*exp(-\Ktz*x) +(\ma^2*\mb^2*24*\Kaz)/(\Kaz - \Ktz)*exp(-\Kaz*x)- 16*(\mb^4 + 3*\ma^2*\mb^2 + \ma^4))*\mb^2
};
\addlegendentry{$k_T/k_A = 0.3$}

\end{axis}
\end{tikzpicture}
\end{subfigure}

\vspace{0.1 cm}
\begin{subfigure}[t]{0.25\textwidth}
\begin{tikzpicture}
\node [] at (1.3,2.7) {$\mu_i=\mu_j$};
\node [] at (-0.5,2.9) {c};
\begin{axis}[
    width=0.95\textwidth,
    height=0.9\textwidth,
    xlabel={\footnotesize Time delay [ps]},
    ylabel={\footnotesize  Fifth-order signal},
    tick label style={font=\footnotesize},
    legend style={
        font=\scriptsize,
        draw=none,
        fill=none,
        at={(1.0,0.55)},
        anchor=north east,
        /tikz/every even column/.append style={column sep=2pt},
    },
    legend image post style={xscale=0.6},
    domain=0:15,
    samples=800,
    ymax=85,
]

\pgfmathsetmacro{\Kax}{1}
\pgfmathsetmacro{\Kay}{1}
\pgfmathsetmacro{\Kaz}{1}
\pgfmathsetmacro{\Ktx}{1.7}
\pgfmathsetmacro{\Kty}{1}
\pgfmathsetmacro{\Ktz}{0.3}
\pgfmathsetmacro{\ma}{1.0}
\pgfmathsetmacro{\mb}{1.0}

\pgfmathsetmacro{\Ka}{1/1}

\pgfmathsetmacro{\fOneMax}{
1485/4 + (135/2)*2/exp(2) - 108/exp(2)
}

\pgfmathsetmacro{\fTwoMax}{80}

\pgfmathsetmacro{\fThreeMax}{1765/4}

\addplot[
    very thick,
     myblue
]
{
+((16*\ma^2 + \mb^2*(48-(24*\Kax)/(\Kax - \Ktx)))*exp(-\Ktx*x)-\mb^2*(48-(24*\Kax)/(\Kax - \Ktx))*exp(-\Kax*x))*\ma^4-((16*\ma^4 + \ma^2*\mb^2*(48 -(24*\Kax)/(\Kax - \Ktx)))*exp(-\Ktx*x) +(\ma^2*\mb^2*24*\Kax)/(\Kax - \Ktx)*exp(-\Kax*x)- 16*(\mb^4 + 3*\ma^2*\mb^2 + \ma^4))*\mb^2
};
\addlegendentry{$k_T/k_A = 1.7$}

\addplot[
    very thick,
     myred
]
{
+16*\ma^6*exp(-\Kty*x) -24* \ma^4*\mb^2*\Kty*x*(exp(-\Kty*x))-16*\ma^4*\mb^2*exp(-\Kty*x)-48*\ma^2*\mb^4*exp(-\Kty*x) +\ma^2*\mb^4*24*\Kty*x*exp(-\Kty*x)+ 16*(\mb^6 + 3*\ma^2*\mb^4 + \ma^4*\mb^2)
};
\addlegendentry{$k_T/k_A = 1.0$}

\addplot[
    very thick,
    orange
]
{
+((16*\ma^2 + \mb^2*(48-(24*\Kaz)/(\Kaz - \Ktz)))*exp(-\Ktz*x)-\mb^2*(48-(24*\Kaz)/(\Kaz - \Ktz))*exp(-\Kaz*x))*\ma^4-((16*\ma^4 + \ma^2*\mb^2*(48 -(24*\Kaz)/(\Kaz - \Ktz)))*exp(-\Ktz*x) +(\ma^2*\mb^2*24*\Kaz)/(\Kaz - \Ktz)*exp(-\Kaz*x)- 16*(\mb^4 + 3*\ma^2*\mb^2 + \ma^4))*\mb^2
};
\addlegendentry{$k_T/k_A = 0.3$}

\end{axis}
\end{tikzpicture}
\end{subfigure}%
\begin{subfigure}[t]{0.25\textwidth}
\begin{tikzpicture}
\node [] at (1.3,2.7) {$\mu_i<\mu_j$};
\node [] at (-0.5,2.9) {d};

\begin{axis}[
    width=0.95\textwidth,
    height=0.9\textwidth,
    xlabel={\footnotesize Time delay [ps]},
    ylabel={\footnotesize  Fifth-order signal},
    tick label style={font=\footnotesize},
    legend style={
        font=\scriptsize,
        draw=none,
        fill=none,
        at={(1.0,0.55)},
        anchor=north east,
        /tikz/every even column/.append style={column sep=2pt},
    },
    legend image post style={xscale=0.6},
    domain=0:15,
    samples=800,
    ymin=150.0,
    ymax=480,
]

\pgfmathsetmacro{\Kax}{1}
\pgfmathsetmacro{\Kay}{1}
\pgfmathsetmacro{\Kaz}{1}
\pgfmathsetmacro{\Ktx}{1.7}
\pgfmathsetmacro{\Kty}{1}
\pgfmathsetmacro{\Ktz}{0.3}
\pgfmathsetmacro{\ma}{1.0}
\pgfmathsetmacro{\mb}{1.5}

\pgfmathsetmacro{\Ka}{1/1}

\pgfmathsetmacro{\fOneMax}{
1485/4 + (135/2)*2/exp(2) - 108/exp(2)
}

\pgfmathsetmacro{\fTwoMax}{80}

\pgfmathsetmacro{\fThreeMax}{1765/4}

\addplot[
    very thick,
     myblue
]
{
+((16*\ma^2 + \mb^2*(48-(24*\Kax)/(\Kax - \Ktx)))*exp(-\Ktx*x)-\mb^2*(48-(24*\Kax)/(\Kax - \Ktx))*exp(-\Kax*x))*\ma^4-((16*\ma^4 + \ma^2*\mb^2*(48 -(24*\Kax)/(\Kax - \Ktx)))*exp(-\Ktx*x) +(\ma^2*\mb^2*24*\Kax)/(\Kax - \Ktx)*exp(-\Kax*x)- 16*(\mb^4 + 3*\ma^2*\mb^2 + \ma^4))*\mb^2
};
\addlegendentry{$k_T/k_A = 1.7$}

\addplot[
    very thick,
     myred
]
{
+16*\ma^6*exp(-\Kty*x) -24* \ma^4*\mb^2*\Kty*x*(exp(-\Kty*x))-16*\ma^4*\mb^2*exp(-\Kty*x)-48*\ma^2*\mb^4*exp(-\Kty*x) +\ma^2*\mb^4*24*\Kty*x*exp(-\Kty*x)+ 16*(\mb^6 + 3*\ma^2*\mb^4 + \ma^4*\mb^2)
};
\addlegendentry{$k_T/k_A = 1.0$}

\addplot[
    very thick,
    orange
]
{
+((16*\ma^2 + \mb^2*(48-(24*\Kaz)/(\Kaz - \Ktz)))*exp(-\Ktz*x)-\mb^2*(48-(24*\Kaz)/(\Kaz - \Ktz))*exp(-\Kaz*x))*\ma^4-((16*\ma^4 + \ma^2*\mb^2*(48 -(24*\Kaz)/(\Kaz - \Ktz)))*exp(-\Ktz*x) +(\ma^2*\mb^2*24*\Kaz)/(\Kaz - \Ktz)*exp(-\Kaz*x)- 16*(\mb^4 + 3*\ma^2*\mb^2 + \ma^4))*\mb^2
};
\addlegendentry{$k_T/k_A = 0.3$}

\end{axis}
\end{tikzpicture}
\end{subfigure}
}

\caption{\justifying Example of spectrally-integrated signal of weakly coupled dimer. (a) Energy-level scheme of a weakly coupled heterodimer model, with energy transfer rate $k_T$ from molecule $i$ to molecule $j$, and EEA rate $k_A/2$ from two-exciton state $|ij\rangle$ to single-excited states $|i \rangle$ and $|j \rangle$. (b-d) Dependence of spectrally-integrated fifth-order signal on the ratio between transfer and annihilation rates $k_T/k_A$ for three different settings of transition dipole moments: (b) $\mu_i= 1.5$ and $\mu_j=1.0$, (c) $\mu_i= 1.0$ and $\mu_j=1.0$, (d) $\mu_i= 1.0$ and $\mu_j=1.5$.}
\label{Fig:dimer_integrated5PP}

\end{figure}
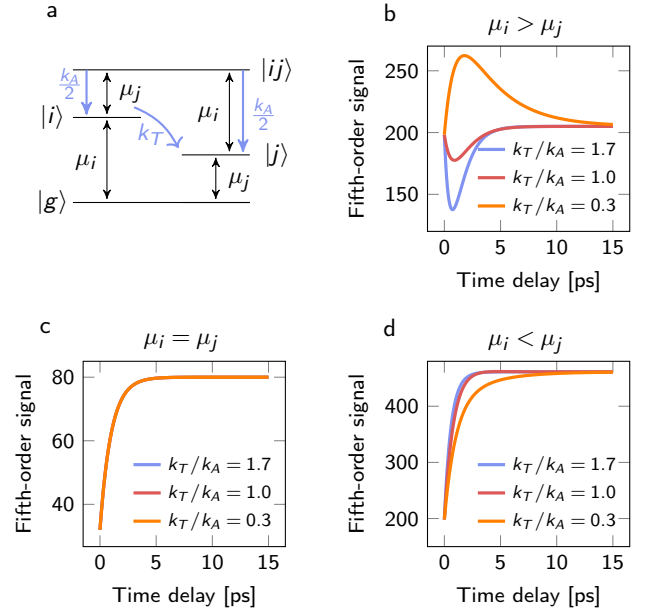

Furthermore, since the dimer has no three-exciton state and we neglect the contribution of higher-excited states to the spectrum, ESA$_2^{(5)}$(2E) is not present in the fifth-order pump--probe signal of this system. The dynamics of the system thus reduces to two possible processes (as depicted in Fig. \ref{Fig:dimer_integrated5PP}a): energy transfer from the single excited state $|i\rangle$ with higher energy to that with lower energy $|j\rangle$ with rate $k_T$, and annihilation of the two-exciton state $|ij\rangle$ to states $|i\rangle$ or $|j\rangle$, with rate $\frac{k_A}{2}$. For the form of the rate matrix, see Eq. 34 in the SI.

Considering all possible pathways (see Fig. \ref{Fig:FD_EEI2DES}) and summing them with coefficients according to Eq. \ref{EEI2D_pathways}, we get an analytical expression for the fifth-order pump--probe spectrum
\begin{equation} 
\begin{split} 
&\mbox{PP}^{(5)}(\omega,t)=16\mu_i^4 e^{-k_Tt}(\mu_i^2\mathcal{L}_{i,g}(\omega)-\mu_j^2\mathcal{L}_{j,g}(\omega)) \\
+&\mu_i^2\mu_j^2\left(48-\tfrac{24k_A}{k_A - k_T}\right)\left(e^{-k_Tt}-e^{-k_At}\right)\left(\mu_i^2\mathcal{L}_{i,g}(\omega)-\mu_j^2\mathcal{L}_{j,g}(\omega)\right) \\
+&16\left(\mu_j^6+3\mu_i^2 \mu_j^4 +\mu_i^4\mu_j^2-3\mu_i^2 \mu_j^4 e^{-k_At}\right)\mathcal{L}_{j,g}(\omega).
\end{split} 
\label{eq_dimer_weak_EEI2D} 
\end{equation} 

We start by comparing the spectrally integrated signal for varying relative strengths of the transition dipole moments, as reported in Fig. \ref{Fig:dimer_integrated5PP}b-d.
For equal transition dipole moments ($\mu_i=\mu_j=\mu$, Fig. \ref{Fig:dimer_integrated5PP}c), the frequency integration of Eq. \ref{eq_dimer_weak_EEI2D} gives 
\begin{equation} 
\int d\omega \ \text{PP}^{(5)}_{\mu_i=\mu_j}(\omega,t)=16\mu^6\left(5-3 e^{-k_At}\right)
\label{eq_dimer_weak_EEI2D_mua=mub_integ} 
\end{equation} 
as the first two rows are zero. In this specific case, the spectrally integrated signal exponentially grows due to the annihilation process, while the dependence on the energy transfer rate is removed. This limit was already considered in our previous work.\cite{maly2023separating} 

In the more general scenario, when $\mu_i \neq \mu_j$, the signal depends on both transfer $k_T$ and annihilation $k_A$ rates, as appears from Eq. \ref{eq_dimer_weak_EEI2D}. Even if we perform the spectral integration analogously to Eq. \ref{eq_dimer_weak_EEI2D_mua=mub_integ}, the terms with $k_T$ do not cancel because of the different prefactor associated with the lineshapes: $\mu_i^2$ for $\mathcal{L}_{ig}(\omega)$ and $\mu_j^2$ for $\mathcal{L}_{jg}(\omega)$. In particular, for $\mu_i>\mu_j$, the signal decay is no longer monotonous (see Fig. \ref{Fig:dimer_integrated5PP}b).
This demonstrates that even the simplest case of a heterodimer already violates the common assumption that integrated PP$^{(5)}$ primarily report on annihilation dynamics.\cite{shi2025annihilation,zhang_probing_2025} Indeed, the spectrally integrated signal mixes transfer and annihilation processes. While it is possible to develop a fitting model that includes both annihilation and transfer rates, the most straightforward approach is to consider the spectrally resolved signal, which is furthermore more sensitive to the system parameters than the spectrally integrated one, as later discussed in Sec. \ref{Sec:parameters_sensitivity}.

In the SI, we report the comparison of the analytical model and numerical methodology presented in Sec. \ref{Sec:model}.

\subsection{Squaraine co-polymer: simulations and experiments}
\label{Sec:examples}
\begin{figure*}[t!]
{\sffamily
\sansmath
\begin{subfigure}[t]{0.245\textwidth}
\centering
\vspace{0.7cm} 
  \begin{overpic}[width=\textwidth]{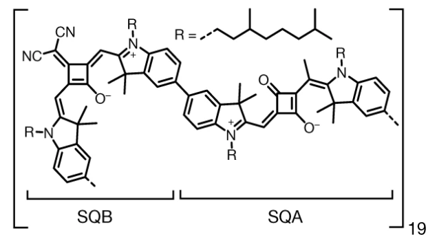}
    \put(-5,75){\footnotesize  a}
    \end{overpic}
    \label{Fig:SQAB19_formula}
\end{subfigure}%
\begin{subfigure}[t]{0.09\textwidth}
\begin{tikzpicture}
    \node [white] at (0.0,0) {$1$};
\end{tikzpicture}
\end{subfigure}%
\begin{subfigure}[t]{0.215\textwidth}
\vspace{0pt}
  \begin{overpic}[width=0.9\textwidth]{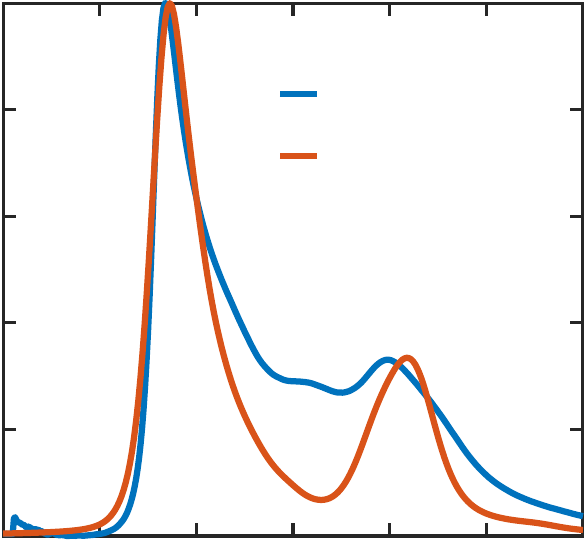}
      \put(-35,96){\footnotesize  b}
  \put(-4,-8){\footnotesize 11}
  \put(13,-8){\footnotesize  12}
  \put(30,-8){\footnotesize  13}
  \put(46,-8){\footnotesize 14}
  \put(62,-8){\footnotesize  15}
  \put(79,-8){\footnotesize  16}
  \put(96,-8){\footnotesize  17}
  \put(5,-18.5){\footnotesize  Wavenumber [$10^{3}$ cm$^{-1}$]}
    \put(-26,1){\footnotesize \rotatebox{90}{Normalized absorption}}
    \put(-12,89){\footnotesize  \rotatebox{0}{$1.0$}}
    \put(-12,71){\footnotesize  \rotatebox{0}{$0.8$}}
    \put(-12,53){\footnotesize  \rotatebox{0}{$0.6$}}
    \put(-12,34){\footnotesize  \rotatebox{0}{$0.4$}}
    \put(-12,16){\footnotesize  \rotatebox{0}{$0.2$}}
    \put(-12,-1){\footnotesize  \rotatebox{0}{$0.0$}}
      \put(60,75){\footnotesize  exp.}
      \put(60,64){\footnotesize  sim.}
      \end{overpic}
    \label{Fig:SQAB19_Abs}
\end{subfigure}%
\begin{subfigure}[t]{0.1\textwidth}
\begin{tikzpicture}
    \node [white] at (0.0,0) {$1$};
\end{tikzpicture}
\end{subfigure}%
\begin{subfigure}[t]{0.215\textwidth}
\vspace{0pt}                   
  \begin{overpic}[width=0.9\textwidth]{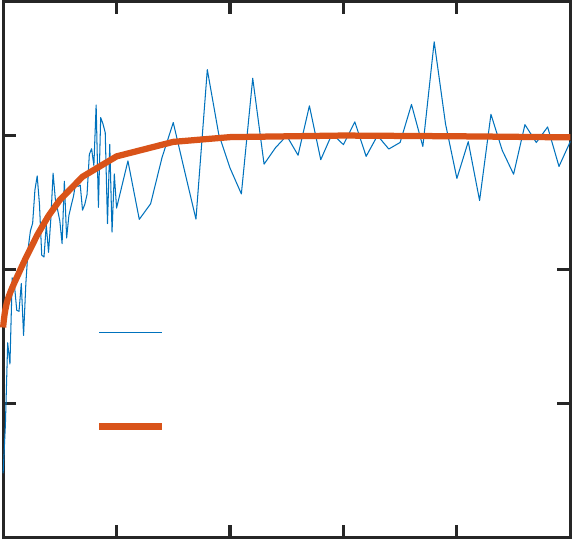}
        \put(-40,97){\footnotesize  c}
         \put(32,34){\footnotesize  exp.}
      \put(32,18){\footnotesize  sim.}
        \put(-1,-8){\footnotesize  0}
  \put(18,-8){\footnotesize  1}
  \put(38,-8){\footnotesize  2}
  \put(58,-8){\footnotesize 3}
  \put(78,-8){\footnotesize  4}
  \put(96,-8){\footnotesize  5}
  \put(23,-18.5){\footnotesize  Time delay [ps]}
      \put(-34,10){\footnotesize  \rotatebox{90}{Integrated fifth-order }}
  \put(-25,6){\footnotesize  \rotatebox{90}{Transient Signal [rel.u.]}}
    \put(-12,90.){\footnotesize  \rotatebox{0}{$1.5$}}
    \put(-12,68){\footnotesize  \rotatebox{0}{$1.0$}}
    \put(-12,45){\footnotesize  \rotatebox{0}{$0.5$}}
    \put(-12,21){\footnotesize  \rotatebox{0}{$0.0$}}
    \put(-18.5,-0.5){\footnotesize  \rotatebox{0}{$-0.5$}}
   \end{overpic}
    \label{Fig:SQAB19_dynamics}
\end{subfigure}
\vspace{1.1cm}

\begin{subfigure}[t]{0.04\textwidth}
\begin{tikzpicture}
    \node [white] at (0.0,0) {$1$};
\end{tikzpicture}
\end{subfigure}%
\begin{subfigure}[t]{0.215\textwidth}
\centering
\begin{overpic}[width=0.95\textwidth]{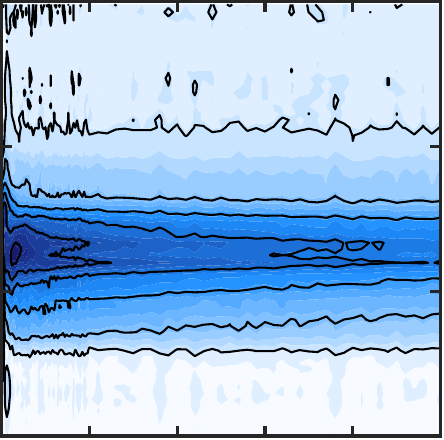}
      \put(-25,107){\footnotesize  d}
  \put(25,103){\footnotesize PP$^{(3)}$ - experiment}
    \put(-1,-8){\footnotesize  0}
  \put(18,-8){\footnotesize  1}
  \put(38,-8){\footnotesize  2}
  \put(58,-8){\footnotesize 3}
  \put(78,-8){\footnotesize  4}
  \put(96,-8){\footnotesize  5}
  \put(23,-18.5){\footnotesize  Time delay [ps]}
      \put(-31,6){\footnotesize  \rotatebox{90}{Wavenumber [$10^{3}$ cm$^{-1}$]}}
    \put(-18.5,95.75){\footnotesize  \rotatebox{0}{$13.5$}}
    \put(-18.5,63){\footnotesize  \rotatebox{0}{$13.0$}}
    \put(-18.5,30.25){\footnotesize  \rotatebox{0}{$12.5$}}
    \put(-18.5,0){\footnotesize  \rotatebox{0}{$12.0$}}
\end{overpic}
\label{Fig:SQAB_3PP_exp}
\end{subfigure}%
\begin{subfigure}[t]{0.215\textwidth}
\centering
\begin{overpic}[width=0.95\textwidth]{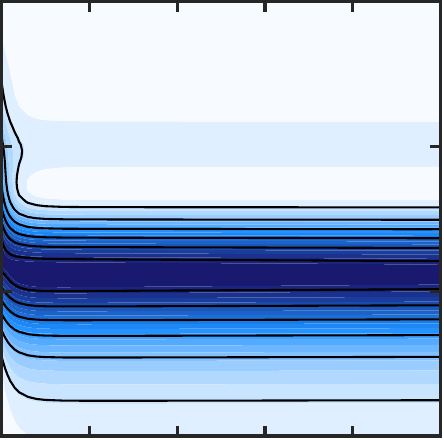}
  \put(15,103){\footnotesize PP$^{(3)}$, $N=10$, set 1}
    \put(-1,-8){\footnotesize  0}
  \put(18,-8){\footnotesize  1}
  \put(38,-8){\footnotesize  2}
  \put(58,-8){\footnotesize 3}
  \put(78,-8){\footnotesize  4}
  \put(96,-8){\footnotesize  5}
  \put(23,-18.5){\footnotesize  Time delay [ps]}
\end{overpic}
\label{Fig:SQAB_3PP}
\end{subfigure}%
\begin{subfigure}[t]{0.215\textwidth}
\centering
\begin{overpic}[width=0.95\textwidth]{data/SQAB_5PP_Exp.pdf}
      \put(-1,107){\footnotesize  e}
  \put(25,103){\footnotesize PP$^{(5)}$ - experiment}
    \put(-1,-8){\footnotesize  0}
  \put(18,-8){\footnotesize  1}
  \put(38,-8){\footnotesize  2}
  \put(58,-8){\footnotesize 3}
  \put(78,-8){\footnotesize  4}
  \put(96,-8){\footnotesize  5}
  \put(23,-18.5){\footnotesize  Time delay [ps]}
\end{overpic}
\label{Fig:SQAB_5PP_exp}
\end{subfigure}%
\begin{subfigure}[t]{0.215\textwidth}
\centering
\begin{overpic}[width=0.95\textwidth]{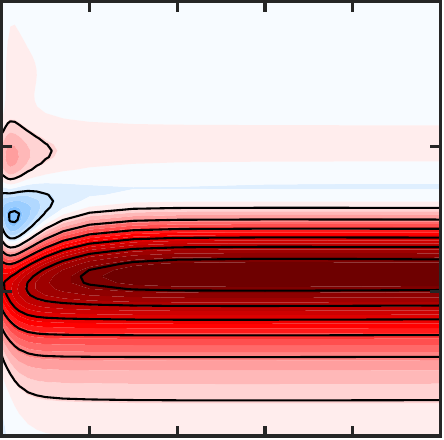}
  \put(15,103){\footnotesize PP$^{(5)}$, $N=10$, set 1}
      \put(-1,-8){\footnotesize  0}
  \put(18,-8){\footnotesize  1}
  \put(38,-8){\footnotesize  2}
  \put(58,-8){\footnotesize 3}
  \put(78,-8){\footnotesize  4}
  \put(96,-8){\footnotesize  5}
  \put(23,-18.5){\footnotesize  Time delay [ps]}
    \put(101.,0){\includegraphics[width=0.0455\textwidth]{data/colorbar_cropped-2.pdf}}
  \put(111,96){\footnotesize   1.0}
  \put(111,71.5){\footnotesize   $0.5$}
  \put(111,47){\footnotesize  $0.0$}
  \put(106,22.5){\footnotesize  $-0.5$}
  \put(106,-2){\footnotesize  $-1.0$} 
\end{overpic}
\label{Fig:SQAB_5PP}
\end{subfigure}
\vspace{0.55cm}
}
\caption{\justifying Experimental and simulated linear, third- and fifth-order signals of $[\text{SQA}-\text{SQB}]_{19}$ polymer. (a) Chemical structure of the polymer. (b) Simulated and experimental linear absorption spectra. Notice that the simulated spectrum underestimates the intensity in the region around 14000 cm$^{-1}$, suggesting the presence of shorter chain segments in the sample (see Sec. \ref{Sec:N_dependence}). (c) Comparison between simulated and experimental spectrally integrated PP$^{(5)}$ signal as a function of the waiting time. This provides a first validation of the parameters of higher-excited states. d) Comparison of experimental and simulated spectra shows that the negative peak near 13000 cm$^{-1}$ originates mainly from SE$_2^{(5)}$(2E), whereas the main positive peak corresponds to NGSB$^{(5)}$, NSE$^{(5)}$, and SE$_2^{(5)}$(EEA).}
\label{Fig:SQAB19_spectra}
\end{figure*}

In the following, we consider a more complex system such as a squaraine co-polymer $[\text{SQA}-\text{SQB}]_{19}$, which consists of nineteen dimeric units, each composed of a squaraine A (SQA) and a squaraine B (SQB) molecule (Fig. \ref{Fig:SQAB19_spectra}a).
To study spectral properties of the fifth-order signal of realistic samples, we compare our simulations with experimental pump--probe spectra. The third- and fifth-order data presented here were isolated by intensity cycling and previously published in Ref. \citenum{maly2023separating}. Aided by the numerical simulations, we now provide a detailed interpretation of the spectrally resolved signal.

For simulations, we used the model presented in Sec. \ref{Sec:model} with parameters reported in Table 2 in the SI. 
Although the actual geometry of $[\text{SQA}-\text{SQB}]_{19}$ is not known, it is expected to form a chain-like structure.\cite{ress2023time} Therefore, we model the polymer as a structurally disordered chain with correlated random sampling of angles with Gaussian distribution (see Sec. 1.2 in the SI for a detailed explanation).

To reliably reproduce the experimental data and avoid the risk of overfitting, we first set the energy of first-excited states and transition dipole moment strengths of SQA and SQB based on previous studies.\cite{schreck2018synthesis} Consequently, we adopted a sequential fitting procedure: first, we fitted the absorption and fluorescence spectra of SQA and SQB monomers by choosing ideal bath parameters; next, absorption and emission spectra of polymer by optimizing the energetic and structural disorder parameters of a polymer model loosely based on previous findings;\cite{schreck2018synthesis,ress2023time} and finally, we adjusted parameters related to the higher-excited $\ket{f_{i}}$ states to achieve an agreement with fifth-order pump--probe spectra.
As will be discussed later in Sec. \ref{Sec:N_dependence}, we assume that, due to structural disorder, the $[\text{SQA}-\text{SQB}]_{19}$ chain is effectively divided into smaller segments.\cite{Malyshev1999, malyshev_channels_2000, Ryzhov2001} For this reason, and to lower the computational demands, we decided to simulate a polymer with five dimeric units, namely $[SQA-SQB]_{5}$. We note, nevertheless, that simulation of the full 19-mer is computationally feasible with the present approach, merely time-demanding when considering averaging over disorder.

The absorption spectrum of the squaraine copolymer from both experiments and simulations is shown in Fig. \ref{Fig:SQAB19_spectra}b.
In the experimental one, we observe three absorption peaks (blue line). The two edge peaks at $13000$\,cm$^{-1}$ and $15000$\,cm$^{-1}$ are well captured by our simulations. These correspond to the two main excitonic states arising from the coupling-induced splitting of first-excited states of SQA and SQB in the polymer chain.\cite{maly2020wavelike}
Instead, our model is not able to reproduce the middle peak around $14000$\,cm$^{-1}$. As we will discuss later in Sec.\ref{Sec:N_dependence}, this feature may originate from shorter chain segments or structural defects.

Nevertheless, only the negative peak around $13000$\,cm$^{-1}$ appears in third-order pump--probe spectra (see Fig. \ref{Fig:SQAB19_spectra}d), since the other transitions fall outside the frequency window covered by the probe pulse in the experiment. In particular, we observe a prominent frequency shift of the peak during the waiting time due to relaxation dynamics within the single-exciton manifold. In contrast to the experiment, the main peak in simulated PP$^{(3)}(\omega)$ is not decaying due to improper cancellation of SE and ESA pathways, as will be further discussed in Sec. \ref{Sec:Discussion}. 

\begin{figure*}[t!]
{\sffamily
\sansmath

\begin{subfigure}[t]{0.05\textwidth}
\begin{tikzpicture}
    \node [white] at (0.0,0) {$1$};
\end{tikzpicture}
\end{subfigure}%
\begin{subfigure}[t]{0.215\textwidth}
\centering
\begin{overpic}[width=0.95\textwidth]{data/10_5PP.pdf}
  \put(15,103){\footnotesize  PP$^{(5)}$, $N=10$, set 1}
        \put(-31,6){\footnotesize  \rotatebox{90}{Wavenumber [$10^{3}$ cm$^{-1}$]}}
    \put(-18.5,95.75){\footnotesize  \rotatebox{0}{$13.5$}}
    \put(-18.5,63){\footnotesize  \rotatebox{0}{$13.0$}}
    \put(-18.5,30.25){\footnotesize  \rotatebox{0}{$12.5$}}
    \put(-18.5,0){\footnotesize  \rotatebox{0}{$12.0$}}
\end{overpic}
\label{Fig:SQAB_5PP_GSB}
\end{subfigure}%
\begin{subfigure}[t]{0.215\textwidth}
\centering
\begin{overpic}[width=0.95\textwidth]{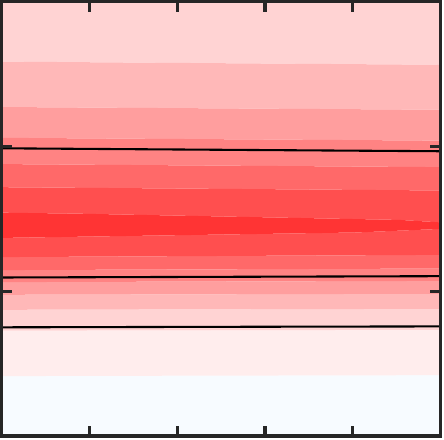}
  \put(40,103){\footnotesize NGSB$^{(5)}$}
\end{overpic}
\label{Fig:SQAB_5PP_total}
\end{subfigure}%
\begin{subfigure}[t]{0.215\textwidth}
\centering
\begin{overpic}[width=0.95\textwidth]{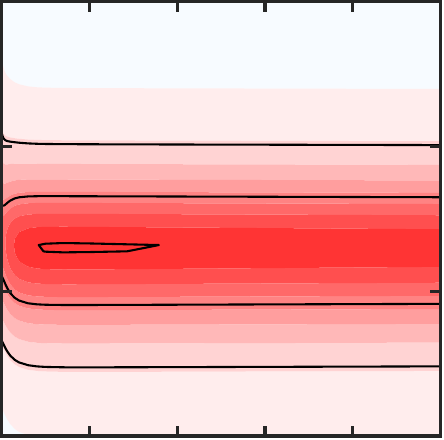}
  \put(40,103){\footnotesize NSE$^{(5)}$}
\end{overpic}
\label{Fig:SQAB_5PP_SE1}
\end{subfigure}%
\begin{subfigure}[t]{0.215\textwidth}
\centering
\begin{overpic}[width=0.95\textwidth]{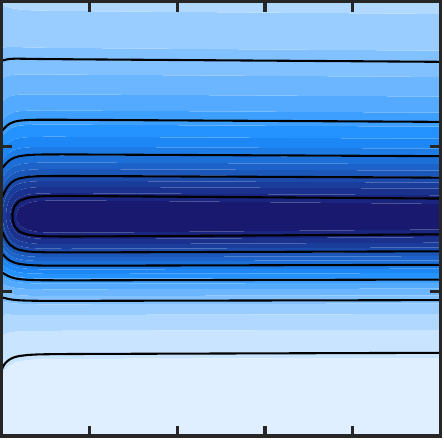}
  \put(40,103){\footnotesize NESA$^{(5)}$}
  \put(101.,-69){\includegraphics[width=0.056\textwidth]{data/colorbar_cropped-2.pdf}}
  \put(111,49){\footnotesize   1.0}
  \put(111,19){\footnotesize   $0.5$}
  \put(111,-11){\footnotesize  $0.0$}
  \put(106.5,-41){\footnotesize  $-0.5$}
  \put(106.5,-71){\footnotesize  $-1.0$} 
\end{overpic}
\label{Fig:SQAB_5PP_ESA1}
\end{subfigure}
\vspace{0.55cm}

\begin{subfigure}[t]{0.05\textwidth}
\begin{tikzpicture}
    \node [white] at (0.0,0) {$1$};
\end{tikzpicture}
\end{subfigure}%
\begin{subfigure}[t]{0.215\textwidth}
\centering
\begin{overpic}[width=0.95\textwidth]{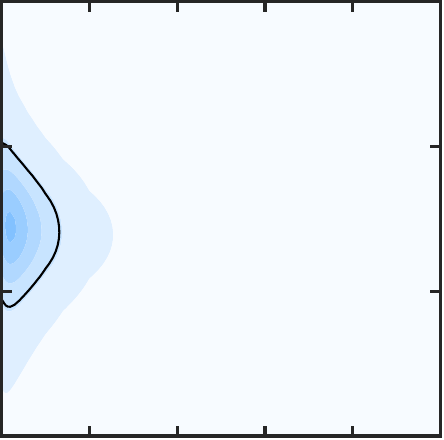}
  \put(35,103){\footnotesize SE$_2^{(5)}$ (2E)}
    \put(-1,-8){\footnotesize  0}
  \put(18,-8){\footnotesize  1}
  \put(38,-8){\footnotesize  2}
  \put(58,-8){\footnotesize 3}
  \put(78,-8){\footnotesize  4}
  \put(96,-8){\footnotesize  5}
  \put(23,-18.5){\footnotesize  Time delay [ps]}
      \put(-31,6){\footnotesize  \rotatebox{90}{Wavenumber [$10^{3}$ cm$^{-1}$]}}
    \put(-18.5,95.75){\footnotesize  \rotatebox{0}{$13.5$}}
    \put(-18.5,63){\footnotesize  \rotatebox{0}{$13.0$}}
    \put(-18.5,30.25){\footnotesize  \rotatebox{0}{$12.5$}}
    \put(-18.5,0){\footnotesize  \rotatebox{0}{$12.0$}}
\end{overpic}
\label{Fig:SQAB_5PP_SE2}
\end{subfigure}%
\begin{subfigure}[t]{0.215\textwidth}
\centering
\begin{overpic}[width=0.95\textwidth]{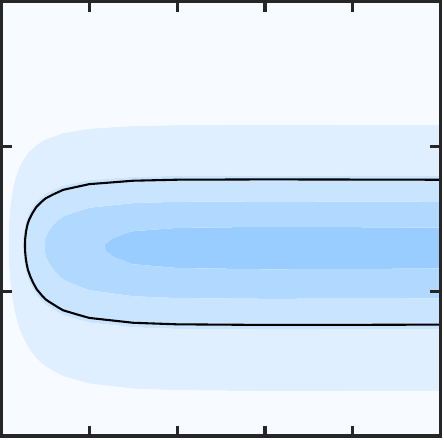}
  \put(35,103){\footnotesize SE$_2^{(5)}$(EEA)}
    \put(-1,-8){\footnotesize  0}
  \put(18,-8){\footnotesize  1}
  \put(38,-8){\footnotesize  2}
  \put(58,-8){\footnotesize 3}
  \put(78,-8){\footnotesize  4}
  \put(96,-8){\footnotesize  5}
  \put(23,-18.5){\footnotesize  Time delay [ps]}
\end{overpic}
\label{Fig:SQAB_5PP_SE_EEA}
\end{subfigure}%
\begin{subfigure}[t]{0.215\textwidth}
\centering
\begin{overpic}[width=0.95\textwidth]{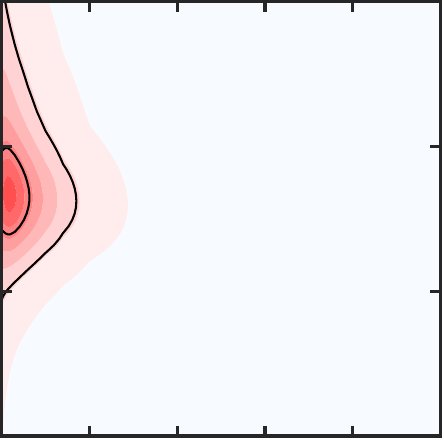}
  \put(35,103){\footnotesize ESA$_2^{(5)}$(2E)}
    \put(-1,-8){\footnotesize  0}
  \put(18,-8){\footnotesize  1}
  \put(38,-8){\footnotesize  2}
  \put(58,-8){\footnotesize 3}
  \put(78,-8){\footnotesize  4}
  \put(96,-8){\footnotesize  5}
  \put(23,-18.5){\footnotesize  Time delay [ps]}
\end{overpic}
\label{Fig:SQAB_5PP_ESA2}
\end{subfigure}%
\begin{subfigure}[t]{0.215\textwidth}
\centering
\begin{overpic}[width=0.95\textwidth]{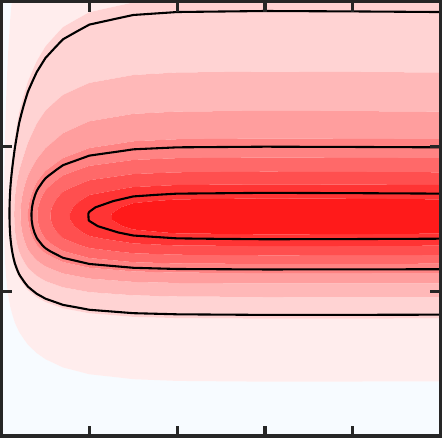}
  \put(35,103){\footnotesize ESA$_2^{(5)}$(EEA)}
    \put(-1,-8){\footnotesize  0}
  \put(18,-8){\footnotesize  1}
  \put(38,-8){\footnotesize  2}
  \put(58,-8){\footnotesize 3}
  \put(78,-8){\footnotesize  4}
  \put(96,-8){\footnotesize  5}
  \put(23,-18.5){\footnotesize  Time delay [ps]}
\end{overpic}
\label{Fig:SQAB_5PP_ESA_EEA}
\end{subfigure}
\vspace{0.55cm}
}
\caption{\justifying Decomposition of the simulated fifth-order pump--probe signalinto the various excitation pathways. The negative decaying signal around $12750$ cm$^{-1}$ in the total PP$^{(5)}$ corresponds to the SE$_2^{(5)}$(2E) pathway.}
\label{Fig:SQAB19_spectra_pathways}
\end{figure*}

Instead, the experimental fifth-order spectrum shows a prominent positive rising peak and a weaker negative decaying feature, which promises insights into double-excited state dynamics (see Fig. \ref{Fig:SQAB19_spectra}e). 
In the simulations, the total signal can be decomposed into contributions of multiple excitation pathways (see Fig. \ref{Fig:FD_EEI2DES}), whose spectra are presented in Fig. \ref{Fig:SQAB19_spectra_pathways}. 
NGSB$^{(5)}$, NSE$^{(5)}$, and NESA$^{(5)}$ contributions each exhibit a broad peak and correspond to the negation of the third-order processes, but with opposite sign.\cite{rose2023interpretations}
During the waiting time, the former reflects ground-state dynamics and remains approximately constant. Instead, the latter two report on one-exciton dynamics and increase due to population transfer from states lying outside the detection frequency window, to the lower-exciton one, which appears in the spectra.
In contrast, SE$_2^{(5)}$- and ESA$_2^{(5)}$-type pathways report on two-exciton dynamics during the waiting time (Fig. \ref{Fig:FD_EEI2DES}).
Due to EEA from the double- to single-excited states, SE$_2^{(5)}$(2E) and ESA$_2^{(5)}$(2E) signals decay, while SE$_2^{(5)}$(EEA) and ESA$_2^{(5)}$(EEA) signals grow during the waiting time.

Therefore, the total PP$^{(5)}$ spectrum results from a delicate balance between different spectral contributions and encodes both single- and double-exciton dynamics (see Fig. \ref{Fig:SQAB19_spectra_pathways}).
However, one- and two-exciton dynamics are intrinsically mixed in the fifth-order pump--probe spectra, because SE$_2^{(5)}$(EEA) and ESA$_2^{(5)}$(EEA) pathways share the same lineshape with NSE$^{(5)}$ and NESA$^{(5)}$, respectively. Together with NGSB$^{(5)}$, these contributions result in the main positive peak, which rises during the waiting time. We point out that the partial cancellation of those pathways leads to a significantly sharper peak compared to those of individual excitation pathways, as further discussed in Sec. \ref{Sec:N_dependence}. Instead, the negative decaying peak, whose description is one of the main objectives of this work, can be attributed to the SE$_2^{(5)}$(2E). In the simulations, an additional weak positive peak is present around $13000$\,cm$^{-1}$, which corresponds to the decaying ESA$_2^{(5)}$(2E). This interpretation is supported by comparison with the spectra of each pathway in Fig. \ref{Fig:SQAB19_spectra_pathways}. 
Although the negative peak decays as a result of EEA, the rate obtained from the fitting does not directly correspond to the annihilation rate. This is because the peak overlaps with contributions from other pathways, which slightly alter the dynamics. For this analysis, it is best to choose the peak whose overlap with other pathways is minimal, as previously done by Heshmatpour et al.\cite{Heshmatpour2020} 

The fifth-order pump--probe data thus reflect both single-and two-exciton state transitions, excitonic relaxation and annihilation. The spectral resolution does provide a two-exciton-specific feature in the form of the negative-signed peak, albeit still partially overlapping with other contributions. 

\section{Discussion}
\label{Sec:Discussion}
The simulations presented above provide a microscopic interpretation of the experimental fifth-order spectra. In the following, we discuss which information can be extracted from spectrally resolved PP$^{(5)}$ signal, how this differs from the spectrally integrated one, and the main limitations of the present model.

\subsection{Spectral features and information encoded in them}
The main advantage of spectrally resolved fifth-order pump–probe spectroscopy is that it separates contributions that are mixed in spectrally integrated signals. While the integrated signal combines energy transfer and EEA, spectral resolution reveals spectral features of individual excitation pathways. Even though one- and two-exciton-manifold contributions overlap even in the fifth-order pump--probe spectra, by fitting the spectrally resolved signal, we can get access to individual excitation pathway contributions and, in this way, disentangle them.
This can be directly seen in the $[\text{SQA}-\text{SQB}]_{19}$ polymer, where the additional negative feature around 13000$^{-1}$\;cm originates predominantly from the SE$_2^{(5)}$(2E) pathway. This negative peak, which is mirrored in higher-order spectra but not present in PP$^{(3)}$,\cite{maly2023separating} proves that fifth-order spectra contain richer information than third order. Indeed, it cannot be simply reduced to a saturating exponential rise of the third-order signal, as previously proposed for the spectrally integrated case, $\text{PP}^{(5)}(t)\propto\text{PP}^{(3)}(t)\cdot(1-e^{-k_At})$.\cite{maly2023separating}

\subsection{Sensitivity of fifth-order pump--probe to higher-excited state properties}
\label{Sec:parameters_sensitivity}
\begin{figure}[t!]
{\sffamily
\sansmath
\begin{subfigure}[t]{0.25\textwidth}
\hspace{2.cm}
\vspace{-0.4cm}
\begin{overpic}[width=0.7\textwidth]{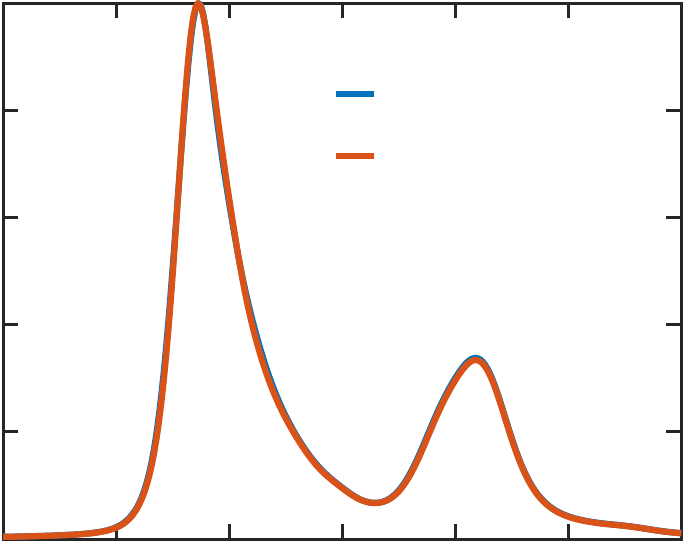}
  \put(2,87){\footnotesize  \textcolor{white}{short chain}}
  \put(94,-7){\footnotesize 18}
  \put(80,-7){\footnotesize 17}
  \put(66.5,-7){\footnotesize 16}
  \put(52,-7){\footnotesize 15}
  \put(38,-7){\footnotesize 14}
  \put(24,-7){\footnotesize 13}
  \put(10,-7){\footnotesize 12}
  \put(-4,-7){\footnotesize 11}
  \put(2,-16){\footnotesize Wavenumber [$10^{3}$\;cm$^{-1}$]}
  \put(-22,-7){\footnotesize  \rotatebox{90}{Normalized absorption}}
  \put(-13,-1){\footnotesize  0.0}
  \put(-13,14.5){\footnotesize  0.2}
  \put(-13,30.5){\footnotesize  0.4}
  \put(-13,45){\footnotesize  0.6}
  \put(-13,60){\footnotesize  0.8}
  \put(-13,74){\footnotesize  1.0}
   \put(60,65){\footnotesize  set 2}
   \put(60,55){\footnotesize  set 1}
      \put(-15,92.9){a}
   
\end{overpic}
\label{Fig:abs_correct_vs_wrong}
\end{subfigure}%
\begin{subfigure}[t]{0.25\textwidth}
\hspace{2cm}
\begin{overpic}[width=0.7\textwidth]{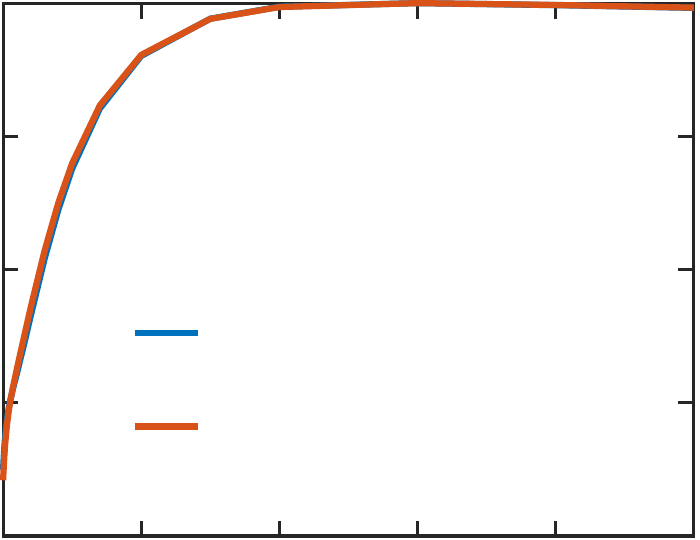}
  \put(96,-7){\footnotesize 5}
  \put(77.3,-7){\footnotesize 4}
  \put(57.5,-7){\footnotesize 3}
  \put(38,-7){\footnotesize 2}
  \put(18.,-7){\footnotesize 1}
  \put(-1,-7){\footnotesize 0}
  \put(25,-16){\footnotesize Time delay [ps]}
  \put(-32,-3){\footnotesize  \rotatebox{90}{Integrated fifth-order }}
  \put(-23,-7){\footnotesize  \rotatebox{90}{Transient Signal [rel.u.]}}
  \put(-13,-1){\footnotesize  0.0}
  \put(-13,15){\footnotesize  0.2}
  \put(-13,31){\footnotesize  0.4}
  \put(-13,46){\footnotesize  0.6}
  \put(-13,62){\footnotesize  0.8}
  \put(-13,77){\footnotesize  1.0}
     \put(30,28){\footnotesize  set 2}
   \put(30,15){\footnotesize  set 1}
   \put(-15,92.9){b}
\end{overpic}
\label{Fig:trace_correct_vs_wrong}
\end{subfigure}

\vspace{1.2cm}

\begin{subfigure}[t]{0.25\textwidth}
\RaggedLeft
\begin{overpic}[width=0.7\textwidth]{data/10_3PP.pdf}
\put(15,103){\footnotesize PP$^{(3)}$, $N=10$, set 1}
   \put(-1,-8){\footnotesize  0}
  \put(18,-8){\footnotesize  1}
  \put(38,-8){\footnotesize  2}
  \put(58,-8){\footnotesize 3}
  \put(78,-8){\footnotesize  4}
  \put(96,-8){\footnotesize  5}
  \put(23,-18.5){\footnotesize  Time delay [ps]}
      \put(-31,-2){\footnotesize  \rotatebox{90}{Wavenumber [$10^{3}$ cm$^{-1}$]}}
    \put(-18.5,95.75){\footnotesize  \rotatebox{0}{$13.5$}}
    \put(-18.5,63){\footnotesize  \rotatebox{0}{$13.0$}}
    \put(-18.5,30.25){\footnotesize  \rotatebox{0}{$12.5$}}
    \put(-18.5,0){\footnotesize  \rotatebox{0}{$12.0$}}
       \put(-35,105){c}
\end{overpic}
\label{Fig:3PP_correct}
\end{subfigure}%
\begin{subfigure}[t]{0.25\textwidth}
\RaggedRight
\begin{overpic}[width=0.7\textwidth]{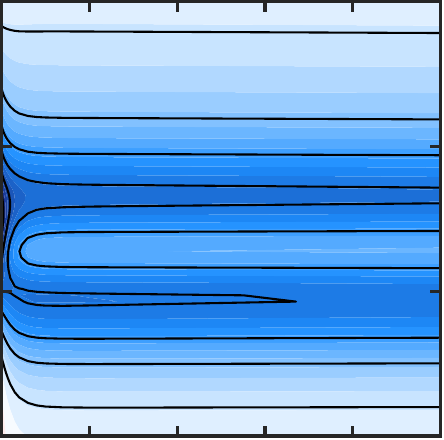}
\put(15,103){\footnotesize PP$^{(3)}$, $N=10$, set 2}
    \put(-1,-8){\footnotesize  0}
  \put(18,-8){\footnotesize  1}
  \put(38,-8){\footnotesize  2}
  \put(58,-8){\footnotesize 3}
  \put(78,-8){\footnotesize  4}
  \put(96,-8){\footnotesize  5}
  \put(23,-18.5){\footnotesize  Time delay [ps]}
     \put(-2,105){d}
\end{overpic}
\label{Fig:3PP_wrong}
\end{subfigure} 
\vspace{1.cm}

\begin{subfigure}[t]{0.25\textwidth}
\RaggedLeft
\begin{overpic}[width=0.7\textwidth]{data/10_5PP.pdf}
\put(15,103){\footnotesize PP$^{(5)}$, $N=10$, set 1}
    \put(-1,-8){\footnotesize  0}
  \put(18,-8){\footnotesize  1}
  \put(38,-8){\footnotesize  2}
  \put(58,-8){\footnotesize 3}
  \put(78,-8){\footnotesize  4}
  \put(96,-8){\footnotesize  5}
  \put(23,-18.5){\footnotesize  Time delay [ps]}
      \put(-31,-2){\footnotesize  \rotatebox{90}{Wavenumber [$10^{3}$ cm$^{-1}$]}}
    \put(-18.5,95.75){\footnotesize  \rotatebox{0}{$13.5$}}
    \put(-18.5,63){\footnotesize  \rotatebox{0}{$13.0$}}
    \put(-18.5,30.25){\footnotesize  \rotatebox{0}{$12.5$}}
    \put(-18.5,0){\footnotesize  \rotatebox{0}{$12.0$}}
       \put(-35,105){e}
\end{overpic}
\label{Fig:5PP_correct}
\end{subfigure}%
\begin{subfigure}[t]{0.25\textwidth}
\RaggedRight
\begin{overpic}[width=0.7\textwidth]{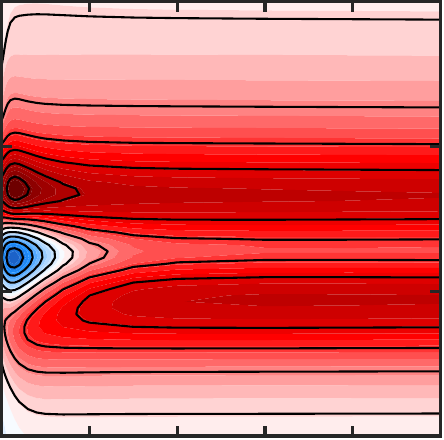}
\put(15,103){\footnotesize PP$^{(5)}$, $N=10$, set 2}
    \put(-1,-8){\footnotesize  0}
  \put(18,-8){\footnotesize  1}
  \put(38,-8){\footnotesize  2}
  \put(58,-8){\footnotesize 3}
  \put(78,-8){\footnotesize  4}
  \put(96,-8){\footnotesize  5}
  \put(23,-18.5){\footnotesize  Time delay [ps]}  
  \put(101.5,60.5){\includegraphics[width=0.04\textwidth]{data/colorbar_cropped-2.pdf}}
  \put(111,175){\footnotesize   1.0}
  \put(111,146){\footnotesize   0.5}
  \put(111,117){\footnotesize   0.0}
  \put(108,89){\footnotesize  -0.5}
  \put(108,61){\footnotesize  -1.0}
     \put(-2,105){f}
\end{overpic}
\label{Fig:5PP_wrong}
\end{subfigure}
\vspace{0.5cm}
}

\caption{\justifying Sensitivity to the higher excited state properties. (Parameter set 1) $F_J=1.25$, $k_{IC}=1/30$ fs$^{-1}$ and (parameter set 2) $F_J=1.50$, $k_{IC}=1/55$ fs$^{-1}$, for both we used $N=10$ squaraine monomer units. (a) The absorption spectrum is identical in both cases, as higher-excited state parameters do not influence linear absorption. (b) The integrated fifth-order signal is also match in both cases. Instead, both (c, d) third- and (e, f) fifth-order spectra are completely different for the two sets of parameters.}
\label{Fig:Main_correct_vs_wrong}

\end{figure}

Compared to a spectrally integrated signal, which can be fitted with several different parameter settings, the spectrally resolved $\text{PP}^{(5)}(\omega,t)$ signal is sensitive to polymer structure, properties of the higher-excited states, and other parameters influencing the dynamics of the double-excited manifold. To demonstrate this, we compare the simulated fifth-order pump--probe data discussed above with those obtained for a different set of parameters (sets 1 and 2 in Fig. \ref{Fig:Main_correct_vs_wrong}, respectively). The structure and parameters of the first-excited states are identical; thus, no difference is observed in their absorption spectra (Fig. \ref{Fig:Main_correct_vs_wrong}a). However, the couplings between two-exciton states and higher-excited states differ significantly, as controlled by the $F_J$ scaling parameter. Since this increases transfer and annihilation rates, we accordingly decreased the internal conversion rate from higher- to first-excited states, to achieve the same dynamics. In this way, the integrated fifth-order signal matches the reference signal (Fig. \ref{Fig:Main_correct_vs_wrong}b). Nevertheless, both third- and fifth-order pump--probe spectra are strikingly different for the two sets of parameters. The reference third- and fifth-order spectra (set 1) show one main positive peak at around 12500$^{-1}$\;cm and very weak signal around 13000$^{-1}$\;cm, whereas in the variant spectra (set 2), the rising peak at 12500$^{-1}$\;cm is much weaker than the peak around 13000$^{-1}$\;cm. This example clearly demonstrates that spectrally integrated signals suffer from parameter degeneracy, whereas spectrally resolved spectra mostly eliminate such uncertainty.

\subsection{Chain-length dependence}
\label{Sec:N_dependence}

\begin{figure}[t!]
{\sffamily
\sansmath
\begin{subfigure}[t]{0.25\textwidth}
\hspace{2.cm}
\vspace{-0.4cm}
\begin{overpic}[width=0.7\textwidth]{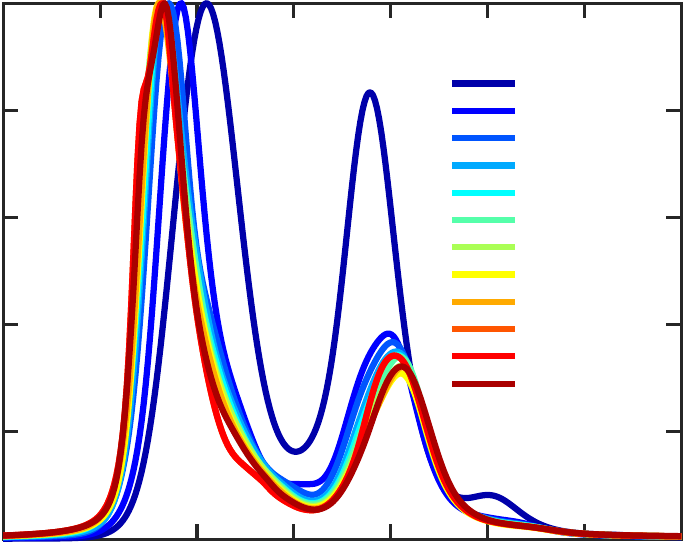}
  \put(2,87){\footnotesize  \textcolor{white}{short chain}}
  \put(94,-7){\footnotesize 18}
  \put(80,-7){\footnotesize 17}
  \put(66.5,-7){\footnotesize 16}
  \put(52,-7){\footnotesize 15}
  \put(38,-7){\footnotesize 14}
  \put(24,-7){\footnotesize 13}
  \put(10,-7){\footnotesize 12}
  \put(-4,-7){\footnotesize 11}
  \put(2,-16){\footnotesize Wavenumber [$10^{3}$\;cm$^{-1}$]}
  \put(-22,-7){\footnotesize  \rotatebox{90}{Normalized absorption}}
  \put(-13,-1){\footnotesize  0.0}
  \put(-13,14.5){\footnotesize  0.2}
  \put(-13,30.5){\footnotesize  0.4}
  \put(-13,45){\footnotesize  0.6}
  \put(-13,60){\footnotesize  0.8}
  \put(-13,74){\footnotesize  1.0}
  \put(67,68){\footnotesize  $N=2$}
  \put(67,15){\footnotesize  $N=24$}
  \put(77,64){\vector(0,-1){37}}
  \put(82,57){\footnotesize  \rotatebox{270}{length}}
        \put(-15,93.9){a}
\end{overpic}
\label{Fig:abs_N_dependence}
\end{subfigure}%
\begin{subfigure}[t]{0.25\textwidth}
\hspace{2cm}
\begin{overpic}[width=0.7\textwidth]{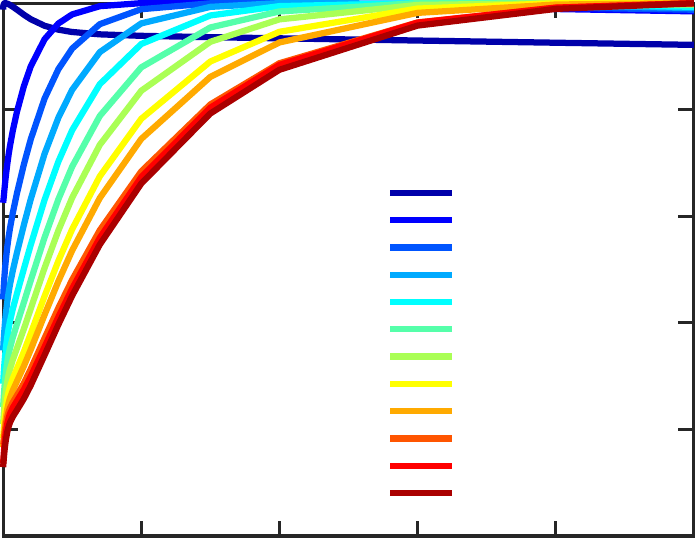}
  \put(35,58){\footnotesize slow dynamics}
  \put(2,80){\footnotesize fast dynamics}
  \put(5,78){\vector(2,-1){30}}
  \put(96,-7){\footnotesize 5}
  \put(77.3,-7){\footnotesize 4}
  \put(57.5,-7){\footnotesize 3}
  \put(38,-7){\footnotesize 2}
  \put(18.,-7){\footnotesize 1}
  \put(-1,-7){\footnotesize 0}
  \put(25,-16){\footnotesize Time delay [ps]}
  \put(-32,-3){\footnotesize  \rotatebox{90}{Integrated fifth-order }}
  \put(-23,-7){\footnotesize  \rotatebox{90}{Transient Signal [rel.u.]}}
  \put(-13,-1){\footnotesize  0.0}
  \put(-13,14){\footnotesize  0.2}
  \put(-13,29){\footnotesize  0.4}
  \put(-13,44){\footnotesize  0.6}
  \put(-13,59){\footnotesize  0.8}
  \put(-13,74){\footnotesize  1.0}
  \put(70,48){\footnotesize  $N=2$}
  \put(70,4){\footnotesize  $N=24$}
  \put(69,45){\vector(0,-1){32}}
  \put(72,42){\footnotesize  \rotatebox{270}{length}}  
        \put(-32,89){b}
\end{overpic}
\label{Fig:trace_N_dependence}
\end{subfigure}

\vspace{1.3cm}

\begin{subfigure}[t]{0.25\textwidth}
\RaggedLeft
\begin{overpic}[width=0.7\textwidth]{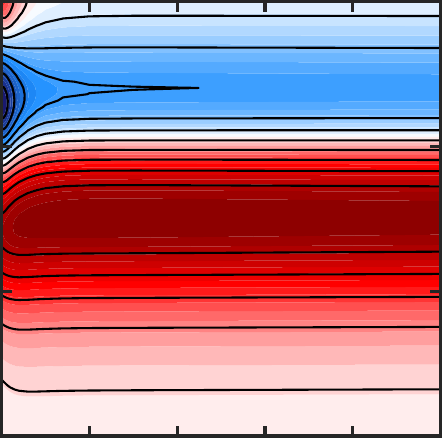}
\put(10,103){\footnotesize PP$^{(5)}$, $N=4$, set 1}
   \put(-1,-8){\footnotesize  0}
  \put(18,-8){\footnotesize  1}
  \put(38,-8){\footnotesize  2}
  \put(58,-8){\footnotesize 3}
  \put(78,-8){\footnotesize  4}
  \put(96,-8){\footnotesize  5}
  \put(23,-18.5){\footnotesize  Time delay [ps]}
      \put(-31,-2){\footnotesize  \rotatebox{90}{Wavenumber [$10^{3}$ cm$^{-1}$]}}
    \put(-18.5,95.75){\footnotesize  \rotatebox{0}{$13.5$}}
    \put(-18.5,63){\footnotesize  \rotatebox{0}{$13.0$}}
    \put(-18.5,30.25){\footnotesize  \rotatebox{0}{$12.5$}}
    \put(-18.5,0){\footnotesize  \rotatebox{0}{$12.0$}}
         \put(-35,105){c}
\end{overpic}
\label{Fig:5PP_N=4_N_dependence}
\end{subfigure}%
\begin{subfigure}[t]{0.25\textwidth}
\RaggedRight
\begin{overpic}[width=0.7\textwidth]{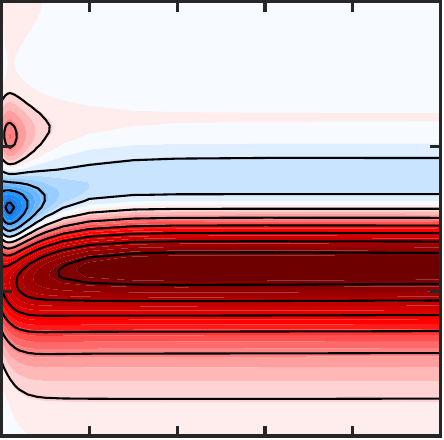}
\put(10,103){\footnotesize PP$^{(5)}$, $N=8$, set 1}
    \put(-1,-8){\footnotesize  0}
  \put(18,-8){\footnotesize  1}
  \put(38,-8){\footnotesize  2}
  \put(58,-8){\footnotesize 3}
  \put(78,-8){\footnotesize  4}
  \put(96,-8){\footnotesize  5}
  \put(23,-18.5){\footnotesize  Time delay [ps]}
       \put(-2,105){d}
\end{overpic}
\label{Fig:5PP_N=8_N_dependence}
\end{subfigure} 
\vspace{1.0cm}

\begin{subfigure}[t]{0.25\textwidth}
\RaggedLeft
\begin{overpic}[width=0.7\textwidth]{data/10_5PP.pdf}
\put(10,103){\footnotesize PP$^{(5)}$, $N=10$, set 1}
    \put(-1,-8){\footnotesize  0}
  \put(18,-8){\footnotesize  1}
  \put(38,-8){\footnotesize  2}
  \put(58,-8){\footnotesize 3}
  \put(78,-8){\footnotesize  4}
  \put(96,-8){\footnotesize  5}
  \put(23,-18.5){\footnotesize  Time delay [ps]}
      \put(-31,-2){\footnotesize  \rotatebox{90}{Wavenumber [$10^{3}$ cm$^{-1}$]}}
    \put(-18.5,95.75){\footnotesize  \rotatebox{0}{$13.5$}}
    \put(-18.5,63){\footnotesize  \rotatebox{0}{$13.0$}}
    \put(-18.5,30.25){\footnotesize  \rotatebox{0}{$12.5$}}
    \put(-18.5,0){\footnotesize  \rotatebox{0}{$12.0$}}
         \put(-35,105){e}
\end{overpic}
\label{Fig:5PP_N=10_N_dependence}
\end{subfigure}%
\begin{subfigure}[t]{0.25\textwidth}
\RaggedRight
\begin{overpic}[width=0.7\textwidth]{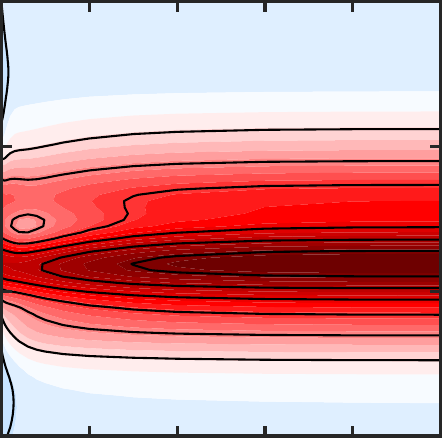}
\put(10,103){\footnotesize PP$^{(5)}$, $N=20$, set 1}
    \put(-1,-8){\footnotesize  0}
  \put(18,-8){\footnotesize  1}
  \put(38,-8){\footnotesize  2}
  \put(58,-8){\footnotesize 3}
  \put(78,-8){\footnotesize  4}
  \put(96,-8){\footnotesize  5}
  \put(23,-18.5){\footnotesize  Time delay [ps]}  
  \put(101.,60.5){\includegraphics[width=0.04\textwidth]{data/colorbar_cropped-2.pdf}}
  \put(111,175){\footnotesize   1.0}
  \put(111,146){\footnotesize   0.5}
  \put(111,117){\footnotesize   0.0}
  \put(108,89){\footnotesize  -0.5}
  \put(108,61){\footnotesize  -1.0}
       \put(-2,105){f}
\end{overpic}
\label{Fig:5PP_N=20_N_dependence}
\end{subfigure}
\vspace{0.5cm}
}

\caption{\justifying Length dependence of spectroscopic signal of $\text{SQA}-\text{SQB}$ chain.  Simulations were performed for former parameter set (set 1) for chains of N=2,4,6,..,24 squaraine monomer units. a) Absorption spectra. b) Spectrally integrated fifth-order signal shows slower dynamics for longer chains. Fifth-order pump--probe spectrum is very length-sensitive. For shorter lengths the negative SE$_2^{(5)}$ peak is very prominent, as is visible on spectra for 4 monomer units (c). With increasing length its relative intensity decreases (see subfigures (d) signal for chain of 8 monomer units and (e) for chain of 10 molecules). For chain with 20 molecules (f) the negative peak disappears.}
\label{Fig:N_dependence}

\end{figure}

As previously mentioned, the simulated absorption spectrum for 5 dimer units ($N=10$) is weaker compared to the experimental one, in the region around $14000$ cm$^{-1}$ (see Fig. \ref{Fig:SQAB19_spectra}b).
We assume that, due to structural disorder, the polymer is divided into smaller segments for the presence of kinks.\cite{Malyshev1999, malyshev_channels_2000, Ryzhov2001} This fact can be supported by analyzing how the absorption spectra depends on the polymer length (see Fig. \ref{Fig:N_dependence}a). For increasing $N$, the two peaks in the absorption spectrum move further apart and the higher-energy one decreases in intensity, due to excitonic delocalization and redistribution of the transition dipole moments between the states.\cite{knapp_1984} Therefore, this gap can be filled, in principle, by including even shorter segments in the simulations, as they have higher amplitude in that spectral region. 

The chain-length dependence of the spectrally integrated signal is considered in Fig.\ref{Fig:N_dependence}b.
Malý et al. previously showed that the spectrally integrated signal approaches the behavior of a saturating exponential rise PP$^{(5)}(t)\propto(1-e^{-k_A T})$, for an increasing number of equivalent states.\cite{maly2023separating} Although, in the general case, the spectrally integrated signal includes both EEA and relaxation dynamics, the normalized spectrally integrated fifth-order signal decreases towards zero for $T=0$ as the polymer length increases, resembling $\propto(1-e^{-k_A T})$ profile (see Fig.\ref{Fig:N_dependence}b). Notice that in the case of a single SQA-SQB unit ($N=2$), the spectrally integrated signal is not rising, as expected based on results from Sec.\ref{Sec:dimer}, because the two first-excited states are not equivalent in the present case. Therefore, to disentangle different contributions to the dynamics and correctly describe the spectrally integrated fifth-order signal, we need a generalized fitting model, which includes not only EEA but also relaxation within the manifold. 

Both third- and fifth-order pump--probe spectra are also very sensitive to the length of the polymer, as can be seen from the comparison of pump--probe spectra for $N=6$, $8$, $10$, and $20$ (see Fig.\ref{Fig:N_dependence}c-f for the fifth-order and Fig. S10 in the SI for the third-order pump--probe). 
In particular, the visibility of the SE$_2^{(5)}$(2E) in the total fifth-order spectrum, contributing as a negative decaying peak, decreases with increasing polymer length. This can be justified based on the different scaling of the number of pathways with $N$, namely, $\mathcal{O}(N^2)$ for SE$_2^{(5)}$(2E), and $\mathcal{O}(N^3)$ for ESA$_2^{(5)}$(2E) (see Eq. \ref{EEI2D_pathways}). Furthermore, shorter segments show a broader main positive peak and the lineshape of SE$_2^{(5)}$(2E) and ESA$_2^{(5)}$(2E) are less overlapped. For increasing $N$, the ESA$_2^{(5)}$(2E) contribution approaches the main peak, canceling the SE$_2^{(5)}$ (2E) contribution (see Fig. S10 and S11 in the SI for all spectra of all pathways).

With increasing chain length, the spectral peaks become narrower due to exchange narrowing.\cite{knapp_1984} An additional narrowing, accompanied by a shift of the main peak to lower energies, arises from the partial cancellation between the ESA and SE contributions. As the polymer length increases, the relative contribution of ESA becomes stronger and red-shifted towards SE, leading to a more pronounced cancellation of the high-energy side of the SE peak. This results in both a red shift and further narrowing of the main negative peak in the third-order pump--probe spectrum as demonstrated in Fig. S10 in the SI. Similar narrowing can be observed in fifth-order pump--probe spectra. 
This behavior is clearly illustrated by comparing the individual pathway contributions with the total third- and fifth-order pump--probe spectra for chains containing $N=10$ and $N=20$ monomer units ($5$ and $10$ SQA--SQB pairs, respectively; see Fig. S10 and S11 in the SI). Although the peaks associated with individual pathways also become narrower for longer chains, the effect is substantially more pronounced in the total spectra, demonstrating that the observed narrowing originates from the combined effects of exchange narrowing and pathway cancellation.
This also highlights the importance of a realistic description of spectral lineshapes. Since the observed spectra result from the cancellation of several broad excitation pathway contributions, even subtle changes in individual lineshapes strongly affect the final spectral features. A realistic treatment of line broadening is therefore essential for reproducing the experimentally observed spectra and correctly interpreting the origin of individual spectral features.

\subsection{Approximations and model limitations}
For the derivation of spectroscopic response, we used the impulsive limit and neglected the reversed time ordering of pump–probe pulses. This can be justified by the fact that the dynamics we are studying is in the picosecond timescale, compared to a pulse duration in the tens of femtoseconds\cite{maly2023separating}.
For the dynamics, we have used secular Redfield theory. This choice is reasonable since the squaraine chain presents strong coupling between chromophores, with average localization length of $3.6$ monomer units for a chain of 20 monomer units (see Eq. 5 in the SI). For the same reason, approximation of dephasing due to population transfer is here also valid, despite its expected inaccuracy in Förster regime. 

\begin{figure}[t!]
{\sffamily
\sansmath
\begin{subfigure}[t]{0.25\textwidth}
\hspace{2.0cm}
\vspace{-0.4cm}
\begin{overpic}[width=0.7\textwidth]{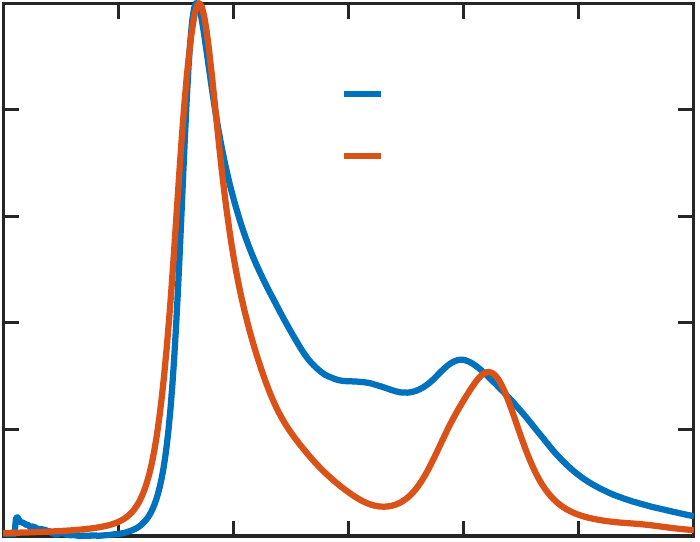}
\put(-4,-8){\footnotesize 11}
  \put(12.75,-8){\footnotesize  12}
  \put(29.5,-8){\footnotesize  13}
  \put(45.25,-8){\footnotesize 14}
  \put(61,-8){\footnotesize  15}
  \put(77.75,-8){\footnotesize  16}
  \put(94.5,-8){\footnotesize  17}
  \put(5,-18.5){\footnotesize  Wavenumber [$10^{3}$ cm$^{-1}$]}
  \put(-13,-1){\footnotesize  0.0}
  \put(-13,14){\footnotesize  0.2}
  \put(-13,29){\footnotesize  0.4}
  \put(-13,44){\footnotesize  0.6}
  \put(-13,59){\footnotesize  0.8}
  \put(-13,74){\footnotesize  1.0}
      \put(60,62){\footnotesize  exp.}
      \put(60,53){\footnotesize  sim.}
  \put(-22,-7){\footnotesize  \rotatebox{90}{Normalized absorption}}
        \put(-15,93.9){a}
\end{overpic}
\label{Fig:abs_set3}
\end{subfigure}%
\begin{subfigure}[t]{0.25\textwidth}
\hspace{2cm}
\begin{overpic}[width=0.7\textwidth]{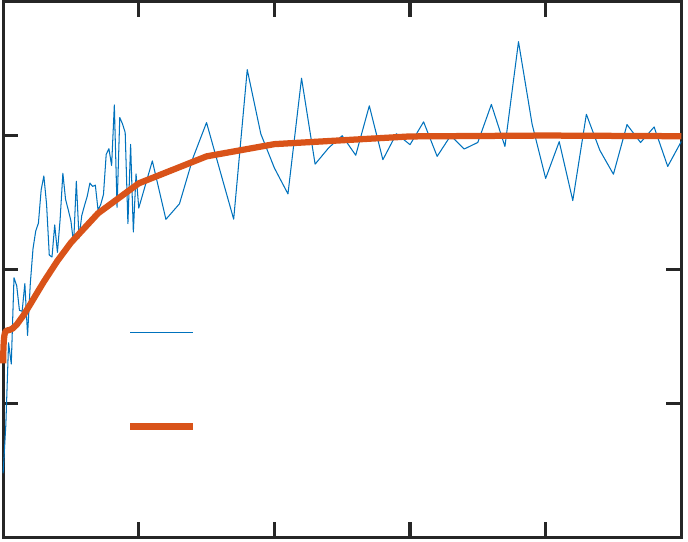}
  \put(96,-7){\footnotesize 5}
  \put(77.3,-7){\footnotesize 4}
  \put(57.5,-7){\footnotesize 3}
  \put(38,-7){\footnotesize 2}
  \put(18.,-7){\footnotesize 1}
  \put(-1,-7){\footnotesize 0}
  \put(25,-16){\footnotesize Time delay [ps]}
  \put(-34,-3){\footnotesize  \rotatebox{90}{Integrated fifth-order }}
  \put(-25,-7){\footnotesize  \rotatebox{90}{Transient Signal [rel.u.]}}
        \put(-32,89){b}
    \put(-13,75.5){\footnotesize  \rotatebox{0}{$1.5$}}
    \put(-13,56){\footnotesize  \rotatebox{0}{$1.0$}}
    \put(-13,36.5){\footnotesize  \rotatebox{0}{$0.5$}}
    \put(-13,17){\footnotesize  \rotatebox{0}{$0.0$}}
    \put(-15.5,-0.5){\footnotesize  \rotatebox{0}{-$0.5$}}
      \put(30,28){\footnotesize  exp.}
      \put(30,15){\footnotesize  sim.}
\end{overpic}
\label{Fig:trace_set3}
\end{subfigure}

\vspace{1.3cm}

\begin{subfigure}[t]{0.25\textwidth}
\RaggedLeft
\begin{overpic}[width=0.7\textwidth]{data/SQAB_3PP_Exp.pdf}
\put(10,103){\footnotesize PP$^{(3)}$, experiment}

      \put(-31,-2){\footnotesize  \rotatebox{90}{Wavenumber [$10^{3}$ cm$^{-1}$]}}
    \put(-18.5,95.75){\footnotesize  \rotatebox{0}{$13.5$}}
    \put(-18.5,63){\footnotesize  \rotatebox{0}{$13.0$}}
    \put(-18.5,30.25){\footnotesize  \rotatebox{0}{$12.5$}}
    \put(-18.5,0){\footnotesize  \rotatebox{0}{$12.0$}}
         \put(-35,105){c}
\end{overpic}
\label{Fig:3PP_N=20_set3_exp}
\end{subfigure}%
\begin{subfigure}[t]{0.25\textwidth}
\RaggedRight
\begin{overpic}[width=0.7\textwidth]{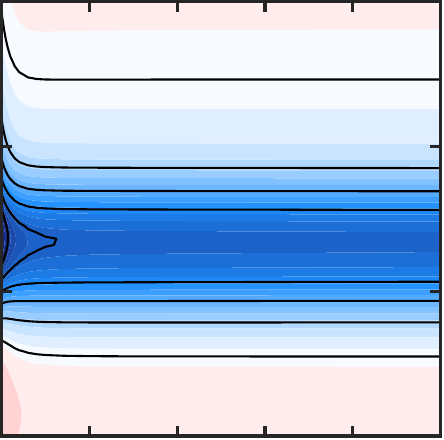}
\put(10,103){\footnotesize PP$^{(3)}$, $N=20$, set 3}

       \put(-2,105){d}
\end{overpic}
\label{Fig:3PP_N=20_N_dependence_set3}
\end{subfigure}
\vspace{0.2cm}

\begin{subfigure}[t]{0.25\textwidth}
\RaggedLeft
\begin{overpic}[width=0.7\textwidth]{data/SQAB_5PP_Exp.pdf}
\put(10,103){\footnotesize PP$^{(5)}$, experiment}
    \put(-1,-8){\footnotesize  0}
  \put(18,-8){\footnotesize  1}
  \put(38,-8){\footnotesize  2}
  \put(58,-8){\footnotesize 3}
  \put(78,-8){\footnotesize  4}
  \put(96,-8){\footnotesize  5}
  \put(23,-18.5){\footnotesize  Time delay [ps]}
      \put(-31,-2){\footnotesize  \rotatebox{90}{Wavenumber [$10^{3}$ cm$^{-1}$]}}
    \put(-18.5,95.75){\footnotesize  \rotatebox{0}{$13.5$}}
    \put(-18.5,63){\footnotesize  \rotatebox{0}{$13.0$}}
    \put(-18.5,30.25){\footnotesize  \rotatebox{0}{$12.5$}}
    \put(-18.5,0){\footnotesize  \rotatebox{0}{$12.0$}}
         \put(-35,105){g}
\end{overpic}
\label{Fig:5PP_N=20_set3_exp}
\end{subfigure}%
\begin{subfigure}[t]{0.25\textwidth}
\RaggedRight
\begin{overpic}[width=0.7\textwidth]{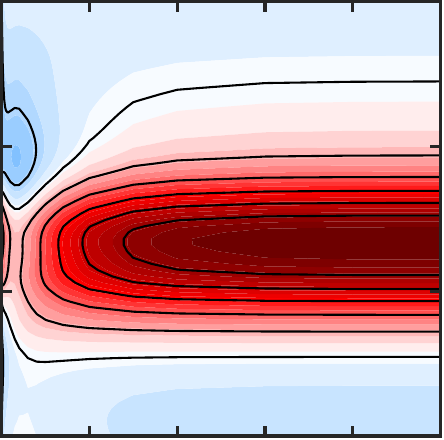}
\put(10,103){\footnotesize PP$^{(5)}$, $N=20$, set 3}
    \put(-1,-8){\footnotesize  0}
  \put(18,-8){\footnotesize  1}
  \put(38,-8){\footnotesize  2}
  \put(58,-8){\footnotesize 3}
  \put(78,-8){\footnotesize  4}
  \put(96,-8){\footnotesize  5}
  \put(23,-18.5){\footnotesize  Time delay [ps]}  
\put(101.,60.5){\includegraphics[width=0.04\textwidth]{data/colorbar_cropped-2.pdf}}
\put(111,175){\footnotesize   1.0}
\put(111,146){\footnotesize   0.5}
\put(111,117){\footnotesize   0.0}
\put(108,89){\footnotesize  -0.5}
\put(108,61){\footnotesize  -1.0}
       \put(-2,105){h}
\end{overpic}
\label{Fig:5PP_N=20_N_dependence_set3}
\end{subfigure}
\vspace{0.5cm}
}

\caption{\justifying Sensitivity of the spectra to coherence dephasing of the two-exciton states. Simulations were performed for modified parameter set (set 3) for $N=20$ monomer units. (a) Absorption spectra and (b) spectrally integrated fifth-order pump–probe signal. (c,d) Experimental and simulated PP$^{(3)}$ spectra, respectively. (g,h) Experimental and simulated PP$^{(5)}$ spectra, respectively. The simulation includes an increased lifetime dephasing rate for the ESA$_2^{(5)}$(2E) pathway, which reduces its spectral contribution and enhances the negative SE$_2^{(5)}$(2E) feature.}
\label{Fig:deph_dependence_set3}

\end{figure}

Although the simulated pump--probe data for $[\text{SQA}-\text{SQB}]_{19}$ reproduce the main experimental features, there are slight differences. As was demonstrated in Sec. \ref{Sec:parameters_sensitivity}, these stem from the pump--probe spectral features being sensitive to the parameters of the higher excited states. This sensitivity represents a challenge for our model, pushing it to its limits. 

In the simulated third-order spectra in Fig. \ref{Fig:SQAB19_spectra}, we do not see two decaying peaks on background of main negative peak, instead we have only one negative peak which decays only later due to relaxation of first excited state. Moreover, in fifth-order spectra, we see slight shift compared to experimental-one and, especially, SE$_2^{(5)}$(2E) peak is weak and shifted down by quite prominent positive ESA$_2^{(5)}$(2E) peak. These differences indicate that we are reaching the limits of our quite general excitonic model. To improve the description of the experimental data, further information would be needed, possibly along the three following lines. 

First, the exact structure of the squaraine co-polymer is not known. For this reason, our model incorporates Gaussian sampling of the angles between different molecules, as described in Sec. \ref{Sec:examples}. Although the absorption spectrum (see Fig. \ref{Fig:SQAB19_spectra}b) agrees with the experiment in the position of the two peaks at $13000$ and $15000$ cm$^{-1}$, the peak at $14000$ cm$^{-1}$ is missing. This suggests the presence of kinks in the polymer structure, which effectively divide the polymer into shorter segments.\cite{Malyshev1999, malyshev_channels_2000, Ryzhov2001} The dispersion in the polymer length manifests itself also in features of third- and fifth-order spectra. While a decaying main peak in third-order pump--probe is a feature corresponding to longer chains, in fifth-order pump--probe the negative peak related to SE$_2^{(5)}$(2E) is more prominent for shorter segments (see Fig. S10 in the SI for comparison of third-order pump--probe spectra for different lengths of the polymer).

A second uncertainty stems from the parameters of higher-excited states. In the simulations, we included a single effective higher-excited state, while a dense manifold of states is expected for real systems. The presence of higher-excited states at higher energy would, among other things, result in blue shift of SE$_2^{(5)}$ and SE$_2^{(5)}$, whereas lower-energy higher-excited states would facilitate efficient EEA. Although higher-excited states play an important role in determining the peak positions, their influence typically diminishes with increasing $N$, as their number scales linearly with $N$ compared to the quadratic scaling of two-exciton states $\propto N^2$. 

Third, an improved theory might be needed for the population decay-induced dephasing, which affects the linewidth of the spectral lines of transitions to multiply excited states. We performed the simulation again, for $N=20$ with new set of parameters (set 3) and with two times increased lifetime dephasing $\Gamma_{\xi \beta}$ for ESA$_2^{(5)}$ (2E), where we assume the largest effect (see Fig. \ref{Fig:deph_dependence_set3} for comparison with experiment). The different linewidth led not only to a less visible ESA$_2^{(5)}$ (2E) peak but also the SE$_2^{(5)}$ (2E) peak was much more pronounced, due to improper cancellation of pathways. As we used longer chain for this simulation, also third-order pump--probe spectrum imporoved in agreement with experiment having now one main negative decaying peak. Similar dependence of fifth-order pump--probe spectra on lifetime dephasing can be observed also for shorter chains, see Fig. S12 in SI for simulation of chain of length $N=10$ with parameters from set 1.
The extent of the influence of this phenomenon on the spectra can be tested in the future by comparison with spectra of the same system, which will be calculated with an improved theory we are working on.     

For our simulations, we consider only 5 $\text{SQA}-\text{SQB}$ units, even though our sample has 19 squaraine pairs. As many factors convincingly indicate that there are shorter segments then expected in the sample and exact distribution is unknown, we have chosen $5$ dimeric units ($N=10$) as an example showing all important features.

Our results demonstrate that spectrally resolved fifth-order pump–probe spectroscopy provides more detailed information than spectrally integrated signals. While integrated signals mix energy transfer and annihilation dynamics, spectral resolution enables separation of individual pathway contributions and determines microscopic parameters that cannot be isolated from integrated signals alone.

\section{Conclusions}
We have developed a theoretical description of fifth-order pump--probe spectra, with realistic lineshapes and exciton dynamics, as well as microscopic description of exciton--exciton annihilation. Using this model, we have shown that the $\text{PP}^{(5)}$ spectra contain complex information about exciton--exciton annihilation, exciton relaxation, and single- and double-excited state energies. Although spectral features characteristic of one- and two-exciton states can be distinguished, their interpretation is complicated by substantial spectral overlap. Consequently, a microscopic theoretical model, such as the one presented here, is required to disentangle the underlying contributions, particularly in larger systems.

The pump--probe spectra are very sensitive to changes in system parameters. As we have shown, even when two sets of parameters produce the same spectrally integrated fifth-order signal, they can lead to very different third- and fifth-order pump--probe spectra. Therefore, our results demonstrate that models used to interpret $\text{PP}^{(5)}$ signals should incorporate spectral information to capture the interplay between exciton--exciton annihilation and exciton transport in the experimental data. We further show that the total fifth-order signal arises from an interference of distinct excitation pathways. Therefore apparently narrow peaks can originate from overlapping broader features and minor shifts or broadening effects significantly reshape the spectral features and realistic description of lineshapes, such as the one presented in our model, is essential for correct simulation and interpretation of experimental pump--probe spectra. 

Contrary to previous claims in the literature, the spectrally unresolved fifth-order signal cannot be, in general, directly used to read off the effective exciton--exciton annihilation rate. On the other hand, supported by a proper analysis, the $\text{PP}^{(5)}$ spectra can be used to disentangle the exciton relaxation and annihilation, tracking the excitons as they move in space and energy. 

\section*{Acknowledgements}

We acknowledge Tom\'a\v{s} Man\v{c}al, Vladislav Sl\'ama and Franti\v{s}ek \v{S}anda for useful discussions. The authors acknowledge funding by Charles University (grant no. PRIMUS/24/SCI/007, to P.M., GA UK grant No. 170326 to K.Ch.) and by Czech Science Foundation (GA\v{C}R No. 26-23570S to P.M.).   M.B. acknowledges financial support from the Programme Johannes Amos Comenius under the MSCA Fellowships CZ-UK4 project 'MEET-UP' (registration No. CZ.02.01.01/00/22\_010/00013392).

\section*{Author declarations}
\noindent Conflict of interest: The authors have no conflicts to disclose.

\medskip

\noindent Author contributions:

\noindent \underline{Kate\v{r}ina Charv\'atov\'a}: Conceptualization, Data Curation, Formal Analysis, Methodology, Software, Visualization, Writing - Original Draft Preparation, Review \& Editing

\noindent \underline{Matteo Bruschi}: Conceptualization, Methodology, Supervision, Writing - Original Draft Preparation, Review \& Editing

\noindent \underline{Pavel Mal\'y}: Conceptualization, Funding Acquisition, Methodology, Resources, Supervision, Writing - Original Draft Preparation, Review \& Editing

\section*{Data availability}

All experimental data have been published in Ref. \citenum{maly2023separating}, for availability see therein. Remaining data that support the findings of this study are available from the corresponding author upon reasonable request. 

\bibliography{references}

\end{document}


\maketitle


\section{Theoretical model for simulations of fifth-order pump--probe spectra} 
\subsection{Fifth-order signal Feynman diagrams}
In the main text in Fig. 4 we present only one representative diagram for each class of fifth-order pathways. Here we provide the complete set of rephasing and nonrephasing diagrams contributing to the PP$^{(5)}$ signal\cite{maly2020wavelike,maly2023separating,rose2023interpretations}: negated ground-state bleach (NGSB$^{(5)}$) in Fig.\ref{Fig:SI_FD_NGSB}, negated stimulated emission (NSE$^{(5)}$) in Fig. \ref{Fig:SI_FD_NSE}, negated excited-state absorption (NESA$^{(5)}$) in Fig. \ref{Fig:SI_FD_NESA}, two-exciton stimulated emission (SE$_2^{(5)}$) in Fig. \ref{Fig:SI_FD_SE2} and two-exciton excited state absorption (ESA$_2^{(5)}$) in Fig. \ref{Fig:SI_FD_ESA2}. Numbers of diagrams of each type of excitation pathway determines the prefactors used in Eq. 7 of the main text.
\label{Sec:allFD}
\input{pictures/FD_GSB_pathways}
\input{pictures/FD_SE1_pathways}
\input{pictures/FD_ESA1_pathways}
\input{pictures/FD_SE2_SE_EEA_pathways}
\input{pictures/FD_ESA2_ESA_EEA_pathways}

\newpage

\subsection{Structure of the Hamiltonian}
This section describes the construction of the Hamiltonian used in the simulations, following the workflow summarized in Fig. S6. We first generate a structurally disordered molecular aggregate in the site basis and construct the corresponding Hamiltonian. After diagonalization, the multiexcitonic states are used for calculation of evolution operator in Sec. \ref{sec:rates} and spectral lineshapes in Sec. \ref{sec:lineshapes}. Transition dipole moments are also transformed to multiexciton basis and further undergo rotational averaging in Sec. \ref{sec:average}.
\begin{figure}
\centering
\begin{tikzpicture}[
node distance=0.5cm,
box/.style={
  draw,
  rounded corners,
  align=center,
  minimum width=1.2cm,
  minimum height=0.9cm,
  font=\small
},
->, >=Stealth
]

\node[box] (g) {g(t)};
\node[box, below=of g] (h) {new H, $\mu$};
\node[box, below=of h] (d) {diagonalization of H};
\node[box, below=of d] (dyn) {$\mathcal{U}(t)$};
\node[box, below=of dyn] (comb) {GSB,SE,... response};
\node[box, below=of comb] (avg) {totsig=totsig+sig};
\node[box, below=of avg] (tot) {total signal};

\node[box, left=0.6cm of dyn] (rot) {$\langle \mu_1 ... \mu_6 \rangle_\Omega$};
\node[box, right=0.6cm of dyn] (line) {$\mathcal{L}(\omega)$};

\draw (g) -- (h);
\draw (h) -- (d);
\draw (d) -- (dyn);
\draw (dyn) -- (comb);
\draw (comb) -- (avg);
\draw (avg) -- (tot);

\draw (d) -| (rot);
\draw (rot) |- (comb);

\draw (d) -| (line);
\draw (line) |- (comb);

\draw[rounded corners] (avg.west) -| ++(-2.3,0) |- (h.west);

\end{tikzpicture}
\caption{Signal construction workflow – First, we generate the system Hamiltonian. In addition, we calculate the rates and lineshapes using separate functions (see Sections \ref{sec:rates} and \ref{sec:lineshapes}). The input of the main function also includes the light polarization (the pump and probe polarizations can be either parallel, perpendicular, or at the magic angle). The transition dipole moments are then processed using rotational averaging. Rotational averaging and the matrices used for this transformation are discussed further in Section \ref{sec:average}. Finally, the signal for each pathway is constructed by combining the lineshape, transition dipole moments, and evolution operator according to the Eq. 7 in the main text.}
\label{Fig:workflow}
\end{figure}
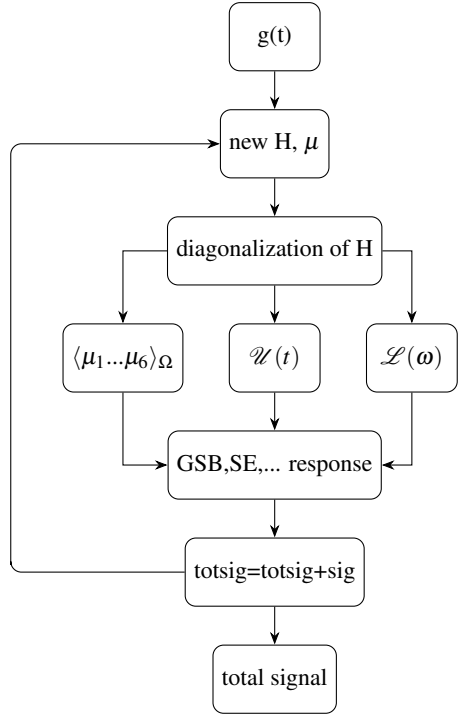

First, we built the single exciton block. To simulate a signal from a real sample with energetic disorder, we vary the energies of first excited states by Gaussian random sampling for different realizations. Then, we average signals from all realizations according to the workflow shown in Fig.\ref{Fig:workflow}. In each realization, the energies are as follows 
\begin{equation}
\begin{split}
    E_{2i-1}=\bar{E}_{\text{SQA}} + \delta E_{\text{SQA}}, i \in 1,..,\frac{N}{2}\\
    E_{2i}=\bar{E}_{\text{SQB}} + \delta E_{\text{SQB}}, i \in 1,..,\frac{N}{2}
\end{split}
\end{equation}
Energies are set such that the resulting squaraine chain will consist of two different molecules SQA and SQB regularly alternating within the chain with average energies $\bar{E}_{\text{SQA}}$ and $\bar{E}_{\text{SQB}}$ respectively. Energy of each molecule differs from the average value by small static disorder $\delta{E}_{\text{SQA/B}}$ with Gaussian distribution. The ground state is our reference state, and we set its energy to zero $E_g=0$.

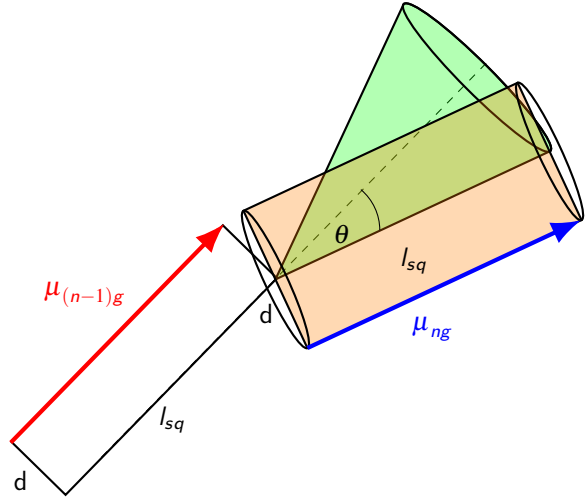
\begin{figure}
\centering
{\sffamily
\sansmath
\begin{tikzpicture}

\drawMyRectangle{white}{0.345,-0.345}{4}{1}{59.7}
\draw[ultra thick, -{Latex[length=4mm, width=3mm]},color=red]  (-0.38,0.345) -- (2.45,3.24);

\draw[dashed] (0.82,0.153) -- (5.95,5.4);
\drawMyCone{green}{(3.12,2.5)}{225}{40}{3.77}

\drawMyCylinder{orange}{1}{4}{(3.12,2.5)}{25}
\draw[ultra thick, -{Latex[length=4mm, width=3mm]},color=blue] (3.543,1.594) -- (7.168,3.284);

\node[] at (4,3.1) {$\theta$};
\node[] at (-0.25,-0.15) {d};
\node[] at (3.0,2.05) {d};
\node[] at (1.75,0.65) {$l_{sq}$};
\node[] at (4.95,2.75) {$l_{sq}$};
\node[text=red] at (0.6,2.3) {$\boldsymbol{\mu}_{(n-1)g}$};
\node[text=blue] at (5.2,1.85) {$\boldsymbol{\mu}_{ng}$};

\draw (4.5,3.15) arc (0:50:0.7cm);
\end{tikzpicture}
}
\caption{Scheme of structure of structurally disordered polymer.}
\label{fig:structure_polymer}
\end{figure}

The positions and directions of the transition dipole moments of each molecule are derived from the positions and directions of the molecules in the polymer. For a polymer with structural disorder, as the one used in our simulations, the position and direction of the transition dipole moment and the axis of the $i$-th molecule depend only on the position and direction of the transition dipole moment and the axis of the previous $(i-1)$-th molecule. Each molecule has its axis and length $l_{sq}$, the transition dipole moment of the molecule in our model is parallel to the molecular axis, but shifted by offset $d$. The following molecule starts at the point where the previous molecule ends and the direction of its axis is, with the highest probability, the same as the direction of the axis of the preceding molecule. However, the angle between these two axes is perturbed with a normal distribution from $-\theta_{max}$ to $\theta_{max}$, the azimuthal angle $\phi$ has an even distribution from 0 to $2\pi$ forming a cone of possible directions of the axis (see the green cone in Fig. \ref{fig:structure_polymer}). The transition dipole moment is, as mentioned earlier, offset from the molecular axis. Because the molecule can rotate along its axis, position of the transition dipole moment is on the shell of a cylinder with radius r and axis formed by molecular axis (red cylinder in the picture \ref{fig:structure_polymer}).  Due to electrostatic repelling of side chains, position of transition dipole moment on the cylinder shell is not random and has the highest probability on the side furthest from the previous dipole moment. In the model, we make a plane formed by transition dipole moment of the previous molecule and axis of the following molecule. This plane intersects the cylinder in two lines parallel with the axis of the molecule (axis of the cylinder). The center of the transition dipole moment of the following molecule has the highest probability of lying on one of these two intersections, which is further from the center of the transition dipole moment of the preceding molecule. The distribution of angles on the cylinder with center at this point is Gaussian. This ensures that molecules are connected so that they form a zig-zag structure, as squaraine polymers are expected. \cite{ress2023time}

\bigskip

The coupling is calculated using the dipole-dipole coupling
\begin{equation}
J_{i,j}=\frac{1}{4\pi\epsilon_0r_{ij}^3}\left(\mathbf{\mu}_i\cdot\boldsymbol{\mu}_j-3\frac{(\boldsymbol{\mu}_i\cdot{\mathbf{r}}_{ij})(\boldsymbol{\mu}_i\cdot\mathbf{r}_{ij})}{r_{ij}^2}\right).
\label{J_dipole-dipole}
\end{equation}
where $\boldsymbol{\mu}_{i,g}$ and $\boldsymbol{\mu}_{j,g}$ are the transition dipole moments of the two molecules, whose centers are distanced by $r_{ij}$.
The coupling matrix is then added to the Hamiltonian with which the single-exciton block is obtained
\begin{equation}
    H_{i,j}=E_{i}\delta_{ij}+J_{i,j}.
\end{equation}
Now we can proceed to the creation of a double- and triple-excited state blocks of the Hamiltonian. Here, energies, couplings, and transition dipole moments are derived from properties of single-excited states. See Table \ref{tab:2Eand3Eproperties}.

\begin{table}
\renewcommand{\arraystretch}{1.6} 
\begin{tabular}{|p{8cm}|p{8cm}|}
\hline
    energy of two-exciton state  $|ij \rangle = |i \rangle|j \rangle$ & $E_{ij}=E_i+E_j$\\
    \hline
    coupling between two two-exciton states $|ij \rangle $ and $|kl \rangle$ &  $J_{ij,kl}=J_{i,k}\delta_{jl}+J_{i,l}\delta_{jk}+J_{j,k}\delta_{il}+J_{j,l}\delta_{ik}$\\
    \hline
     energy of higher-excited state $f_i$  &   $E_{f_i}=E_i + \Delta E_{f_i} + \delta E_{f_i}$ \\
    \hline
    coupling between two-exciton $|ij\rangle$ state and higher-excited $f_k$ state & $ J_{ij,f_k}=F_J J_{i,j} (\delta_{ik}+ \delta_{jk})$ \\
    \hline
    energy of three-exciton state energy of two-exciton state  $|ijk \rangle = |i \rangle|j \rangle|k \rangle$ & $ E_{ijk}=E_i+E_j+E_k$ \\
    \hline
    coupling between three-exciton states $|ijk \rangle$ and $|lmn\rangle$  & $J_{ijk,lmn}=J_{i,l}\delta_{jm}\delta_{kn}+J_{j,m}\delta_{il}\delta_{kn}+J_{k,n}\delta_{il}\delta_{jm}+J_{i,m}\delta_{jl}\delta_{kn} +J_{i,n}\delta_{jl}\delta_{km}+J_{j,l}\delta_{im}\delta_{kn} +J_{j,n}\delta_{il}\delta_{km}+J_{k,l}\delta_{im}\delta_{jn}+J_{k,m}\delta_{il}\delta_{jn}$ \\
    \hline
    energy of triple-excited state $|f_ij\rangle=|f_i\rangle|j\rangle$ state & $E_{f_ij}=E_{f_i}+E_j$ \\
        \hline
    coupling between triple-excited states $|f_ij\rangle$ and $|f_kl\rangle$ & $J_{f_ij,f_kl}=\delta_{ik}J_{j,l}$ \\
            \hline
    coupling between triple-excited states $|f_ij\rangle$ and $|klm\rangle$ states is defined as  & $J_{f_i j,klm}=J_{f_i,lm}\delta_{jk} +J_{f_i,km}\delta_{jl} +J_{f_i,kl}\delta_{jm}$ \\
    \hline
    transition dipole moment between states $|k\rangle$ and $|ij\rangle$ & 
$\boldsymbol{\mu}_{k,ij}=\delta_{jk}\boldsymbol{\mu}_{g,i}+\delta_{ik}\boldsymbol{\mu}_{g,j}$\\
         \hline
     transition dipole moments from the first-excited state  $|i\rangle$ to higher-excited state  $|f_i\rangle$ & $\boldsymbol{\mu}_{i,f_i}=F_\mu\boldsymbol{\mu}_{g,i}$ \\
              \hline
     transition dipole moment between the three-exciton state $|ijk\rangle$ and two-exciton state $|lm\rangle$ & $\boldsymbol{\mu}_{lm,ijk}=\delta_{jl}\delta_{km}\boldsymbol{\mu}_{i,g}+\delta_{il}\delta_{km}\boldsymbol{\mu}_{j,g}+\delta_{il}\delta_{jm}\boldsymbol{\mu}_{k,g}$ \\
                   \hline
    transition dipole moment between the triple-excited state $|f_kl\rangle$ and two-exciton state $|ij\rangle$ & $\boldsymbol{\mu}_{ij,f_kl}=f_\mu(\delta_{ki}\delta_{lj}+\delta_{kj}\delta_{li})\boldsymbol{\mu}_{k,g}$ \\
                             \hline
    transition dipole moment between higher-excited state $|f_k\rangle$ and triple-excited state $|f_kl\rangle$ & $\boldsymbol{\mu}_{f_k,f_kl}=\boldsymbol{\mu}_{l,g} $ \\
                                 \hline
\end{tabular}
\caption{Energies, transition dipole moments and couplings between single-, double- and triple-excited states.}
\label{tab:2Eand3Eproperties}
\end{table}

The dimension of the system's Hamiltonian is 
\begin{equation}
    \mbox{dim}_H=\underbrace{1}_{g}+\underbrace{N}_{e}+\underbrace{N_f}_{f}+\underbrace{\frac{N\cdot(N-1)}{2}}_{ee}+\underbrace{\frac{N\cdot (N-1)\cdot (N-2)}{6}}_{eee}+\underbrace{N\cdot N_f}_{ef}
\end{equation}

The delocalization length is here calculated as\cite{maly2020wavelike}
\begin{equation}
    l=\frac{N}{\sum_{i,\alpha=1}^N|c_{i}^\alpha|^4}.
\end{equation}
obtained from the coefficients in the transformation matrix c.

\subsection{Dynamics}
\label{sec:rates}
For the dynamics, we used Redfield theory in secular approximation.\cite{may2023charge} The transition rate between single-excited states $|\alpha\rangle$ and $|\tilde{\alpha}\rangle$ is
\begin{equation}k_{\tilde{\alpha}\alpha}=\sum_i|c^\alpha_i|^2|c^{\tilde{\alpha}}_i|^2\nu_i \tilde{C}(\omega_{\alpha \tilde{\alpha}}).
\end{equation}
where $\tilde{C}(\omega)$ is the spectral density, $\nu_{i}$ is a scaling factor for molecule $i$, and $\omega_{\alpha\tilde{\alpha}}$ is the frequency difference between the two states.
Instead, the transition rate between double-excited states $|\tilde{\beta} \rangle$ and $|\beta \rangle$ is
\begin{equation}
\begin{split}
    k_{\tilde{\beta}\beta}=& k_{\tilde{\beta}\beta}^{ee}+k_{\tilde{\beta}\beta}^{fee}+k_{\tilde{\beta}\beta}^{eef}+k_{\tilde{\beta}\beta}^{f}=k_{\tilde{\beta}\beta}^{ee}+2k_{\tilde{\beta}\beta}^{fee}+k_{\tilde{\beta}\beta}^{f} = \\
    =&\sum_{i<j,k<l}(\delta_{ik}\nu_i+\delta_{il}\nu_i+\delta_{jk}\nu_j+\delta_{jl}\nu_j)c_{ij}^{\tilde{\beta}} c_{ij}^\beta c_{kl}^{\tilde{\beta}} c_{kl}^\beta \tilde{C}(\omega_{\beta\tilde{\beta}})+\\
    &+2\Phi_{\text{corr}}\Phi\sum_{i<j,k}(\delta_{ik}\nu_i+\delta_{jk}\nu_j)c_{ij}^{\tilde{\beta}} c_{ij}^\beta c_{fk}^{\tilde{\beta}} c_{fk}^\beta \tilde{C}(\omega_{\beta\tilde{\beta}})+\\
    &+\Phi^2\sum_i |c^{\tilde{\beta}}_{fi}|^2|c^\beta_{fi}|^2 \nu_i \tilde{C}(\omega_{\tilde{\beta}\beta}).
\end{split}
\label{Eq:Dynamics_double_ex}
\end{equation}
where $\Phi$ and $\Phi_{\text{corr}}$ are scaling factors, as explained in the main text. 
The process of exciton--exciton annihilation can be described in two stages. First, the energy transfer between two-exciton state and higher-excited state occurs, which corresponds to the transition between two double-excited states in exciton basis with a rate given in Eq.\ref{Eq:Dynamics_double_ex}. Second, the internal conversion of higher-excited state with rate $k_{\text{IC}}$ can be described using the Lindblad theory in the exciton basis\cite{bruggemann2003exciton}
\begin{equation}
    k_{\alpha\beta}=\sum_i k_{\text{IC}}|c^{\alpha}_{i}|^2|c^{\beta}_{f_i}|^2 
\end{equation}
for the transfer from double-excited state $|\beta\rangle$ to single-excited state $|\alpha\rangle$.
The decay of triple-excited states is given by
\begin{equation}
\begin{split}
    k_{\xi{\tilde{\xi}}}=& k_{\xi{\tilde{\xi}}}^{eee}+k_{\xi{\tilde{\xi}}}^{fe}+k_{\xi{\tilde{\xi}}}^{fe\rightarrow eee}+k_{\xi{\tilde{\xi}}}^{eee\rightarrow fe}=k_{\xi{\tilde{\xi}}}^{eee}+k_{\xi{\tilde{\xi}}}^{fe}+2k_{\xi{\tilde{\xi}}}^{fe\rightarrow eee}=\\
    =&\sum_{\substack{i<j<k \\ m<n<l}}c_{ijk}^{\tilde{\xi}}  c_{ijk}^\xi  c_{mnl}^{\tilde{\xi}}  c_{mnl}^\xi \left(\nu_m\left(\delta_{mi}+\delta_{mj}+\delta_{mk}\right)+\nu_n\left(\delta_{ni}+\delta_{nj}+\delta_{nk}\right)+\nu_l\left(\delta_{li}+\delta_{lj}+\delta_{lk}\right)\right)  \tilde{C}(\omega_{{\tilde{\xi}}\xi})\\
    &+\sum_{\substack{i \neq j \\ m \neq n}}c_{f_ij}^{\tilde{\xi}}  c_{f_ij}^\xi  c_{f_mn}^{\tilde{\xi}}  c_{f_mn}^\xi \left(\delta_{im}\Phi^2\nu_m+\delta_{in}\Phi\Phi_{\text{corr}}\nu_n+\delta_{jm}\Phi\Phi_{\text{corr}}\nu_m+\delta_{jn}\nu_n\right)  \tilde{C}(\omega_{{\tilde{\xi}}\xi})\\
    &+2\sum_{\substack{i \neq j \\ m<n<l}}c_{f_ij}^{\tilde{\xi}}  c_{f_ij}^\xi  c_{mnl}^{\tilde{\xi}}  c_{mnl}^\xi \left(\Phi\Phi_{\text{corr}}\nu_i(\delta_{im}+\delta_{in}+\delta_{il})+\nu_j(\delta_{jm}+\delta_{jn}+\delta_{jl})\right)  \tilde{C}(\omega_{{\tilde{\xi}}\xi}).\\
\end{split}
\end{equation}
The annihilation rate from triple-excited state $|\xi\rangle$ to double-excited state $|\beta\rangle$ can also be obtained using Lindblad theory
\begin{equation}
    k_{\beta \xi}=\sum_{i>j} |c^{\beta}_{ij}|^2(|c^\xi_{f_ji}|^2+|c^\xi_{f_i j}|^2) k_{\text{IC}}.
\end{equation}

\subsection{Spectral lineshapes}
\label{sec:lineshapes}
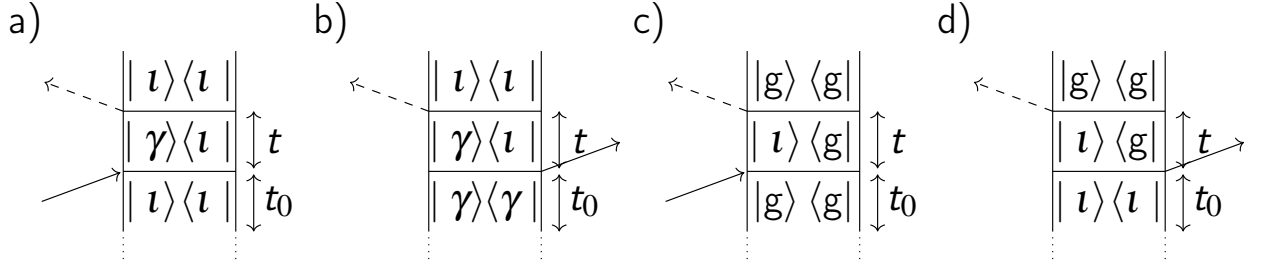
\begin{figure}
{\sffamily
\sansmath
\setlength{\tabcolsep}{0mm} 
\def\arraystretch{0.0} 
\resizebox{\linewidth}{!}{%
\begin{tabular}{ c c c c }
\begin{tikzpicture}[every node/.style={font=\Large}]
\draw[] (0,0.737) -- (0,2.948);
\draw[dotted] (0,0.4) -- (0,0.737);
\draw[] (1.38,0.737) -- (1.38,2.948);
\draw[dotted] (1.38,0.4) -- (1.38,0.737);
\draw[] (0,1.474) -- (1.38,1.474);
\draw[] (0,2.211) -- (1.38,2.211);
\draw[->, shorten > = 1pt] (-1,1.1055) -- (0,1.474); 
\draw[dashed, ->, shorten > = 1pt] (0, 2.211) -- (-1,2.5795); 

\draw[<->, shorten > = 1pt] (1.6,0.737) -- (1.6,1.474); 
\node at (1.9,1.105) {$t_0$};
\draw[<->, shorten > = 1pt] (1.6,2.211) -- (1.6,1.474); 
\node at (1.85,1.842) {$t$};
\node at (0.35,2.579) {$\mid\iota\rangle$};
\node at (0.35,1.842) {$\mid\gamma\rangle$};
\node at (0.35,1.105) {$\mid\iota\rangle$};
\node at (1.02,2.579) {$\langle\iota\mid$};
\node at (1.02,1.842) {$\langle\iota\mid$};
\node at (1.02,1.105) {$\langle\iota\mid$}; 
\node at (-1.2,3.3) {a)}; 
\end{tikzpicture}

&\begin{tikzpicture}[every node/.style={font=\Large}]
\draw[] (0,0.737) -- (0,2.948);
\draw[dotted] (0,0.4) -- (0,0.737);
\draw[] (1.38,0.737) -- (1.38,2.948);
\draw[dotted] (1.38,0.4) -- (1.38,0.737);
\draw[] (0,1.474) -- (1.38,1.474);
\draw[] (0,2.211) -- (1.38,2.211);
\draw[->, shorten > = 1pt] (1.38,1.474) -- (2.38,1.8425); 
\draw[dashed, ->, shorten > = 1pt] (0, 2.211) -- (-1,2.5795); 

\draw[<->, shorten > = 1pt] (1.6,0.737) -- (1.6,1.474); 
\node at (1.9,1.105) {$t_0$};
\draw[<->, shorten > = 1pt] (1.6,2.211) -- (1.6,1.474); 
\node at (1.85,1.842) {$t$};
\node at (0.35,2.579) {$\mid\iota\rangle$};
\node at (0.35,1.842) {$\mid\gamma\rangle$};
\node at (0.35,1.105) {$\mid\gamma\rangle$};
\node at (1.02,2.579) {$\langle\iota\mid$};
\node at (1.02,1.842) {$\langle\iota\mid$};
\node at (1.02,1.105) {$\langle\gamma\mid$}; 
\node at (-1.2,3.3) {b)}; 
\end{tikzpicture}
&\begin{tikzpicture}[every node/.style={font=\Large}]
\draw[] (0,0.737) -- (0,2.948);
\draw[dotted] (0,0.4) -- (0,0.737);
\draw[] (1.38,0.737) -- (1.38,2.948);
\draw[dotted] (1.38,0.4) -- (1.38,0.737);
\draw[] (0,1.474) -- (1.38,1.474);
\draw[] (0,2.211) -- (1.38,2.211);
\draw[->, shorten > = 1pt] (-1,1.1055) -- (0,1.474); 
\draw[dashed, ->, shorten > = 1pt] (0, 2.211) -- (-1,2.5795); 

\draw[<->, shorten > = 1pt] (1.6,0.737) -- (1.6,1.474); 
\node at (1.9,1.105) {$t_0$};
\draw[<->, shorten > = 1pt] (1.6,2.211) -- (1.6,1.474); 
\node at (1.85,1.842) {$t$};
\node at (0.35,2.579) {$\mid$g$\rangle$};
\node at (0.35,1.842) {$\mid\iota\rangle$};
\node at (0.35,1.105) {$\mid$g$\rangle$};
\node at (1.02,2.579) {$\langle$g$\mid$};
\node at (1.02,1.842) {$\langle$g$\mid$};
\node at (1.02,1.105) {$\langle$g$\mid$}; 
\node at (-1.2,3.3) {c)}; 
\end{tikzpicture}

&\begin{tikzpicture}[every node/.style={font=\Large}]
\draw[] (0,0.737) -- (0,2.948);
\draw[dotted] (0,0.4) -- (0,0.737);
\draw[] (1.38,0.737) -- (1.38,2.948);
\draw[dotted] (1.38,0.4) -- (1.38,0.737);
\draw[] (0,1.474) -- (1.38,1.474);
\draw[] (0,2.211) -- (1.38,2.211);
\draw[->, shorten > = 1pt] (1.38,1.474) -- (2.38,1.8425); 
\draw[dashed, ->, shorten > = 1pt] (0, 2.211) -- (-1,2.5795); 

\draw[<->, shorten > = 1pt] (1.6,0.737) -- (1.6,1.474); 
\node at (1.9,1.105) {$t_0$};
\draw[<->, shorten > = 1pt] (1.6,2.211) -- (1.6,1.474); 
\node at (1.85,1.842) {$t$};
\node at (0.35,2.579) {$\mid$g$\rangle$};
\node at (0.35,1.842) {$\mid\iota\rangle$};
\node at (0.35,1.105) {$\mid\iota\rangle$};
\node at (1.02,2.579) {$\langle$g$\mid$};
\node at (1.02,1.842) {$\langle$g$\mid$};
\node at (1.02,1.105) {$\langle\iota\mid$}; 
\node at (-1.2,3.3) {d)}; 
\end{tikzpicture}
\end{tabular}}
}
\caption{Feynman diagrams related to the absorption and emission lineshapes in the doorway-window approximation.  a) Absorption lineshape for the coherence $|\gamma\rangle\langle\iota|$, b) emission lineshape for the coherence $|\gamma\rangle\langle\iota|$, c) absorption lineshape for the coherence $|\iota\rangle\langle$g$|$, and d) emission lineshape for the coherence $|\iota\rangle\langle$g$|$.}
\label{FD_lineshape}
\end{figure}

To define the lineshapes, we proceed with derivation of the "window" function in the doorway-window approximation.\cite{mukamel}
As depicted in Fig. \ref{FD_lineshape}a, we start from the population state $|\iota\rangle\langle \iota|$, which evolves for time $t_0$. The interaction with the electric field results in coherence state $|\gamma\rangle\langle \iota|$, which evolves for time $t$. Finally, the system emits the signal returning in the population state $|\iota\rangle\langle\iota|$. This can be described using the following correlation function:
\begin{align}
    &C(t_0,t)=Tr_B\{\mathcal{U}_\gamma(t)\mathcal{U}_\iota(t_0)\hat{w}_g^{eq}\mathcal{U}_\iota^+(t_0)\mathcal{U}_\iota^+(t)\}=Tr_B\{\mathcal{U}_\iota^+(t_0+t)\mathcal{U}_\gamma(t)\mathcal{U}_\iota(t_0)\hat{w}_g^{eq}\} \notag \\ \notag
    &=Tr_B\{\mathcal{U}_\iota^+(t_0+t)\mathcal{U}_g(t_0+t)\mathcal{U}_g^+(t_0+t)\mathcal{U}_\gamma(t_0+t)\mathcal{U}_\gamma^+(t_0)\mathcal{U}_g(t_0)\mathcal{U}_g^+(t_0)\mathcal{U}_\iota(t_0)\hat{w}_g^{eq}\}\\ \notag
    &=Tr_B\{\mathcal{U}_{\iota g}(t_0+t)\mathcal{U}_{g\gamma}(t_0+t)\mathcal{U}_{\gamma g}(t_0)\mathcal{U}_{g\iota}(t_0)\hat{w}_g^{eq}\}, \notag
\end{align}
where $\hat{w}_{g}^{eq}$ is the bath density matrix in the ground-state equilibrium, while $\mathcal{U}_{gx}(t)$ are the coherence Green functions.\cite{mukamel} By taking the fast evolving part of the evolution out of the trace, we get:
\begin{align}
    C(t,t_0)&=e^{-i\omega_{\gamma g}t}e^{i\omega_{\iota g}t}Tr_B\{\widetilde{\mathcal{U}}_{\iota g}(t_0+t)\widetilde{\mathcal{U}}_{g\gamma}(t_0+t)\widetilde{\mathcal{U}}_{\gamma g}(t_0)\widetilde{\mathcal{U}}_{g \iota}(t_0)\hat{w}_g^{eq}\} \notag \\
    &=e^{-i\omega_{\gamma \iota}t}\widetilde{C}(t_0,t)=e^{-i(\omega_{\gamma \iota}^0+\lambda_{\gamma\gamma}-\lambda_{\iota\iota})t}\widetilde{C}(t_0,t)=e^{-i(\omega_{\gamma\iota}^0+\lambda_{\gamma\gamma}-\lambda_{\iota\iota})t}e^{\Gamma(t_0,t)}. \notag
\end{align}
The correlation function can be evaluated exactly for a Gaussian bath using cumulant expansion. This procedure is conveniently implemented in the Quantarhei package\cite{manvcal2020quantarhei}, resulting in
\begin{align}
    \Gamma(t_0,t)=&-g_{\iota\iota}(t)-g_{\gamma\gamma}(t)+g_{\iota \gamma}(t)+g_{\gamma\iota}(t)-g_{\iota\iota}+(t_0)+g^*_{\iota\iota}(t_0)+g_{\iota\iota}(t_0+t) \notag \\
    &-g_{\iota\iota}^*(t_0+t)+g_{\gamma\iota}(t_0)-g_{\iota \gamma}^*(t_0)-g_{\gamma\iota}(t_0+t)+g^*_{\iota \gamma}(t_0+t) \notag \\
    =&-g_{\iota\iota}(t)-g_{\gamma\gamma}(t)+2g_{\iota\gamma}(t)-i2\Im\{g_{\iota\iota}(t_0)-g_{\iota\iota}(t_0+t)\} \notag \\
    &+i2\Im\{g_{\iota\gamma}(t_0)-g_{\iota\gamma}(t_0+t)\},\notag    
\end{align}
where $g_{\iota\gamma}(t)$ is the lineshape function.
Assuming that coherences quickly dephase, it is possible to perform a Taylor expansion to the first order in $t$, following Ref. \citenum{cho2009two}
\begin{equation}
    g(t_0+t)\simeq g(t_0) + \dot{g}(t_0)t, 
\end{equation}
where $\dot{g}(t)$ represents the time derivative of the lineshape function. We obtain 
\begin{align}
    \Gamma(t_0,t) \simeq -g_{\iota\iota}(t)-g_{\gamma\gamma}(t)+2g_{\iota\gamma}(t)+i2Im\{\dot{g}_{\iota\iota}(t_0)t\}-i2Im\{\dot{g}_{\iota\gamma}(t_0)t\}\notag. 
\end{align}
Taking the long-time limit in $t_0$ and using the definition of reorganization energy\cite{mukamel}
\begin{equation}
    \lambda=-\lim_{t_0\to\infty} \Im\{\dot{g}(t_0)\},
\end{equation}
the expression simplifies to 
\begin{align}
      \widetilde{C}(t_0,t) \simeq -g_{\iota\iota}(t)-g_{\gamma\gamma}(t)+2g_{\iota\gamma}(t)-i2\lambda_{\iota\iota}t+i2\lambda_{\iota\gamma}t\notag.
\end{align}

By taking the Fourier transform, we obtain the expression for the absorption lineshape
\begin{equation}
  \mathcal{L}^{abs}_{\iota\gamma}(\omega)=\int dt\ e^{i\omega t}  e^{-i(\omega_{\iota\gamma}+2\lambda_{\iota\iota}-2\lambda_{\iota\gamma}) t-(g_{\iota\iota}(t)+g_{\gamma\gamma}(t)-2g_{\iota\gamma}(t))-\Gamma_{\iota\gamma}t},
  \end{equation}
as reported in the literature.\cite{novoderezhkin2004energy} $\Gamma_{\iota\gamma}$ is lifetime dephasing due to population decay, and as was discussed in the main text, we approximated it as with $\Gamma_{\iota\gamma} \approx \frac{k_{\iota\iota}+k_{\gamma\gamma}}{2}$.\cite{chang_accuracy_2015}

\bigskip

  For the emission lineshape, we follow the same steps but starting with the Feynman diagram in Fig. \ref{FD_lineshape}b, to get
%
 \begin{equation}
  \mathcal{L}^{\text{em}}_{\iota\gamma}(\omega)=\int_{-\infty}^{+\infty} dt\ e^{i\omega t}  e^{-i(\omega_{\iota\gamma}-2\lambda_{\gamma\gamma}+2\lambda_{\iota\gamma}) t-(g^*_{\iota\iota}(t)+g^*_{\gamma\gamma}(t)-2g^*_{\iota\gamma}(t))-\Gamma_{\iota\gamma}t}.
  \end{equation}
The lineshape functions $g_{\iota\gamma}(t)$ can be obtained by integrating the bath correlation function
\begin{equation}
g_{\iota\gamma}(t)=\int^t_0 d\tau \int^\tau_0 dt' C_{\iota\gamma}(t') =\int^t_0 d\tau \int^\tau_0 dt' \langle \Delta V_{\iota\iota}(t')V_{\gamma\gamma}(0) \rangle dt'd\tau
\end{equation}
and the reorganization energy as 
\begin{equation}
    \lambda=\int^\infty_{0}d\omega\frac{1}{\pi\omega}\int^{\infty}_{-\infty} dt\ e^{i\omega t}C(t).
\end{equation}
%

%

To derive the coefficients for $ g_{abcd}(t)$, we first have to make some assumptions about the system and bath of each state to be able to proceed. We assume that for a two-exciton state $ij$ in site basis we can write
\begin{equation}
    \Delta V_{ij,ij}= \Delta V_{ii} + \Delta V_{jj}, 
\end{equation}
in other words, we expect the bath hamiltonian to be diagonal, thus bath of double-excited state is just sum of baths of each single-excited state. The fluctuations of higher-excited state are assumed to be proportional to those of the single-excited state of the same molecule, scaled by the coefficient $\Phi$:
\begin{equation}
\langle \Delta V_{f_af_a}(t')V_{f_bf_b}(0) \rangle = \Phi^2 \langle \Delta V_{aa}(t')V_{bb}(0) \rangle
\end{equation}
For the correlation between fluctuations of higher-excited state $f_a$ and single excited state $b$, we again use the assumption that the bath of $f_a$ state has the same structure as bath of single excited state $a$ except for coefficient $\Phi$, but now there is additional coefficient $\Phi_{\text{corr}}$ scaling correlation between fluctuations of bath of first and higher excited states of the same molecule:
\begin{equation}
\langle \Delta V_{f_af_a}(t')V_{bb}(0) \rangle = \Phi\Phi_{\text{corr}} \langle \Delta V_{aa}(t')V_{bb}(0) \rangle.
\end{equation}
Using these definitions and the transformation coefficients between the site and excitonic bases, we can obtain expressions for the line-shape functions in the excitonic basis. The lineshape functions related to the auto-correlation function of the energy-gap operator of the single-excited state is:
\begin{equation}
g_{\alpha\alpha\alpha\alpha}(t)=\sum_{\substack{i}} |c^\alpha_i|^4\nu_ig(t),
\end{equation}
for the double-excited state is:
\begin{equation}
\begin{split}
g_{\beta\beta\beta\beta}=&\sum_{\substack{i}}|c^{\beta}_{f_i}|^4\Phi^2\nu_i g(t)\\
&+\sum_{\substack{i<j \\ k<l}}(c^{\beta}_{ij}c^{\beta}_{kl})^2((\delta_{ik}+\delta_{il})\nu_i+(\delta_{jk}+\delta_{jl})\nu_j) g(t)\\
&+2\sum_{\substack{i<j \\ k}}(c^{\beta}_{ij}c^{\beta}_{f_k})^2(\delta_{ik}+\delta_{jk})\nu_k\Phi\Phi_{\text{corr}}g(t)
\end{split}
\end{equation}
and for the triple-excited state is:
\begin{equation}
    \begin{split}
        g_{\xi\xi\xi\xi}(t)=&\sum_{\substack{i<j<k \\ l<m<n}}(c^\xi_{ijk}c^\xi_{lmn})^2\left((\delta_{il}+\delta_{im}+\delta_{in})\nu_i+(\delta_{jl}+\delta_{jm}+\delta_{jn})\nu_j+(\delta_{kl}+\delta_{km}+\delta_{kn})\nu_k)\right)g(t)\\
        &+\sum_{\substack{i,j \\ k,l}}(c^\xi_{if_j}c^\xi_{kf_l})^2((\delta_{ik}+\delta_{il}\Phi\Phi_{\text{corr}})\nu_i+(\delta_{jk}\Phi_{\text{corr}}+\delta_{jl}\Phi)\nu_j\Phi) g(t)\\
        &+2\sum_{\substack{i<j<k \\ l,m}}(c^\xi_{ijk}c^\xi_{f_lm})^2((\delta_{li}+\delta_{lj}+\delta_{lk})\Phi\Phi_{\text{corr}}\nu_l+(\delta_{mi}+\delta_{mj}+\delta_{mk})\nu_m)g(t)
    \end{split}
\end{equation}
Instead, the linshape function related to the cross-correlation function of the energy-gap operator for single- and double-excited state is:
\begin{equation}
g_{\beta\beta \alpha\alpha}(t)=\sum_{\substack{i<j \\ k}} \left(c^\beta_{ij}c^\alpha_{k}\right)^2\left(\delta_{jk}\nu_j + \delta_{ik}\nu_i\right)g(t)+\sum\limits_i\left(c^\beta_{f_i}c^\alpha_i\right)^2\Phi_{\text{corr}}\Phi\nu_i g(t)
\end{equation}
and between double- and triple-excited state is:
\begin{equation}
\begin{split}
g_{\xi\xi\beta\beta}(t)=&\sum_{\substack{i<j<k \\ l<m}}\left(c^\xi_{ijk}c^\beta_{lm}\right)^2\lbrack\left(\delta_{il}+\delta_{im}\right)\nu_i+\left(\delta_{jl}+\delta_{jm}\right)\nu_j+\left(\delta_{kl}+\delta_{km}\right)\nu_k\rbrack g(t)\\
    &+\sum_{\substack{i,j \\ k}}\left(c^\xi_{f_ij}c^\beta_{f_k}\right)^2\left(\delta_{ik}\Phi^2\nu_i+\delta_{jk}\Phi_{\text{corr}}\Phi\nu_j\right)g(t)\\
    &+\sum_{\substack{i<j \\ k,l}}\left(c^\xi_{f_kl}c^\beta_{ij}\right)^2\lbrack\left(\delta_{ki}+\delta_{kj}\right)\Phi_{\text{corr}}\Phi\nu_k+\left(\delta_{li}+\delta_{lj}\right)\nu_l\rbrack g(t)\\
    &+\sum_{\substack{i<j<k \\l}}\left(c^\xi_{ijk}c^\beta_{f_l}\right)^2\left(\delta_{il}+\delta_{jl}+\delta_{kl}\right)\Phi_{\text{corr}}\Phi\nu_lg(t)
\end{split}
\end{equation}

\subsection{Rotational averaging}
\label{sec:average}
Using a model for various pulses,\cite{luttig2021anisotropy} the intensity of the signal can be described in the impulsive limit as a multiplication of the system response and pump and probe field envelopes ($\pmb{\mathcal{E}}^{Pu}$ and $\pmb{\mathcal{E}}^{Pr}$, respectively)

\begin{equation}
    I_{\text{signal}} \propto \sum_{A,B,C,D,E,F}S^{(5)}_{ABCDEF} \pmb{\mathcal{E}}^{Pr*}_{A} \pmb{\mathcal{E}}^{Pr}_{B} \pmb{\mathcal{E}}^{Pu}_{C} \pmb{\mathcal{E}}^{Pu}_{D} \pmb{\mathcal{E}}^{Pu*}_{E} \pmb{\mathcal{E}}^{Pu*}_{F}.
\end{equation}

Uppercase letters denote individual indices of the three-dimensional vector of the electric field $ \boldsymbol{E}(t)$ and the six-dimensional tensor for response of the system $\overleftrightarrow{S}^{(5)}$. Five of the dimensions correspond to the matter-field interactions (see Eq. 2 in the main text), producing the polarization (last index) projected on the probe field. For calculation of the signal intensity we need to sum over A,B,C,D,E and F indices in the laboratory frame. This is connected to the molecular frame response by rotational averaging, \cite{luttig2021anisotropy}
\begin{equation}
  S^{(5)}_{ABCDEF} =  \sum_{a,b,c,d,e,f} T_{ABCDEF:abcdef}s^{(5)}_{abcdef}. 
\end{equation}

\noindent The tensor $T_{ABCDEF:abcdef}$ transforms the response of individual molecules  $s^{(5)}_{abcdef}$  from their molecular frame (lowercase letters) into the laboratory frame (uppercase letters) and depends on the angle between the polarization of the incident pulses. \cite{luttig2021anisotropy} To remove orientational anisotropy, we employed a specific experimental configuration where the probe pulse is oriented at the magic angle (54.7°) relative to the pump polarization. The transformation tensor for this setup is \cite{luttig2021anisotropy}

\begin{equation}
    T_{ABCDEF:abcdef}= \frac{1}{45} \delta_{ab} (\delta_{cd}\delta_{ef}+\delta_{ce}\delta_{df}+\delta_{cf}\delta_{de}).
\end{equation}

This tensor is obtained by rotational averaging over all molecular orientations. For fifth-order pump--probe spectroscopy, the response function contains six scalar products between the transition dipole moments and the polarization vectors of the incident light. For an isotropic ensemble, the orientational average is rotationally invariant and therefore can depend only on rotational invariants, namely scalar products between the polarization vectors and between the transition dipole moments. This yields 

\[
\adjustbox{angle=0, max width=\textheight}{
 $\langle (\vec{\mu}_1\cdot \vec{e}_1)...(\vec{\mu}_6\cdot \vec{e}_6) \rangle_\Omega=
\begin{pmatrix}
(\vec{e}_1\cdot \vec{e}_2)(\vec{e}_3\cdot \vec{e}_4)(\vec{e}_5\cdot \vec{e}_6)\\
(\vec{e}_1\cdot \vec{e}_2)(\vec{e}_3\cdot \vec{e}_5)(\vec{e}_4\cdot \vec{e}_6)\\
(\vec{e}_1\cdot \vec{e}_2)(\vec{e}_3\cdot \vec{e}_6)(\vec{e}_4\cdot \vec{e}_5)\\
(\vec{e}_1\cdot \vec{e}_3)(\vec{e}_2\cdot \vec{e}_4)(\vec{e}_5\cdot \vec{e}_6)\\
(\vec{e}_1\cdot \vec{e}_3)(\vec{e}_2\cdot \vec{e}_5)(\vec{e}_4\cdot \vec{e}_6)\\
(\vec{e}_1\cdot \vec{e}_3)(\vec{e}_2\cdot \vec{e}_6)(\vec{e}_4\cdot \vec{e}_5)\\
(\vec{e}_1\cdot \vec{e}_4)(\vec{e}_2\cdot \vec{e}_3)(\vec{e}_5\cdot \vec{e}_6)\\
(\vec{e}_1\cdot \vec{e}_4)(\vec{e}_2\cdot \vec{e}_5)(\vec{e}_3\cdot \vec{e}_6)\\
(\vec{e}_1\cdot \vec{e}_4)(\vec{e}_2\cdot \vec{e}_6)(\vec{e}_3\cdot \vec{e}_5)\\
(\vec{e}_1\cdot \vec{e}_5)(\vec{e}_2\cdot \vec{e}_3)(\vec{e}_4\cdot \vec{e}_6)\\
(\vec{e}_1\cdot \vec{e}_5)(\vec{e}_2\cdot \vec{e}_4)(\vec{e}_3\cdot \vec{e}_6)\\
(\vec{e}_1\cdot \vec{e}_5)(\vec{e}_2\cdot \vec{e}_6)(\vec{e}_3\cdot \vec{e}_4)\\
(\vec{e}_1\cdot \vec{e}_6)(\vec{e}_2\cdot \vec{e}_3)(\vec{e}_4\cdot \vec{e}_5)\\
(\vec{e}_1\cdot \vec{e}_6)(\vec{e}_2\cdot \vec{e}_4)(\vec{e}_3\cdot \vec{e}_5)\\
(\vec{e}_1\cdot \vec{e}_6)(\vec{e}_2\cdot \vec{e}_5)(\vec{e}_3\cdot \vec{e}_4)\\
\end{pmatrix}^{T}
\mathds{A}
\begin{pmatrix}
(\vec{\mu}_1\cdot \vec{\mu}_2)(\vec{\mu}_3\cdot \vec{\mu}_4)(\vec{\mu}_5\cdot \vec{\mu}_6)\\
(\vec{\mu}_1\cdot \vec{\mu}_2)(\vec{\mu}_3\cdot \vec{\mu}_5)(\vec{\mu}_4\cdot \vec{\mu}_6)\\
(\vec{\mu}_1\cdot \vec{\mu}_2)(\vec{\mu}_3\cdot \vec{\mu}_6)(\vec{\mu}_4\cdot \vec{\mu}_5)\\
(\vec{\mu}_1\cdot \vec{\mu}_3)(\vec{\mu}_2\cdot \vec{\mu}_4)(\vec{\mu}_5\cdot \vec{\mu}_6)\\
(\vec{\mu}_1\cdot \vec{\mu}_3)(\vec{\mu}_2\cdot \vec{\mu}_5)(\vec{\mu}_4\cdot \vec{\mu}_6)\\
(\vec{\mu}_1\cdot \vec{\mu}_3)(\vec{\mu}_2\cdot \vec{\mu}_6)(\vec{\mu}_4\cdot \vec{\mu}_5)\\
(\vec{\mu}_1\cdot \vec{\mu}_4)(\vec{\mu}_2\cdot \vec{\mu}_3)(\vec{\mu}_5\cdot \vec{\mu}_6)\\
(\vec{\mu}_1\cdot \vec{\mu}_4)(\vec{\mu}_2\cdot \vec{\mu}_5)(\vec{\mu}_3\cdot \vec{\mu}_6)\\
(\vec{\mu}_1\cdot \vec{\mu}_4)(\vec{\mu}_2\cdot \vec{\mu}_6)(\vec{\mu}_3\cdot \vec{\mu}_5)\\
(\vec{\mu}_1\cdot \vec{\mu}_5)(\vec{\mu}_2\cdot \vec{\mu}_3)(\vec{\mu}_4\cdot \vec{\mu}_6)\\
(\vec{\mu}_1\cdot \vec{\mu}_5)(\vec{\mu}_2\cdot \vec{\mu}_4)(\vec{\mu}_3\cdot \vec{\mu}_6)\\
(\vec{\mu}_1\cdot \vec{\mu}_5)(\vec{\mu}_2\cdot \vec{\mu}_6)(\vec{\mu}_3\cdot \vec{\mu}_4)\\
(\vec{\mu}_1\cdot \vec{\mu}_6)(\vec{\mu}_2\cdot \vec{\mu}_3)(\vec{\mu}_4\cdot \vec{\mu}_5)\\
(\vec{\mu}_1\cdot \vec{\mu}_6)(\vec{\mu}_2\cdot \vec{\mu}_4)(\vec{\mu}_3\cdot \vec{\mu}_5)\\
(\vec{\mu}_1\cdot \vec{\mu}_6)(\vec{\mu}_2\cdot \vec{\mu}_5)(\vec{\mu}_3\cdot \vec{\mu}_4)\\
\end{pmatrix}$
},
\]

with transformation matrix

\[
\adjustbox{angle=0, max width=\textheight}{
 $\mathds{A}=\frac{1}{210}
\begin{pmatrix}
16 & -5 & -5 & -5 & 2 & 2 & -5 & 2 & 2 & 2 & 2 & -5 & 2 & 2 & -5\\
-5 & 16 & -5 & 2 & -5 & 2 & 2 & 2 & -5 & -5 & 2 & 2 & 2 & -5 & 2\\
-5 & -5 & 16 & 2 & 2 & -5 & 2 & -5 & 2 & 2 & -5 & 2 & -5 & 2 & 2\\
-5 & 2 & 2 & 16 & -5 & -5 & -5 & 2 & 2 & 2 &-5 & 2 & 2 & -5 & 2\\
2 & -5 & 2 & -5 & 16 & -5 & 2 & -5 & 2 & -5 & 2 & 2 & 2 & 2 & -5\\
2 & 2 & -5 & -5 & -5 & 16 & 2 & 2 & -5 & 2 & 2 & -5 & -5 & 2 & 2\\
-5 & 2 & 2 & -5 & 2 & 2 & 16 & -5 & -5 & -5 & 2 & 2 & -5 & 2 & 2\\
2 & 2 & -5 & 2 & -5 & 2 & -5 & 16 & -5 & 2 & -5 & 2 & 2 & 2 & -5\\
2 & -5 & 2 & 2 & 2 & -5 & -5 & -5 & 16 & 2 & 2 & -5 & 2 & -5 & 2\\
2 & -5 & 2 & 2 & -5 & 2 & -5 & 2 & 2 & 16 & -5 & -5 & -5 & 2 & 2\\
2 & 2 & -5 & -5 & 2 & 2 & 2 & -5 & 2 & -5 & 16 & -5 & 2 & -5 & 2\\

-5 & 2 & 2 & 2 & 2 & -5 & 2 & 2 & -5 & -5 & -5 & 16 & 2 & 2 & -5\\

2 & 2 & -5 & 2 & 2 & -5 & -5 & 2 & 2 & -5 & 2 & 2 & 16 & -5 & -5\\

2 & -5 & 2 & -5 & 2 & 2 & 2 & 2 & -5 & 2 & -5 & 2 & -5 & 16 & -5\\
-5 & 2 & 2 & 2 & -5 & 2 & 2 & -5 & 2 & 2 & 2 & -5 & -5 & -5 & 16
\end{pmatrix}$
}
\]
For all parallel polarizations, we have $\vec{e}_i\cdot \vec{e}_j=1$, $\forall i,j \in \{1,\cdots,6\}$
and the term for rotational averaging simplifies to (using Einstein's summing notation) 
\begin{equation}
\begin{split}
\langle (\vec{\mu}_1\cdot \vec{e}_1)...(\vec{\mu}_6\cdot \vec{e}_6) \rangle_\Omega = \frac{2}{210}\mu_1^a\mu_2^b\mu_3^c\mu_4^d\mu_5^e\mu_6^f (&\delta_{ab}\delta_{cd}\delta_{ef}+\delta_{ab}\delta_{ce}\delta_{df}+\delta_{ab}\delta_{cf}\delta_{de}+\\
+&\delta_{ac}\delta_{bd}\delta_{ef}+\delta_{ac}\delta_{be}\delta_{df}+\delta_{ac}\delta_{bf}\delta_{de}+\\
+&\delta_{ad}\delta_{bc}\delta_{ef}+\delta_{ad}\delta_{be}\delta_{cf}+\delta_{ad}\delta_{bf}\delta_{ce}+\\
+&\delta_{ae}\delta_{bc}\delta_{df}+\delta_{ae}\delta_{bd}\delta_{cf}+\delta_{ae}\delta_{bf}\delta_{cd}+\\
+&\delta_{af}\delta_{bc}\delta_{de}+\delta_{af}\delta_{bd}\delta_{ce}+\delta_{af}\delta_{be}\delta_{cd})
\end{split}
\end{equation}
For magic angle between polarizations of pump and probe pulses the coefficient reduces to 
\begin{equation}
    \langle (\vec{\mu}_1\cdot \vec{e}_1)...(\vec{\mu}_6\cdot \vec{e}_6) \rangle_\Omega = \frac{1}{45} \mu_1^a\mu_2^b\mu_3^c\mu_4^d\mu_5^e\mu_6^f \delta_{ab} (\delta_{cd}\delta_{ef}+\delta_{ce}\delta_{df}+\delta_{cf}\delta_{de})
\end{equation}

\bigskip

For 4th order response, the procedure is analogous, leading
\[
\adjustbox{angle=0, max width=\textheight}{
 $\langle (\vec{\mu}_1\cdot \vec{e}_1)...(\vec{\mu}_4\cdot \vec{e}_4) \rangle_\Omega=
\begin{pmatrix}
(\vec{e}_1\cdot \vec{e}_2)(\vec{e}_3\cdot \vec{e}_4)\\
(\vec{e}_1\cdot \vec{e}_3)(\vec{e}_2\cdot \vec{e}_4)\\
(\vec{e}_1\cdot \vec{e}_4)(\vec{e}_2\cdot \vec{e}_3)\\
\end{pmatrix}^{T}
\frac{1}{30}
\begin{pmatrix}
4 & -1 & -1\\
-1 & 4 & -1\\
-1 & -1 & 4\\
\end{pmatrix}
\begin{pmatrix}
(\vec{\mu}_1\cdot \vec{\mu}_2)(\vec{\mu}_3\cdot \vec{\mu}_4)\\
(\vec{\mu}_1\cdot \vec{\mu}_3)(\vec{\mu}_2\cdot \vec{\mu}_4)\\
(\vec{\mu}_1\cdot \vec{\mu}_4)(\vec{\mu}_2\cdot \vec{\mu}_3)\\
\end{pmatrix}$
}.
\]
If all four pulses have parallel polarization, we have
\begin{equation}
\langle (\vec{\mu}_1\cdot \vec{e}_1)...(\vec{\mu}_4\cdot \vec{e}_4) \rangle_\Omega = \frac{1}{15}\mu_1^a\mu_2^b\mu_3^c\mu_4^d(\delta_{ab}\delta_{cd}+\delta_{ac}\delta_{bd}+\delta_{ad}\delta_{bc}).
\end{equation}
In case of polarization of pump pulses being perpendicular to polarization of probe pulses the equation reduces to 
\begin{equation}
\langle (\vec{\mu}_1\cdot \vec{e}_1)...(\vec{\mu}_4\cdot \vec{e}_4) \rangle_\Omega = \frac{1}{30}\mu_1^a\mu_2^b\mu_3^c\mu_4^d(4\delta_{ab}\delta_{cd}-\delta_{ac}\delta_{bd}-\delta_{ad}\delta_{bc}).
\end{equation}
If we set magic angle between pump and probe polarizations, the signal stops being sensitive to change of transition dipole moment due to rotation of the molecule during waiting time:
\begin{equation}
\langle (\vec{\mu}_1\cdot \vec{e}_1)...(\vec{\mu}_4\cdot \vec{e}_4) \rangle_\Omega = \frac{1}{9}\mu_1^a\mu_2^b\mu_3^c\mu_4^d(\delta_{ab}\delta_{cd}).
\end{equation}

\section{Signal of simple samples}
\subsubsection{Weakly coupled dimer}

For the weakly coupled dimer described in Sec. III A of the main text, we have population transfer rates matrix 

\begin{gather}
K_{rate}=
   \begin{pmatrix}
0 & 0 & 0 & 0\\
0 & 0 & k_T & \frac{k_A}{2}\\
0 & 0 & -k_T & \frac{k_A}{2}\\
0 & 0 & 0 & -k_A\\
\end{pmatrix}.
\end{gather}

\noindent The Green function $\mathcal{G}(t)$, with which we can evolve the wave function in time as $|\Psi(t)\rangle = \mathcal{G}(t)|\Psi(0)\rangle$, was calculated as an exponential of the $K_{rate}$ matrix

\begin{gather}
\mathcal{G}(t)=
\begin{pmatrix}
1 & 0 & 0 & 0\\
0 & 1 & 1-e^{-k_Tt} &1-e^{-k_At} -\frac{k_A(e^{-k_Tt}-e^{-k_At})}{2(k_A-k_T)}\\
0 & 0 & e^{-k_Tt} & \frac{k_A(e^{-k_Tt}-e^{-k_At})}{2(k_A-k_T)}\\
0 & 0 & 0 & e^{-k_At}\\
\end{pmatrix}
.
\end{gather}

\noindent By counting all possible pathways according to Eq.7 in the main text, we get the following fifth-order pump--probe signal:

\begin{equation} 
\begin{split} 
\mbox{PP}^{(5)}(\omega,t)=&16\mu_i^4 e^{-k_Tt}(\mu_i^2\mathcal{L}_{i,g}(\omega)-\mu_j^2\mathcal{L}_{j,g}(\omega)) \\
&+\mu_i^2\mu_j^2\left(48-\tfrac{24k_A}{k_A - k_T}\right)\left(e^{-k_Tt}-e^{-k_At}\right)\left(\mu_i^2\mathcal{L}_{i,g}(\omega)-\mu_j^2\mathcal{L}_{j,g}(\omega)\right) \\
&+16\left(\mu_j^6+3\mu_i^2 \mu_j^4 +\mu_i^4\mu_j^2-3\mu_i^2 \mu_j^4 e^{-k_At}\right)\mathcal{L}_{j,g}(\omega).
\end{split} 
\label{eq_dimer_weak_EEI2D} 
\end{equation} 
In case of $\mu_i=\mu_j=\mu$ the spectrally integrated signal rises exponentially with annihilation
\begin{equation} 
\begin{split} 
\int\mbox{PP}^{(5)}_{\mu_i=\mu_j}(\omega,t)d\omega=\left(80-48 e^{-k_At}\right)\mu^6.
\end{split} 
\label{eq_dimer_weak_EEI2D_mua=mub_integ} 
\end{equation} 
However, in any other case $\mu_i \neq \mu_j$ the integrated signal depends both on annihilation and transfer rates.
\begin{equation} 
\begin{split} 
&\int\mbox{PP}^{(5)}(\omega,t)d\omega=\\
+&\left[\left(16\mu_i^2 + \mu_j^2\left(48-\frac{24k_A}{k_A - k_T}\right)\right) e^{-k_Tt}-\mu_j^2\left(48-\frac{24k_A}{k_A - k_T}\right) e^{-k_At}\right]\mu_i^4 \\ 
-&\left[\left(16\mu_i^4 + \mu_i^2 \mu_j^2 \left(48 - \frac{24k_A}{k_A - k_T} \right) 
 \right) e^{-k_Tt} +\frac{\mu_i^2 \mu_j^2 24k_A}{k_A - k_T} e^{-k_At}- 16(\mu_j^4 + 3\mu_i^2 \mu_j^2 + \mu_i^4) \right]\mu_j^2
\end{split} 
\label{eq_dimer_weak_EEI2D_integ} 
\end{equation} 
After applying limit $k_T \longrightarrow k_A$, we get for the general integrated signal
\begin{equation} 
\begin{split} 
&\int\mbox{PP}^{(5)}_{k_A=k_T}(\omega,t)d\omega=16(\mu_i^4\mu_j^2+\mu_j^6+3\mu_i^2\mu_j^4)-\mu_i^2\mu_j^2(\mu_j^2-\mu_i^2)24k_Ate^{-k_At}+16(\mu_i^6-\mu_i^4\mu_j^2-3\mu_i^2\mu_j^4)e^{-k_At}.
\end{split} 
\label{eq_dimer_weak_EEI2D_mua>mub_integ} 
\end{equation}

\noindent Let us set $\mu_i=1.5 > \mu_j =1.0 $ and use the limit $k_A=k_T$. The formula for spectrally integrated fifth-order signal is then 
\begin{equation} 
\begin{split} 
\int\mbox{PP}^{(5)}_{\mu_i=1.5,\,\,\mu_j=1.0,\,\,k_A=k_T}(\omega,t)d\omega=\frac{1485}{4}+\frac{135}{2}k_Ate^{-k_At}-\frac{27}{4}e^{-k_At},
\end{split} 
\label{eq_dimer_weak_EEI2D_mua>mub_kA=kT_integ} 
\end{equation} 

\noindent where the first positive term causes the signal to decrease and creates, competing with the negative exponential contribution, a characteristic peak on top of the exponentially rising curve known from the special case $\mu_i = \mu_j$.
For the opposite case $\mu_i=1.0 < \mu_j =1.5 $ and $k_A=k_T$ we have
\begin{equation} 
\begin{split} 
\int\mbox{PP}^{(5)}_{\mu_i=1.0,\,\,\mu_j=1.5,\,\,k_A=k_T}(\omega,t)d\omega=205-\frac{135}{2}k_Ate^{-k_At}-263e^{-k_At}.
\end{split} 
\label{eq_dimer_weak_EEI2D_mua<mub_integ} 
\end{equation} 
Both components with exponentials have negative sign and thus the signal is monotonous, rising the whole time.

\begin{figure*}[t!]
{\sffamily
\sansmath
\begin{subfigure}[t]{0.5\textwidth}
\RaggedLeft
\begin{overpic}[width=0.7\textwidth]{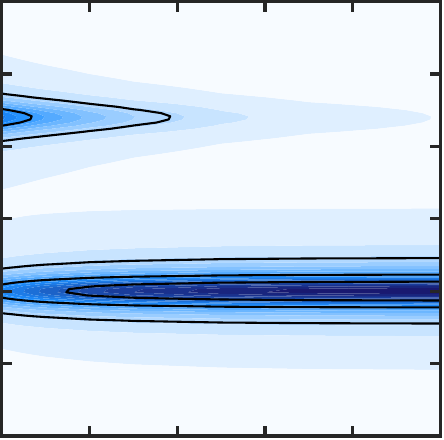}
\put(35,101){\footnotesize PP$^{(3)},\;\;k_A=0$}
   \put(-1,-6){\footnotesize  0}
  \put(17.5,-6){\footnotesize  20}
  \put(37.5,-6){\footnotesize  40}
  \put(57.5,-6){\footnotesize  60}
    \put(77.5,-6){\footnotesize  80}
  \put(95,-6){\footnotesize  100}
  \put(35,-15){\footnotesize  Time delay [ps]}
      \put(-20,20){\footnotesize  \rotatebox{90}{Wavenumber [$10^{3}$ cm$^{-1}$]}}
    \put(-10,97){\footnotesize  \rotatebox{0}{$15.5$}}
    \put(-10,80,67){\footnotesize  \rotatebox{0}{$15.0$}}
    \put(-10,64,3){\footnotesize  \rotatebox{0}{$14.5$}}
    \put(-10,48){\footnotesize  \rotatebox{0}{$14.0$}}
    \put(-10,31.67){\footnotesize  \rotatebox{0}{$13.5$}}
    \put(-10,15.3){\footnotesize  \rotatebox{0}{$13.0$}}
    \put(-10,-1){\footnotesize  \rotatebox{0}{$12.5$}}
       \put(-30,105){a}
\end{overpic}
\label{Fig:dimer_kA0_3PP}
\end{subfigure}%
\hspace{0.2cm}
\begin{subfigure}[t]{0.5\textwidth}
\RaggedRight
\begin{overpic}[width=0.7\textwidth]{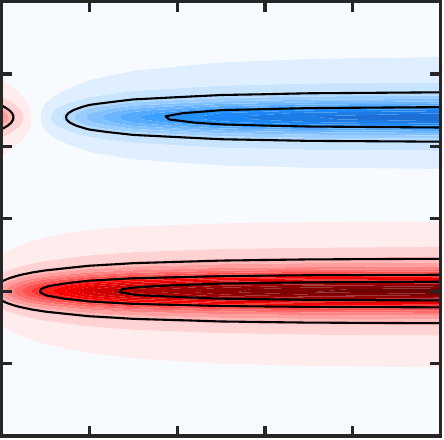}
\put(35,101){\footnotesize PP$^{(5)},\;\;k_A=0$}
   \put(-1,-6){\footnotesize  0}
  \put(17.5,-6){\footnotesize  20}
  \put(37.5,-6){\footnotesize  40}
  \put(57.5,-6){\footnotesize  60}
    \put(77.5,-6){\footnotesize  80}
  \put(95,-6){\footnotesize  100}
  \put(35,-15){\footnotesize  Time delay [ps]}
     \put(-2,105){b}
\end{overpic}
\label{Fig:dimer_kA0_5PP}
\end{subfigure} 
\vspace{1.8cm}

\begin{subfigure}[t]{0.5\textwidth}
\RaggedLeft
\begin{overpic}[width=0.7\textwidth]{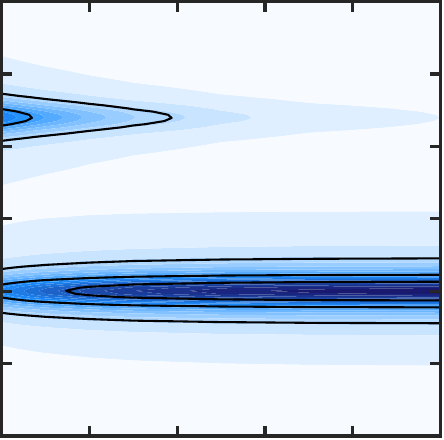}
\put(35,101){\footnotesize PP$^{(3)},\;\;k_A=k_T$}
   \put(-1,-6){\footnotesize  0}
  \put(17.5,-6){\footnotesize  20}
  \put(37.5,-6){\footnotesize  40}
  \put(57.5,-6){\footnotesize  60}
    \put(77.5,-6){\footnotesize  80}
  \put(95,-6){\footnotesize  100}
  \put(35,-15){\footnotesize  Time delay [ps]}
      \put(-20,20){\footnotesize  \rotatebox{90}{Wavenumber [$10^{3}$ cm$^{-1}$]}}
    \put(-10,97){\footnotesize  \rotatebox{0}{$15.5$}}
    \put(-10,80,67){\footnotesize  \rotatebox{0}{$15.0$}}
    \put(-10,64,3){\footnotesize  \rotatebox{0}{$14.5$}}
    \put(-10,48){\footnotesize  \rotatebox{0}{$14.0$}}
    \put(-10,31.67){\footnotesize  \rotatebox{0}{$13.5$}}
    \put(-10,15.3){\footnotesize  \rotatebox{0}{$13.0$}}
    \put(-10,-1){\footnotesize  \rotatebox{0}{$12.5$}}
       \put(-35,105){c}
\end{overpic}
\label{Fig:dimer_kA=kT_3PP}
\end{subfigure}%
\hspace{0.2cm}
\begin{subfigure}[t]{0.5\textwidth}
\RaggedRight
\begin{overpic}[width=0.7\textwidth]{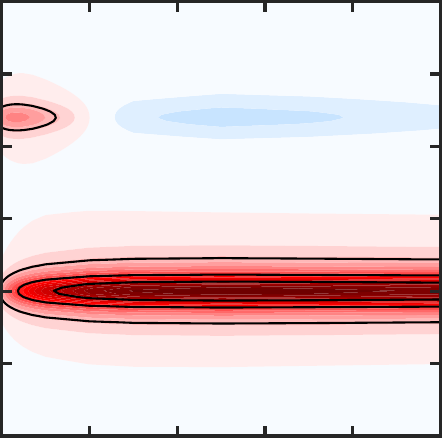}
\put(35,101){\footnotesize PP$^{(5)},\;\;k_A=k_T$}
   \put(-1,-6){\footnotesize  0}
  \put(17.5,-6){\footnotesize  20}
  \put(37.5,-6){\footnotesize  40}
  \put(57.5,-6){\footnotesize  60}
    \put(77.5,-6){\footnotesize  80}
  \put(95,-6){\footnotesize  100}
  \put(35,-15){\footnotesize  Time delay [ps]}
  \put(104,72){\includegraphics[width=0.03\textwidth]{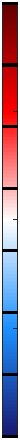}}
  \put(112,165){\footnotesize   1.0}
  \put(112,142){\footnotesize   0.5}
  \put(112,118){\footnotesize   0.0}
  \put(109,95){\footnotesize  -0.5}
  \put(109,72){\footnotesize  -1.0}
     \put(-2,105){f}
\end{overpic}
\label{Fig:dimer_kA=kT_5PP}
\end{subfigure}
\vspace{0.5cm}
}

\caption{\justifying Third- (left column) and fifth-order pump--probe spectra (left column) for weakly coupled heterodimer with $\mu_a=\mu_b=1$ in two settings: without annihilation (top row) and with equal annihilation and transfer rate (bottom row). }
\label{Fig:dimer_spectrum}

\end{figure*}

In order to compare analytical results with simulation, we assume specific case of $k_A=0$ to simplify expression for fifth-order pump--probe
\begin{equation}
\begin{split}
    \mbox{PP}^{(5)}_{\mbox{dim., }\mu_i=\mu_j=1,\,\,k_A=0}=\left(-48+64e^{-k_Tt}\right)\mathcal{L}_{ig}(\omega)+\left(80-64e^{-k_Tt}\right)\mathcal{L}_{jg}(\omega).
\end{split}
\label{eq_dimer_weak_EEI2D_kA=0}
\end{equation}
For $k_A=k_T=k$ we can write
\begin{equation}
\begin{split}
    \mbox{PP}^{(5)}_{\mbox{dim., }\mu_i=\mu_j=1,\,\,k_A=k_T}=(16-24kt)e^{-kt}\mathcal{L}_{ig}(\omega)+\left(80-64e^{-kt}+24kte^{-kt}\right)\mathcal{L}_{jg}(\omega).
\end{split}
\label{eq_dimer_weak_EEI2D_kA=kT}
\end{equation}
For the same system the third-order response can be expressed as 
\begin{equation}
\begin{split}
&\mbox{PP}^{(3)}_{\mbox{dim.}}=2\mu_i^4e^{-k_Tt}\mathcal{L}_{ig}(\omega)+\left(2\mu_j^4+ 2\mu_i^2\mu_j^4(1-e^{-k_Tt})\right)\mathcal{L}_{ig}(\omega).
\label{eq_dimer_weak_PP}
\end{split}
\end{equation}
Based on spectrally integrated signal one could assume that the fifth-order signal can be written in form of a product of third-order signal and annihilation dynamics $\mbox{PP}^{(5)}(\omega,t)=\mbox{PP}^{(3)}(\omega,t)\cdot(1-e^{-k_At})$. But for frequency-resolved fifth-order signal it does not hold anymore, as can be seen from comparison of eq. \ref{eq_dimer_weak_EEI2D} and \ref{eq_dimer_weak_PP}, because the necessary term $e^{-k_At}e^{-k_Tt}$, which would arise by such multiplication, is missing in the fifth-order signal.
In order to check theory and its implementation, we simulated third- and  fifth-order pump--probe spectra for weakly coupled dimer with $\mu_i=\mu_j$. 
In fifth-order pump--probe spectra for $k_A=k_T$ and weak coupling the signal from molecule $j$ is rising meanwhile signal from molecule $i$ is at first positive and decreasing, after about 5 ps it crosses zero and starts rising in negative numbers and then dying to zero (see Fig.\ref{Fig:dimer_spectrum}f). This very well corresponds with equation \ref{eq_dimer_weak_EEI2D_kA=kT}. For the system without annihilation, the fifth-order spectrum looks similar to previous case, except for the blue peak, which does not decrease (see Fig. \ref{Fig:dimer_spectrum}b), which is again in agreement with our prediction in Eq. \ref{eq_dimer_weak_EEI2D_kA=0}.

\clearpage

\section{Experimental example of $[\text{SQA}-\text{SQB}]_{19}$ polymer}
\label{Sec:SQAB_parameters_and_data}
This section presents additional simulations of SQA-SQB co-polymer that complement the results discussed in the main text. We include choice of all three sets of parameters used in microscopic model of this polymer in Table \ref{Tab:parameters SQAB19} (set 1 was used for simulations in Fig. 7, 8, 9 and 10 in the main text and Fig. S10, S11 and S12 here; set 2 in Fig. 9 and set 3 in Fig. 12 of the main text). 

\begin{figure*}[t!]
{\sffamily
\sansmath
\begin{subfigure}[t]{0.07\textwidth}
\begin{tikzpicture}
    \node [white] at (0.0,0) {$1$};
\end{tikzpicture}
\end{subfigure}%
\begin{subfigure}[t]{0.215\textwidth}
\centering
\begin{overpic}[width=0.95\textwidth]{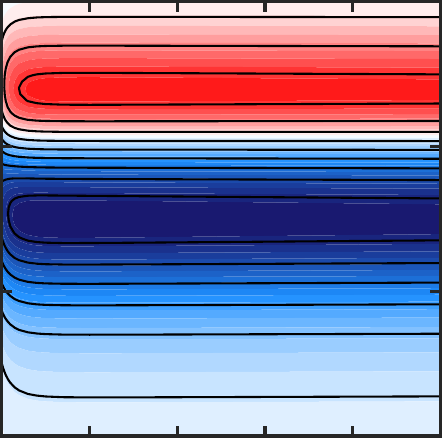}
  \put(5,103){\footnotesize $[SQA-SQB]_{2}\;\;(N=4)$}
    \put(-45,20){\footnotesize \rotatebox{90}{normalized PP$^{(3)}$}}

      \put(-31,6){\footnotesize  \rotatebox{90}{Wavenumber [$10^{3}$ cm$^{-1}$]}}
    \put(-18.5,95.75){\footnotesize  \rotatebox{0}{$13.5$}}
    \put(-18.5,63){\footnotesize  \rotatebox{0}{$13.0$}}
    \put(-18.5,30.25){\footnotesize  \rotatebox{0}{$12.5$}}
    \put(-18.5,0){\footnotesize  \rotatebox{0}{$12.0$}}
\end{overpic}
\end{subfigure}%
\begin{subfigure}[t]{0.215\textwidth}
\centering
\begin{overpic}[width=0.95\textwidth]{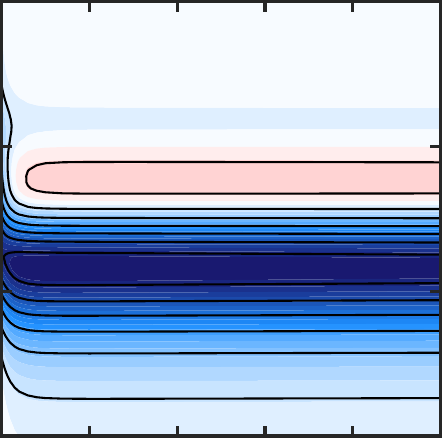}
  \put(5,103){\footnotesize $[SQA-SQB]_{4}\;\;(N=8)$}
\end{overpic}
\end{subfigure}%
\begin{subfigure}[t]{0.215\textwidth}
\centering
\begin{overpic}[width=0.95\textwidth]{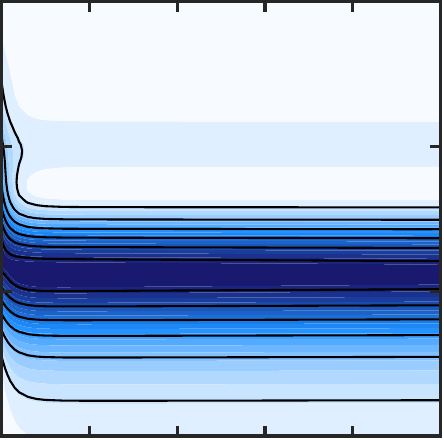}
  \put(5,103){\footnotesize $[SQA-SQB]_{5}\;\;(N=10)$}
\end{overpic}
\end{subfigure}%
\begin{subfigure}[t]{0.215\textwidth}
\centering
\begin{overpic}[width=0.95\textwidth]{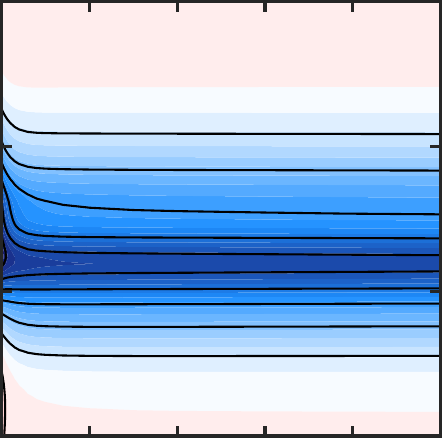}
  \put(5,103){\footnotesize $[SQA-SQB]_{10}\;\;(N=20)$}
\end{overpic}
\end{subfigure}
\vspace{0.2cm}

\begin{subfigure}[t]{0.07\textwidth}
\begin{tikzpicture}
    \node [white] at (0.0,0) {$1$};
\end{tikzpicture}
\end{subfigure}%
\begin{subfigure}[t]{0.215\textwidth}
\centering
\begin{overpic}[width=0.95\textwidth]{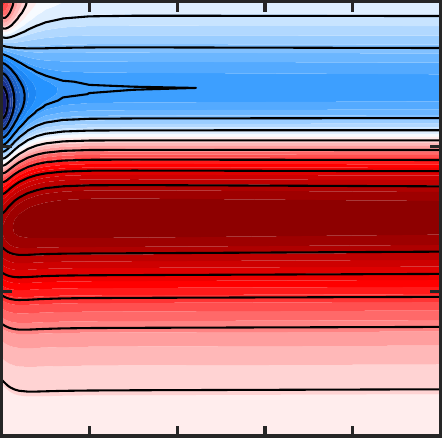}
  \put(-45,40){\footnotesize \rotatebox{90}{PP$^{(5)}$}}

      \put(-31,6){\footnotesize  \rotatebox{90}{Wavenumber [$10^{3}$ cm$^{-1}$]}}
    \put(-18.5,95.75){\footnotesize  \rotatebox{0}{$13.5$}}
    \put(-18.5,63){\footnotesize  \rotatebox{0}{$13.0$}}
    \put(-18.5,30.25){\footnotesize  \rotatebox{0}{$12.5$}}
    \put(-18.5,0){\footnotesize  \rotatebox{0}{$12.0$}}
\end{overpic}
\end{subfigure}%
\begin{subfigure}[t]{0.215\textwidth}
\centering
\begin{overpic}[width=0.95\textwidth]{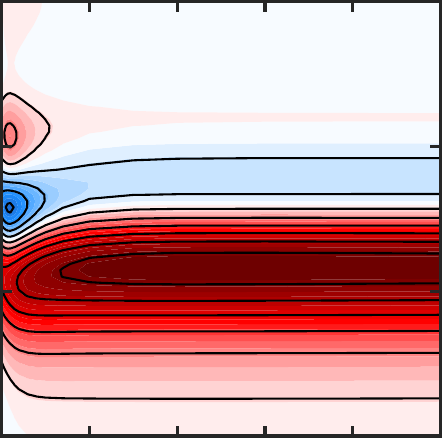}
\end{overpic}
\end{subfigure}%
\begin{subfigure}[t]{0.215\textwidth}
\centering
\begin{overpic}[width=0.95\textwidth]{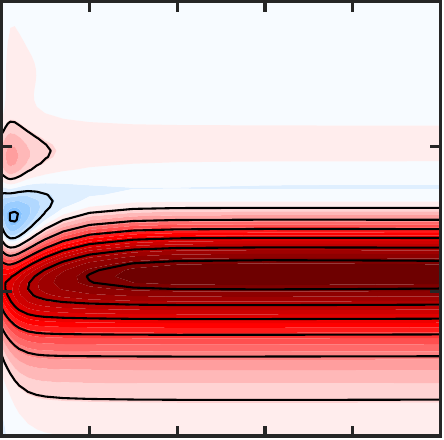}
\end{overpic}
\end{subfigure}%
\begin{subfigure}[t]{0.215\textwidth}
\centering
\begin{overpic}[width=0.95\textwidth]{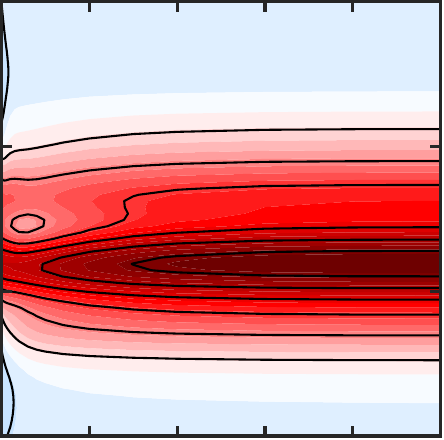}
\end{overpic}
\end{subfigure}
\vspace{0.2cm}

\begin{subfigure}[t]{0.07\textwidth}
\begin{tikzpicture}
    \node [white] at (0.0,0) {$1$};
\end{tikzpicture}
\end{subfigure}%
\begin{subfigure}[t]{0.215\textwidth}
\centering
\begin{overpic}[width=0.95\textwidth]{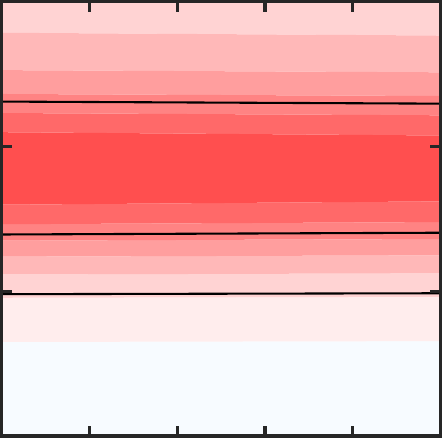}
  \put(-45,35){\footnotesize \rotatebox{90}{NGSB$^{(5)}$}}
      \put(-31,6){\footnotesize  \rotatebox{90}{Wavenumber [$10^{3}$ cm$^{-1}$]}}
    \put(-18.5,95.75){\footnotesize  \rotatebox{0}{$13.5$}}
    \put(-18.5,63){\footnotesize  \rotatebox{0}{$13.0$}}
    \put(-18.5,30.25){\footnotesize  \rotatebox{0}{$12.5$}}
    \put(-18.5,0){\footnotesize  \rotatebox{0}{$12.0$}}
\end{overpic}
\end{subfigure}%
\begin{subfigure}[t]{0.215\textwidth}
\centering
\begin{overpic}[width=0.95\textwidth]{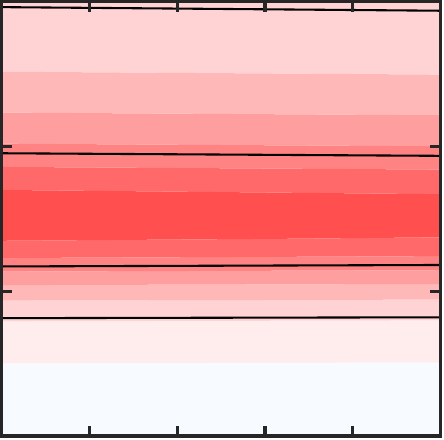}
\end{overpic}
\end{subfigure}%
\begin{subfigure}[t]{0.215\textwidth}
\centering
\begin{overpic}[width=0.95\textwidth]{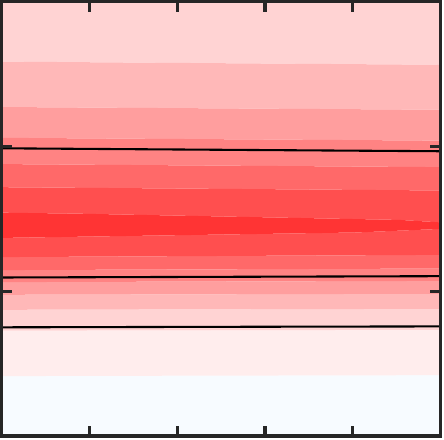}
\end{overpic}
\end{subfigure}%
\begin{subfigure}[t]{0.215\textwidth}
\centering
\begin{overpic}[width=0.95\textwidth]{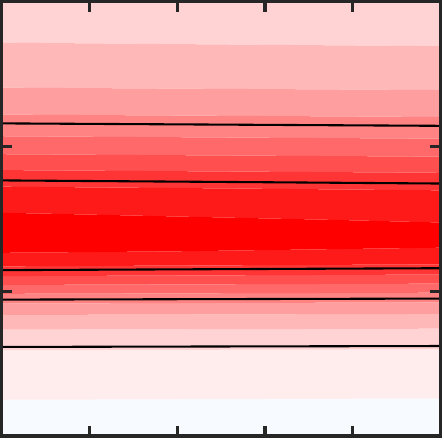}
 \put(101.,45.5){\includegraphics[width=0.0535\textwidth]{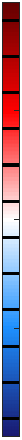}}
  \put(111,159){\footnotesize   1.0}
  \put(111,130){\footnotesize   0.5}
  \put(111,101){\footnotesize   0.0}
  \put(108,73){\footnotesize  -0.5}
  \put(108,45){\footnotesize  -1.0}
\end{overpic}
\end{subfigure}
\vspace{0.2cm}

\begin{subfigure}[t]{0.07\textwidth}
\begin{tikzpicture}
    \node [white] at (0.0,0) {$1$};
\end{tikzpicture}
\end{subfigure}%
\begin{subfigure}[t]{0.215\textwidth}
\centering
\begin{overpic}[width=0.95\textwidth]{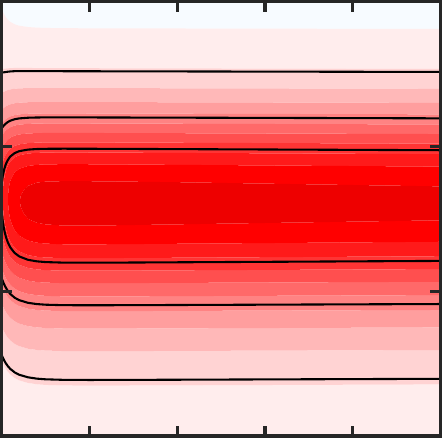}
  \put(-45,35){\footnotesize \rotatebox{90}{NSE$^{(5)}$}}
      \put(-31,6){\footnotesize  \rotatebox{90}{Wavenumber [$10^{3}$ cm$^{-1}$]}}
    \put(-18.5,95.75){\footnotesize  \rotatebox{0}{$13.5$}}
    \put(-18.5,63){\footnotesize  \rotatebox{0}{$13.0$}}
    \put(-18.5,30.25){\footnotesize  \rotatebox{0}{$12.5$}}
    \put(-18.5,0){\footnotesize  \rotatebox{0}{$12.0$}}
\end{overpic}
\end{subfigure}%
\begin{subfigure}[t]{0.215\textwidth}
\centering
\begin{overpic}[width=0.95\textwidth]{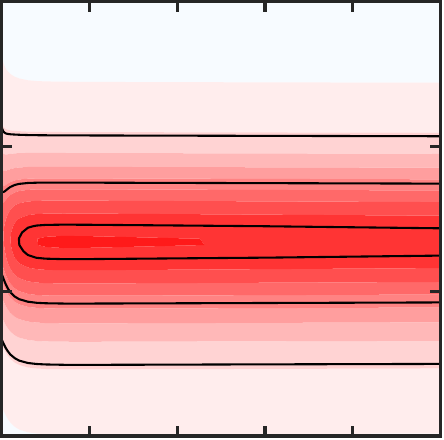}
\end{overpic}
\end{subfigure}%
\begin{subfigure}[t]{0.215\textwidth}
\centering
\begin{overpic}[width=0.95\textwidth]{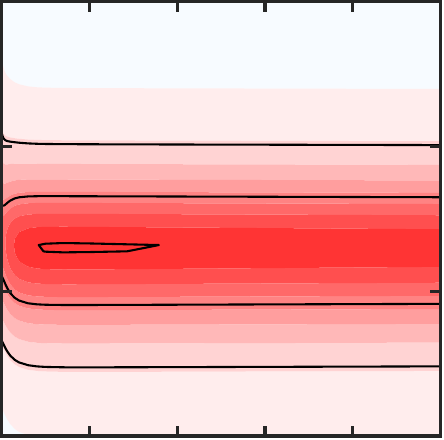}
\end{overpic}
\end{subfigure}%
\begin{subfigure}[t]{0.215\textwidth}
\centering
\begin{overpic}[width=0.95\textwidth]{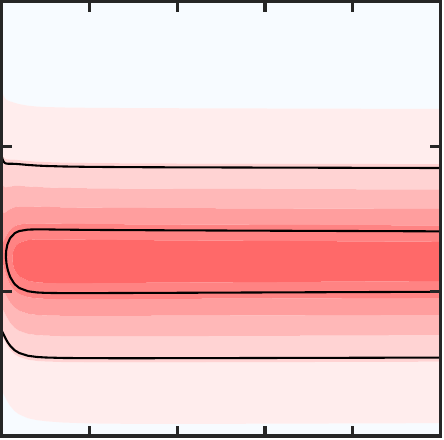}
\end{overpic}
\end{subfigure}
\vspace{0.2cm}

\begin{subfigure}[t]{0.07\textwidth}
\begin{tikzpicture}
    \node [white] at (0.0,0) {$1$};
\end{tikzpicture}
\end{subfigure}%
\begin{subfigure}[t]{0.215\textwidth}
\centering
\begin{overpic}[width=0.95\textwidth]{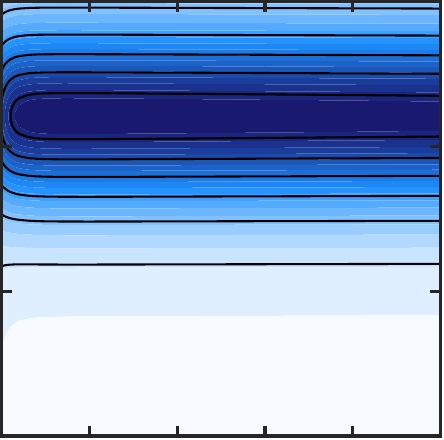}
  \put(-45,35){\footnotesize \rotatebox{90}{NESA$^{(5)}$}}
    \put(-1,-8){\footnotesize  0}
  \put(18,-8){\footnotesize  1}
  \put(38,-8){\footnotesize  2}
  \put(58,-8){\footnotesize 3}
  \put(78,-8){\footnotesize  4}
  \put(96,-8){\footnotesize  5}
  \put(33,-18.5){\footnotesize  Time [ps]}
      \put(-31,6){\footnotesize  \rotatebox{90}{Wavenumber [$10^{3}$ cm$^{-1}$]}}
    \put(-18.5,95.75){\footnotesize  \rotatebox{0}{$13.5$}}
    \put(-18.5,63){\footnotesize  \rotatebox{0}{$13.0$}}
    \put(-18.5,30.25){\footnotesize  \rotatebox{0}{$12.5$}}
    \put(-18.5,0){\footnotesize  \rotatebox{0}{$12.0$}}
\end{overpic}
\end{subfigure}%
\begin{subfigure}[t]{0.215\textwidth}
\centering
\begin{overpic}[width=0.95\textwidth]{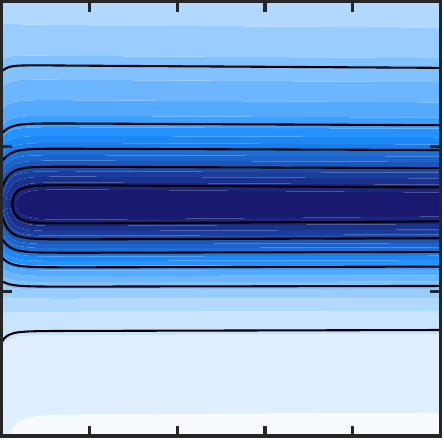}
    \put(-1,-8){\footnotesize  0}
  \put(18,-8){\footnotesize  1}
  \put(38,-8){\footnotesize  2}
  \put(58,-8){\footnotesize 3}
  \put(78,-8){\footnotesize  4}
  \put(96,-8){\footnotesize  5}
  \put(33,-18.5){\footnotesize  Time [ps]}
\end{overpic}
\end{subfigure}%
\begin{subfigure}[t]{0.215\textwidth}
\centering
\begin{overpic}[width=0.95\textwidth]{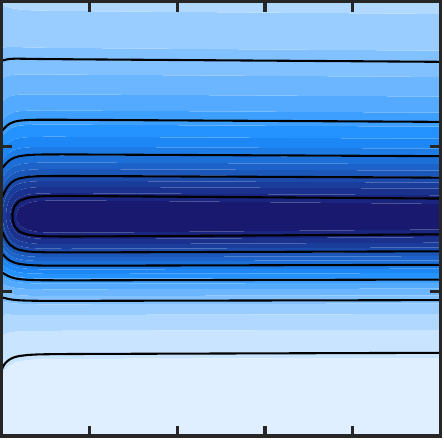}
    \put(-1,-8){\footnotesize  0}
  \put(18,-8){\footnotesize  1}
  \put(38,-8){\footnotesize  2}
  \put(58,-8){\footnotesize 3}
  \put(78,-8){\footnotesize  4}
  \put(96,-8){\footnotesize  5}
  \put(33,-18.5){\footnotesize  Time [ps]}
\end{overpic}
\end{subfigure}%
\begin{subfigure}[t]{0.215\textwidth}
\centering
\begin{overpic}[width=0.95\textwidth]{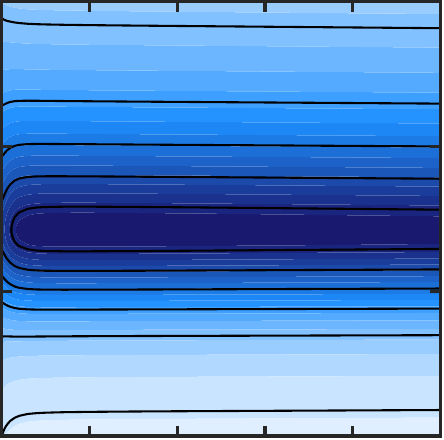}
    \put(-1,-8){\footnotesize  0}
  \put(18,-8){\footnotesize  1}
  \put(38,-8){\footnotesize  2}
  \put(58,-8){\footnotesize 3}
  \put(78,-8){\footnotesize  4}
  \put(96,-8){\footnotesize  5}
  \put(33,-18.5){\footnotesize  Time [ps]}
\end{overpic}
\end{subfigure}
\vspace{0.55cm}
}
\caption{\justifying Dependence of third- and fifth-order pump--probe spectra and its excitation pathways spectra on length of the chain. Narrowing of the lineshapes with increasing number of molecules can be observed. Center of main NESA$^{(5)}$ peak moves with incerasing N towards lower energies, approaching NSE${^{(5)}}$. As spectral features of ESA and SE share the same pump--probe spectral features with its fifth order counterparts. This   leads to disapperance of ESA$^{(5)}$ peak in third order pump--probe for N=10 and for decay of the main peak in third order pump--robe for N=20. }
\label{Fig:SQAB19_spectra_pathways}
\end{figure*}

\begin{figure*}[t!]
{\sffamily
\sansmath
\begin{subfigure}[t]{0.07\textwidth}
\begin{tikzpicture}
    \node [white] at (0.0,0) {$1$};
\end{tikzpicture}
\end{subfigure}%
\begin{subfigure}[t]{0.215\textwidth}
\centering
\begin{overpic}[width=0.95\textwidth]{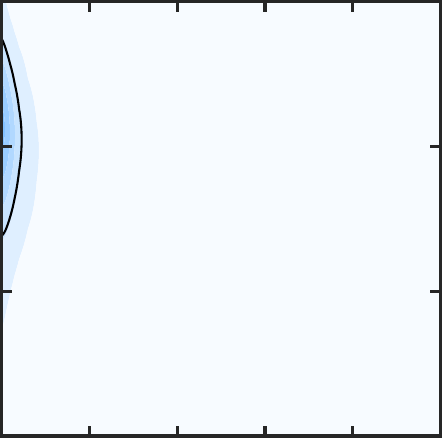}
  \put(5,103){\footnotesize $[SQA-SQB]_{2}\;\;(N=4)$}
    \put(-45,40){\footnotesize \rotatebox{90}{SE$_2^{(5)}$}}

      \put(-31,6){\footnotesize  \rotatebox{90}{Wavenumber [$10^{3}$ cm$^{-1}$]}}
    \put(-18.5,95.75){\footnotesize  \rotatebox{0}{$13.5$}}
    \put(-18.5,63){\footnotesize  \rotatebox{0}{$13.0$}}
    \put(-18.5,30.25){\footnotesize  \rotatebox{0}{$12.5$}}
    \put(-18.5,0){\footnotesize  \rotatebox{0}{$12.0$}}
\end{overpic}
\end{subfigure}%
\begin{subfigure}[t]{0.215\textwidth}
\centering
\begin{overpic}[width=0.95\textwidth]{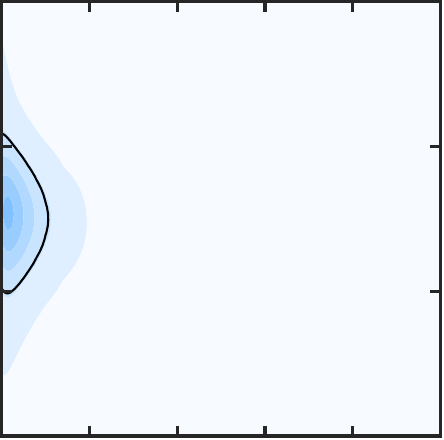}
  \put(5,103){\footnotesize $[SQA-SQB]_{4}\;\;(N=8)$}
\end{overpic}
\end{subfigure}%
\begin{subfigure}[t]{0.215\textwidth}
\centering
\begin{overpic}[width=0.95\textwidth]{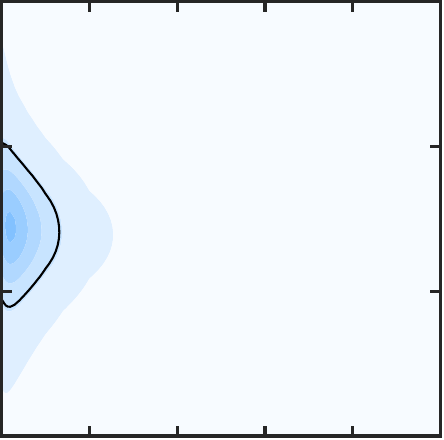}
  \put(5,103){\footnotesize $[SQA-SQB]_{5}\;\;(N=10)$}
\end{overpic}
\end{subfigure}%
\begin{subfigure}[t]{0.215\textwidth}
\centering
\begin{overpic}[width=0.95\textwidth]{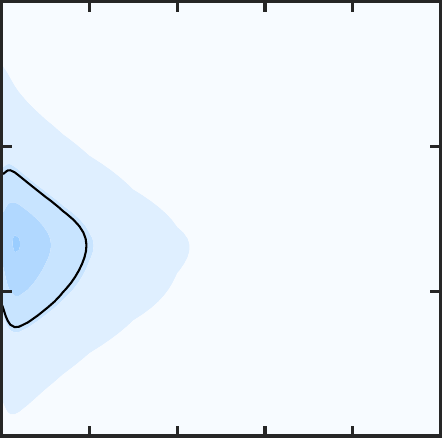}
  \put(5,103){\footnotesize $[SQA-SQB]_{10}\;\;(N=20)$}
\end{overpic}
\end{subfigure}
\vspace{0.2cm}

\begin{subfigure}[t]{0.07\textwidth}
\begin{tikzpicture}
    \node [white] at (0.0,0) {$1$};
\end{tikzpicture}
\end{subfigure}%
\begin{subfigure}[t]{0.215\textwidth}
\centering
\begin{overpic}[width=0.95\textwidth]{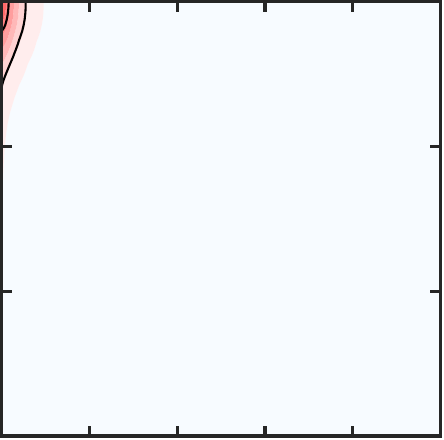}
  \put(-45,40){\footnotesize \rotatebox{90}{ESA$_2^{(5)}$}}

      \put(-31,6){\footnotesize  \rotatebox{90}{Wavenumber [$10^{3}$ cm$^{-1}$]}}
    \put(-18.5,95.75){\footnotesize  \rotatebox{0}{$13.5$}}
    \put(-18.5,63){\footnotesize  \rotatebox{0}{$13.0$}}
    \put(-18.5,30.25){\footnotesize  \rotatebox{0}{$12.5$}}
    \put(-18.5,0){\footnotesize  \rotatebox{0}{$12.0$}}
\end{overpic}
\end{subfigure}%
\begin{subfigure}[t]{0.215\textwidth}
\centering
\begin{overpic}[width=0.95\textwidth]{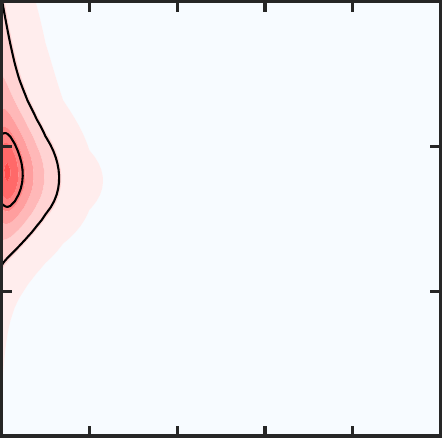}
\end{overpic}
\end{subfigure}%
\begin{subfigure}[t]{0.215\textwidth}
\centering
\begin{overpic}[width=0.95\textwidth]{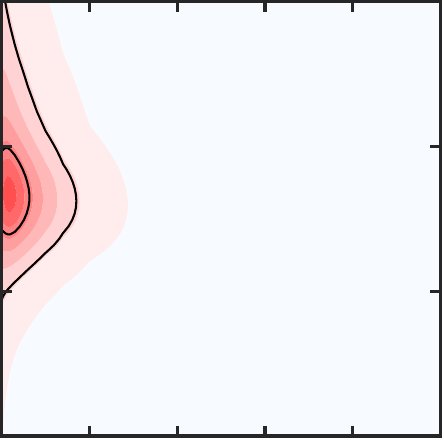}
\end{overpic}
\end{subfigure}%
\begin{subfigure}[t]{0.215\textwidth}
\centering
\begin{overpic}[width=0.95\textwidth]{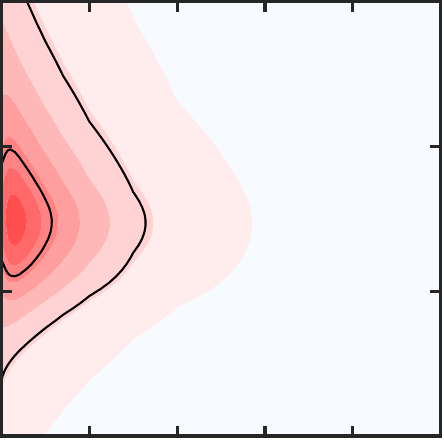}
\end{overpic}
\end{subfigure}
\vspace{0.2cm}

\begin{subfigure}[t]{0.07\textwidth}
\begin{tikzpicture}
    \node [white] at (0.0,0) {$1$};
\end{tikzpicture}
\end{subfigure}%
\begin{subfigure}[t]{0.215\textwidth}
\centering
\begin{overpic}[width=0.95\textwidth]{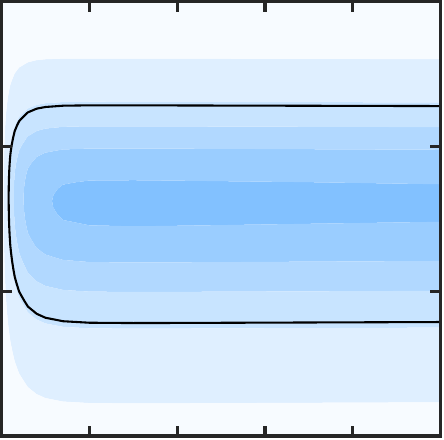}
  \put(-45,35){\footnotesize \rotatebox{90}{SE$_2^{(5)}$ (EEA)}}
      \put(-31,6){\footnotesize  \rotatebox{90}{Wavenumber [$10^{3}$ cm$^{-1}$]}}
    \put(-18.5,95.75){\footnotesize  \rotatebox{0}{$13.5$}}
    \put(-18.5,63){\footnotesize  \rotatebox{0}{$13.0$}}
    \put(-18.5,30.25){\footnotesize  \rotatebox{0}{$12.5$}}
    \put(-18.5,0){\footnotesize  \rotatebox{0}{$12.0$}}
\end{overpic}
\end{subfigure}%
\begin{subfigure}[t]{0.215\textwidth}
\centering
\begin{overpic}[width=0.95\textwidth]{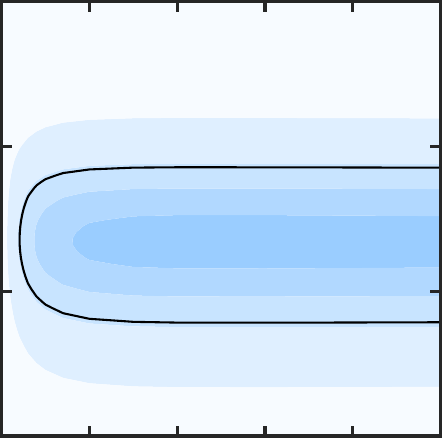}
\end{overpic}
\end{subfigure}%
\begin{subfigure}[t]{0.215\textwidth}
\centering
\begin{overpic}[width=0.95\textwidth]{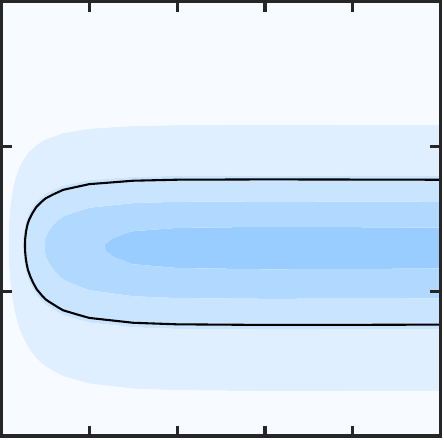}
\end{overpic}
\end{subfigure}%
\begin{subfigure}[t]{0.215\textwidth}
\centering
\begin{overpic}[width=0.95\textwidth]{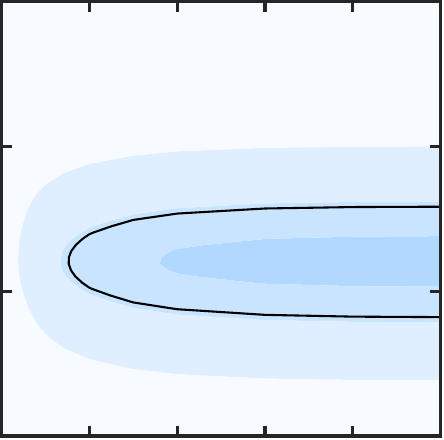}
 \put(101.,45.5){\includegraphics[width=0.0535\textwidth]{pictures/colorbar_cropped-2.pdf}}
  \put(111,159){\footnotesize   1.0}
  \put(111,130){\footnotesize   0.5}
  \put(111,101){\footnotesize   0.0}
  \put(108,73){\footnotesize  -0.5}
  \put(108,45){\footnotesize  -1.0}
\end{overpic}
\end{subfigure}
\vspace{0.2cm}

\begin{subfigure}[t]{0.07\textwidth}
\begin{tikzpicture}
    \node [white] at (0.0,0) {$1$};
\end{tikzpicture}
\end{subfigure}%
\begin{subfigure}[t]{0.215\textwidth}
\centering
\begin{overpic}[width=0.95\textwidth]{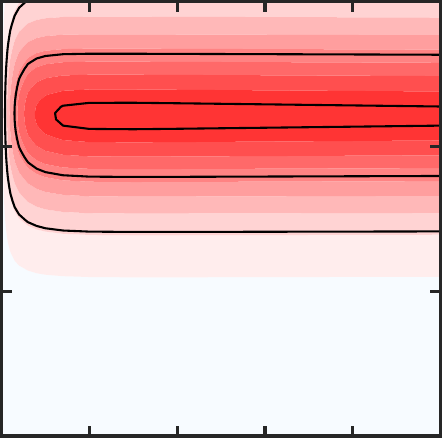}
  \put(-45,35){\footnotesize \rotatebox{90}{ESA$_2^{(5)}$ (EEA)}}
    \put(-1,-8){\footnotesize  0}
  \put(18,-8){\footnotesize  1}
  \put(38,-8){\footnotesize  2}
  \put(58,-8){\footnotesize 3}
  \put(78,-8){\footnotesize  4}
  \put(96,-8){\footnotesize  5}
  \put(33,-18.5){\footnotesize  Time [ps]}
      \put(-31,6){\footnotesize  \rotatebox{90}{Wavenumber [$10^{3}$ cm$^{-1}$]}}
    \put(-18.5,95.75){\footnotesize  \rotatebox{0}{$13.5$}}
    \put(-18.5,63){\footnotesize  \rotatebox{0}{$13.0$}}
    \put(-18.5,30.25){\footnotesize  \rotatebox{0}{$12.5$}}
    \put(-18.5,0){\footnotesize  \rotatebox{0}{$12.0$}}
\end{overpic}
\end{subfigure}%
\begin{subfigure}[t]{0.215\textwidth}
\centering
\begin{overpic}[width=0.95\textwidth]{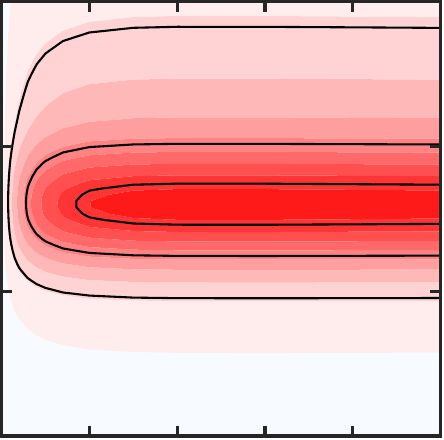}
    \put(-1,-8){\footnotesize  0}
  \put(18,-8){\footnotesize  1}
  \put(38,-8){\footnotesize  2}
  \put(58,-8){\footnotesize 3}
  \put(78,-8){\footnotesize  4}
  \put(96,-8){\footnotesize  5}
  \put(33,-18.5){\footnotesize  Time [ps]}
\end{overpic}
\end{subfigure}%
\begin{subfigure}[t]{0.215\textwidth}
\centering
\begin{overpic}[width=0.95\textwidth]{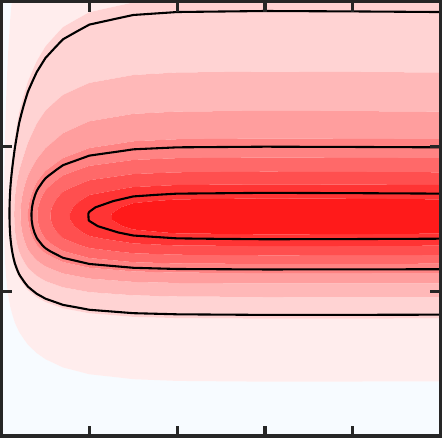}
    \put(-1,-8){\footnotesize  0}
  \put(18,-8){\footnotesize  1}
  \put(38,-8){\footnotesize  2}
  \put(58,-8){\footnotesize 3}
  \put(78,-8){\footnotesize  4}
  \put(96,-8){\footnotesize  5}
  \put(33,-18.5){\footnotesize  Time [ps]}
\end{overpic}
\end{subfigure}%
\begin{subfigure}[t]{0.215\textwidth}
\centering
\begin{overpic}[width=0.95\textwidth]{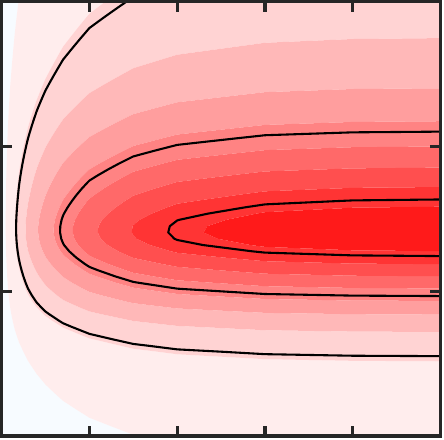}
    \put(-1,-8){\footnotesize  0}
  \put(18,-8){\footnotesize  1}
  \put(38,-8){\footnotesize  2}
  \put(58,-8){\footnotesize 3}
  \put(78,-8){\footnotesize  4}
  \put(96,-8){\footnotesize  5}
  \put(33,-18.5){\footnotesize  Time [ps]}
\end{overpic}
\end{subfigure}
\vspace{0.55cm}
}
\caption{\justifying Dependence of fifth-order pump--probe excitation pathways spectra on length of SQA-SQB chain.}
\label{Fig:SQAB19_spectra_pathways_2E}
\end{figure*}

\begin{figure*}[t!]
{\sffamily
\sansmath
\begin{subfigure}[t]{0.07\textwidth}
\begin{tikzpicture}
    \node [white] at (0.0,0) {$1$};
\end{tikzpicture}
\end{subfigure}%
\begin{subfigure}[t]{0.215\textwidth}
\centering
\begin{overpic}[width=0.95\textwidth]{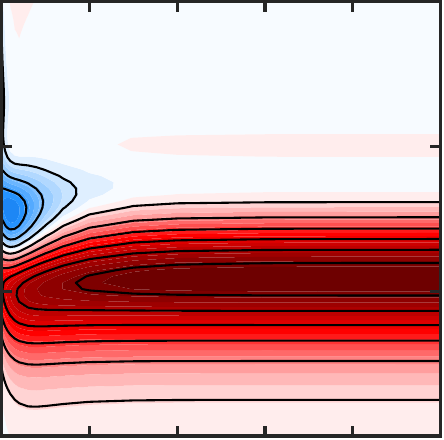}
  \put(15,103){\footnotesize lifetime deph. $2\Gamma_{\beta \xi}$ }
    \put(-1,-8){\footnotesize  0}
  \put(18,-8){\footnotesize  1}
  \put(38,-8){\footnotesize  2}
  \put(58,-8){\footnotesize 3}
  \put(78,-8){\footnotesize  4}
  \put(96,-8){\footnotesize  5}
  \put(33,-18.5){\footnotesize  Time [ps]}
      \put(-31,6){\footnotesize  \rotatebox{90}{Wavenumber [$10^{3}$ cm$^{-1}$]}}
    \put(-18.5,95.75){\footnotesize  \rotatebox{0}{$13.5$}}
    \put(-18.5,63){\footnotesize  \rotatebox{0}{$13.0$}}
    \put(-18.5,30.25){\footnotesize  \rotatebox{0}{$12.5$}}
    \put(-18.5,0){\footnotesize  \rotatebox{0}{$12.0$}}
         \put(1,120){\footnotesize  a}
\end{overpic}
\label{Fig:SQAB_5PP_SE2}
\end{subfigure}%
\begin{subfigure}[t]{0.215\textwidth}
\centering
\begin{overpic}[width=0.95\textwidth]{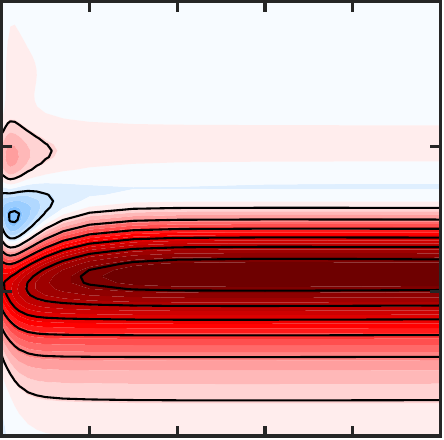}
  \put(15,103){\footnotesize lifetime deph. $\Gamma_{\beta \xi}$ }
    \put(-1,-8){\footnotesize  0}
  \put(18,-8){\footnotesize  1}
  \put(38,-8){\footnotesize  2}
  \put(58,-8){\footnotesize 3}
  \put(78,-8){\footnotesize  4}
  \put(96,-8){\footnotesize  5}
  \put(33,-18.5){\footnotesize  Time [ps]}
       \put(1,120){\footnotesize  b}
\end{overpic}
\label{Fig:SQAB_5PP_SE_EEA}
\end{subfigure}%
\begin{subfigure}[t]{0.215\textwidth}
\centering
\begin{overpic}[width=0.95\textwidth]{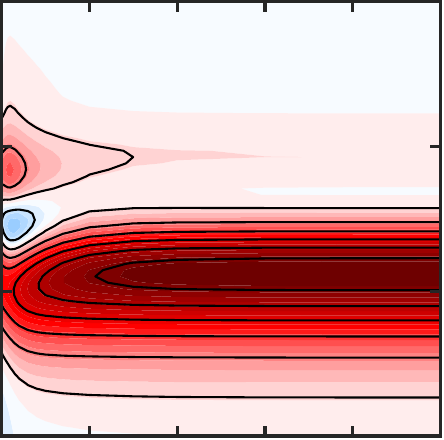}
  \put(10,103){\footnotesize lifetime deph. $0.75\Gamma_{\beta \xi}$ }
    \put(-1,-8){\footnotesize  0}
  \put(18,-8){\footnotesize  1}
  \put(38,-8){\footnotesize  2}
  \put(58,-8){\footnotesize 3}
  \put(78,-8){\footnotesize  4}
  \put(96,-8){\footnotesize  5}
  \put(33,-18.5){\footnotesize  Time [ps]}
       \put(1,120){\footnotesize  c}
        \put(101.,-1){\includegraphics[width=0.047\textwidth]{pictures/colorbar_cropped-2.pdf}}
  \put(111,97){\footnotesize   1.0}
  \put(111,73){\footnotesize   0.5}
  \put(111,47){\footnotesize   0.0}
  \put(108,21){\footnotesize  -0.5}
  \put(108,-3){\footnotesize  -1.0}
\end{overpic}
\label{Fig:SQAB_5PP_ESA2}
\end{subfigure}
\vspace{0.55cm}
}
\caption{\justifying Dependence of fifth-order pump--probe on lifetime dephasing of $|\xi\rangle\langle\beta|$ coherence in ESA$_2^{(5)}$(2E) excitation pathway. Simulation for $N=10$ using parameters from set 1.}
\label{Fig:SQAB19_spectra_pop_deph}
\end{figure*}

To understand the cause of peak narrowing, we investigate the dependence of the third- and fifth-order pump--probe spectra and the individual excitation pathways on the polymer length (Fig. \ref{Fig:SQAB19_spectra_pathways} and \ref{Fig:SQAB19_spectra_pathways_2E}), illustrating the effects of exciton delocalization and pathway interference, which both lead to narrowing of spectral peaks in the total spectrum. 
Finally, we examine the influence of the lifetime dephasing of the $|\xi\rangle\langle\beta|$ coherence on the simulated fifth-order spectra (Fig. \ref{Fig:SQAB19_spectra_pop_deph}) for set 1, $N=10$. Two times bigger lifetime dephasing $\Gamma_{\xi\beta}$ leads to suppression of ESA$_2^{(5)}$(2E) peak enhancing negative SE$_2^{(5)}$(2E) peak (Fig. \ref{Fig:SQAB19_spectra_pop_deph}a) compared to reference situation in Fig. \ref{Fig:SQAB19_spectra_pop_deph}b, where the lifetime dephasing is estimated as $\Gamma_{\xi\beta}=\frac{k_{\xi\xi}+k_{\beta\beta}}{2}$. On the contrary for smaller dephasing of $|\xi\rangle\langle\beta|$ coherence the ESA$_2^{(5)}$(2E) peak is significantly sharper and almost cancels SE$_2^{(5)}$(2E) contribution (Fig.\ref{Fig:SQAB19_spectra_pop_deph}c). We observed similar behavior for simulation with parameter set 2 and $N=20$, see Fig. 12 in the main text.

\begin{table}[htbp]
\centering
\renewcommand{\arraystretch}{1.4} 
\begin{tabular}{|c|c|c|c|}
\hline
parameter & set 1 & set 2 & set 3 \\
\hline 
    length of squaraine monomer unit &  \multicolumn{3}{c|}{1.25}\\
    \hline
    energy of first excited state of molecule SQA $E_\text{SQA}$ & \multicolumn{3}{c|}{$13500\text{ cm}^{-1}$} \\
    \hline
    energy of first excited state of molecule SQB $E_\text{SQB}$ & \multicolumn{3}{c|}{$E_\text{SQA} + 1200\text{ cm}^{-1}$} \\
    \hline
    standard deviation of $\delta E_\text{SQA}$ & \multicolumn{3}{c|}{$148.8\text{ cm}^{-1}$} \\
    \hline
    standard deviation of $\delta E_\text{SQB}$ & \multicolumn{3}{c|}{$240.0\text{ cm}^{-1}$} \\
    \hline
    coefficient for transition dipole moment $\mu_i$ & \multicolumn{3}{c|}{1.37} \\
    \hline
    coefficient for transition dipole moment $\mu_j$ & \multicolumn{3}{c|}{1} \\
    \hline
    coupling to the bath for SQA $\nu_A$ & \multicolumn{3}{c|}{1.25} \\
    \hline
    coupling to the bath for SQB $\nu_B$ & \multicolumn{3}{c|}{1.67} \\
    \hline
    energy of higher excited state f & \multicolumn{3}{c|}{$E_n+\frac{1}{2}\Delta E$} \\
    \hline
    $F_\mu$ & \multicolumn{3}{c|}{0.33} \\
    \hline
    $\Phi$ & 1.5 & 1.5 & 8.0 \\
    \hline
    $\Phi_{\text{corr}}$ & \multicolumn{3}{c|}{0.1} \\
    \hline
    coefficient for coupling between ee and f states $F_J$ & 1.25 & 1.5 & 1.25 \\
    \hline
    rate of internal conversion ($K_\text{IC}$) & $\frac{1}{30\text{ fs}}$ & $\frac{1}{55\text{ fs}}$& $\frac{1}{20\text{ fs}}$\\
    \hline
    decay of first excited state & \multicolumn{3}{c|}{$0.73e^{-0.0007 T}+0.27e^{-0.015 T}$} \\
\hline
\end{tabular}
\caption{Three different sets of parameters used in simulations of SQA-SQB co-polymer.}
\label{Tab:parameters SQAB19}
\end{table}


\clearpage
\bibliography{si_references}